\documentclass[12pt,preprint,amssymb]{aastex}
\pdfoutput=1
\usepackage{graphicx}
\usepackage[figuresright]{rotating}

\def\asca{{\em ASCA}\/}

\shorttitle{Probe Galaxy-ICM Interaction}
\shortauthors{Gu et al.}

\begin{document}
\title{Probing of the Interactions Between the Hot Plasmas and
  Galaxies in Clusters from $z=0.1$ to $0.9$}

\author {Liyi Gu\altaffilmark{1,2}, Poshak Gandhi\altaffilmark{3}, Naohisa Inada\altaffilmark{4}, Madoka Kawaharada\altaffilmark{3}, Tadayuki Kodama\altaffilmark{5}, Saori Konami\altaffilmark{6, 7}, Kazuhiro Nakazawa\altaffilmark{1}, Kazuhiro Shimasaku\altaffilmark{8}, Haiguang Xu\altaffilmark{2}, and Kazuo Makishima\altaffilmark{1, 9, 10}}

\altaffiltext{1}{Department of Physics, The University of Tokyo, 7-3-1
  Hongo, Bunkyo-ku, Tokyo 113-0011, Japan}
\altaffiltext{2}{Department of Physics, Shanghai Jiao Tong University, 800 Dongchuan Road, Shanghai 200240, PRC}
\altaffiltext{3}{Institute of Space and Astronautical Science (ISAS), Japan Aerospace Exploration Agency, 3-1-1 Yoshinodai, Chuo-ku,
Sagamihara, Kanagawa 252-5210}
\altaffiltext{4}{Department of Physics, Nara National College of Technology, Yamatokohriyama, Nara 639-1080, Japan}
\altaffiltext{5}{National Astronomical Observatory of Japan, Mitaka, Tokyo 181-8588, Japan}
\altaffiltext{6}{Department of Physics, Tokyo University of Science, 1-3 Kagurazaka, Shinjuku-ku, Tokyo 162-8601}
\altaffiltext{7}{High Energy Astrophysics Laboratory, Institute of Physical and Chemical Research, 2-1 Hirosawa, Wako, Saitama
351-0198}
\altaffiltext{8}{Department of Astronomy, Graduate School of Science, The University of Tokyo, 7-3-1 Hongo, Bunkyo-ku, Tokyo 113-0033, Japan}
\altaffiltext{9}{Research Center for the Early Universe, School of Science, The University of Tokyo, 7-3-1 Hongo, Bunkyo-ku, Tokyo 113-0033, Japan}
\altaffiltext{10}{MAXI Team, Institute of Physical and Chemical Research, 2-1 Hirosawa, Wako, Saitama 351-0198}

\begin{abstract}
Based on optical and X-ray data for a sample of 34 relaxed rich
clusters of galaxies with redshifts of $0.1-0.9$, we studied relative spatial distributions of
the two major baryon contents, the cluster galaxies and the hot
plasmas. Using multi-band photometric
data taken with the UH88 telescope, we determined the integrated (two dimensional) radial light profiles of member
galaxies in each cluster using two independent approaches, i.e.,
the background subtraction and the color-magnitude filtering. The ICM mass
profile of each cluster in our sample, also integrated in two dimensions, was derived from a
spatially-resolved spectral analysis using {\it XMM-Newton} and {\it
  Chandra} data. Then, the radially-integrated light profile of each cluster was divided by its ICM mass profile, to obtain a profile of 
``galaxy light vs. ICM mass ratio''. When the sample is divided into three subsamples with
redshift intervals of $z=0.11-0.22$, $0.22-0.45$, and $0.45-0.89$, the ratio profiles over the central
$0.65$ $R_{500}$ regions were found to
steepen from the higher- to lower- redshift subsamples, meaning that
the galaxies become more concentrated in the ICM sphere towards lower
redshifts. A K-S test indicates that this evolution in the cluster
structure is significant  
on $\geq 94$\% confidence level. The evolution is also seen in galaxy number vs. 
ICM mass ratio profiles. A range of systematic uncertainties
in the galaxy light measurements, as well as many radius-/redshift-
dependent biases to the galaxy vs. ICM profiles have been assessed, but none of 
them is significant against the observed evolution. Besides, the
galaxy light vs. total mass ratio profiles also exhibit gradual concentration
towards lower redshift. We interpret in the context that the 
galaxies, the ICM, and the dark matter components followed a similar spatial
distribution in the early phase ($z>0.5$), while the galaxies have fallen
towards the center relative to the others at a later phase. Such galaxy infall is likely to be caused by
the drag exerted by the ICM to the galaxies as they move through the
ICM and interact with it, while gravitational drag can
enhance the infall of the most massive galaxies.  
\end{abstract}
\section{INTRODUCTION}
The X-ray emitting hot plasma in galaxy clusters, or the so called
intracluster medium (ICM), with temperature of $T_{\rm ICM} = 2-10$ keV and number
density of $10^{-4}-10^{-2}$ cm$^{-3}$, constitutes about $80-90$\% of
the detected baryon content in clusters. The remaining $10-20$\% resides in member
galaxies, which move through the ICM with transonic speeds, $v \sim
\sqrt{G M/R} \sim \sqrt{kT_{\rm ICM}/\mu m_{\rm p}} \sim 1000$ km
s$^{-1}$, where $G$, $M$, $R$, $\mu$, and $m_{\rm p}$ are the
gravitational constant, cluster mass, radius of the galaxy orbit, mean
molecular weight and the proton mass, respectively. Since two-body
relaxation between dark matter halo and member galaxies occurs on timescales
much longer than the time for a galaxy to cross the cluster (Sarazin
1988), the ICM-filled cluster volume is considered to be rather transparent to the moving
galaxies from a pure gravitational point of view. From a fluid
mechanics/plasma physics 
point of view, however, the member galaxies, which have their own
plasmas (so called inter-stellar medium; ISM),  are not totally
collisionless with the ICM. Given the gas density and relative velocity,
the galaxies have a cross section large enough for the ICM to be
dragged significantly via, e.g., ram pressure (Gunn \& Gott 1972; see
also the review by Roediger 2009) and/or magnetohydrodynamical (MHD)
effects (Asai et al. 2005, Makishima et al. 2001). Recent observations
have revealed filamentary structures in HI associated with the moving
galaxies (e.g., Oosterloo \& van Gorkom 2005), in H$\alpha$
(e.g., Yoshida et al. 2008), and also in X-ray images (e.g. Sun et
al. 2006), which hint for such galaxy-ICM interactions. These urge us
to study whether such interactions proceed efficiently in
general or not, and how such interactions affect the spatial
extents of the two baryon components.

When discussing the radial extents of the baryons, we must consider the
following three important observational facts that are found among
nearby clusters. (1) The stellar component is spatially much more
concentrated than the ICM and the dark matter (e.g., Bahcall 1999). (2) On the central
several tens to hundreds kpc scales, the ICM is considerably metal
enriched, but the metals in the ICM are still much more extended than
the stellar mass which must have produced them (e.g., Kawaharada et
al. 2009; Sato et al. 2009). (3) In outer ($R \sim 2$ Mpc) regions of
the Perseus cluster, the ICM metallicity remains relatively constant
at $\approx 0.3 Z_{\odot}$ (Simionescu et al. 2011). One possible explanation for (1) would be
to assume that galaxies were formed in the past predominantly in core
regions of the ICM and dark matter spheres. However, neither (2) nor
(3) would be readily explained, unless assuming an extensive
wind-driven outflow of
metal-enriched gas at an expense of enormous energies ($> 10^{60}$
erg; Kawaharada et al. 2009), or mechanical transport by active
galactic nuclei (AGNs) of which the efficiency is still controversial
(e.g., Vernaleo \& Reynolds 2006). Alternatively, (1) alone could be
explained if the ICM sphere grew by accretion of primordial gas
after the initial galaxy formation was almost completed. However, this
would also result in the same difficulty in explaining (2) and
(3). Instead, the three observed facts can be consistently explained
by the galaxy infall scenario proposed by Makishima et
al. (2001). That is, the galaxies have been falling to the cluster
center, presumably via galaxy-ICM interactions, while ejecting metals
to the intra-cluster volume.   


The present research on this subject has yet another interesting aspect,
namely, ICM heating. 
The high ICM density ($n_{\rm ICM} \sim 10^{-2}$ cm$^{-3}$) in cluster centers leads to significant energy
losses by X-ray radiation, which can cool the gas down on a very short ($\ll
1/H_{0}$) timescale. However, broad-band X-ray imaging spectroscopy,
starting with {\it ASCA}, found that the
effect of cooling is much weaker than previously predicted (Makishima
et al. 2001; Tamura et al. 2001; Peterson et al. 2001; Xu et
al. 2002), suggesting that some significant heating mechanisms are in
operation. Current solutions include energy injection from the central
AGN (e.g., Churazov et al. 2001), heat inflow by thermal
conduction (e.g., Zakamska \& Narayan 2003), and turbulence and
dynamical friction driven by galaxy motion (Kim \& Narayan 2003;
El-Zant et al. 2004). The numerical results of Ruszkowski \& Oh (2011)
indicate that the interaction between galaxies and ICM may give rise to
subsonic turbulence in the ICM at a velocity of $100-200$ km s$^{-1}$,
and the turbulence can in turn boost heat transfer towards the cluster
center so as to maintain the thermal stability of the ICM halo. Such a turbulent
model gains observational support from the power spectrum analysis of
gas pressure (Schuecker et al. 2004) and Faraday rotation (Vogt \& Ensslin
2005) maps. At the same time, the galaxies, which empower
the turbulent heating, should 
gradually lose their kinetic energy to the 
ICM and fall to the bottom of cluster potential (Makishima et al. 2001)

From the above pieces of evidence, it is suggested that the galaxies used to
be distributed more widely to the edge of the ICM sphere (and the
cluster potential), and have been falling to the cluster center,
presumably over the Hubble time, as they are dragged by the ICM
component. Furthermore, as shown in, e.g., Fujita \& Nagashima (1999) and Quilis et
al. (2000), such galaxy-ICM interactions would induce a burst of star
formation in the onset of galaxy infall, and subsequently strip out most of
the remaining cool disk gas. Hence, the idea of galaxy infall may also shed light on several
observed phenomena of cluster galaxies, e.g., the truncated cool disk gas profiles and severe
reduction of star formation in the center of the Virgo cluster (Koopmann et al. 2001; Koopmann \& 
Kenney 2004), and the positive dependence of blue galaxy fraction on both the
clustocentric distance (e.g., Whitmore et al. 1993) and redshift (e.g., Butcher
\& Oemler 1978).

By stacking the SDSS DR7 data of over 20000
optically-selected groups and clusters, Budzynski et al. (2012) recently
reported that the galaxy density profile  
does not vary in shape over $z=0.15-0.4$. However, by analyzing the
spectroscopic and photometric data of 15 X-ray bright rich clusters at $z=0.18-0.55$, Ellingson
et al. (2001) showed an indication that the galaxy distribution gets
more centrally-concentrated in lower redshifts (see their
Fig. 9). Since the sample in the former study is dominated by groups and
poor clusters (see their Fig. 1), where the hot gas density is much lower than 
that in the rich objects studied in the latter work, the difference in
these two results possibly hints for galaxy infall dragged by the ICM. So far no
systematic study has been made to compare directly the spatial extents of the
galaxies and ICM for different redshifts, so that
the role of galaxy-ICM interaction still remains rather unexplored. To
address this issue, we jointly analyze in the present work high
quality optical and X-ray data, and study the relative spatial
distributions of 
the stellar component vs. the ICM component, for a sample of 34 X-ray bright massive
galaxy clusters with a redshift range of $z=0.1$ to 0.9.

The layout of this paper is as follows. Section 2 gives a brief
description of the sample selection and data
reduction procedure. The data analysis and results are described in \S 3. We discuss the
physical implication of our results in \S 4, and summarize our work in \S5. Throughout the
paper we assume a Hubble constant of $H_0=71$$h_{71}$ km s$^{-1}$
Mpc$^{-1}$, a flat universe with the cosmological parameters of
$\Omega_M=0.27$ and $\Omega_\Lambda=0.73$, and quote errors by the 68\%
confidence level unless stated otherwise. The optical magnitudes used
in this paper are all given in the Vega-magnitude system.






\section{Observation and Data Preparation}

\subsection{Sample Selection}

Our sample was primarily constructed based on previous X-ray flux-limited samples
presented in Croston et al. (2008; 31 clusters), Leccardi \& Molendi
(2008; 48 clusters), and Ettori et al. (2004; 28 clusters), which
contribute to the redshift ranges of $0.05-0.2$, $0.1-0.4$, and
$0.4-1.3$, respectively. All the three samples were originally constructed to study the ICM properties
(e.g., density, temperature) of X-ray bright clusters using high quality {\it Chandra} and {\it XMM-Newton} 
data. This merged primary sample, comprising 107 ($=31
+48+28$) clusters, was further filtered through the following
criteria. First, to study clusters with sufficient member galaxies and
similar extent, the 
sample clusters should be massive and similar in scale. The 107 clusters were for this purpose filtered by 
their $M_{500}$ and $R_{500}$, where $R_{500}$ is defined as a radius corresponding to a
mean overdensity $\Delta = 500$ of the critical density at the cluster
redshift, and $M_{500}$ is the total mass enclosed therein. To account for 
cosmic growth of the cluster halo, we further calculated the expected mass and scale for each cluster 
after experiencing evolution down to $z=0$, i.e., $M_{500}^{z=0}$ and $R_{500}^{z=0}$, respectively,
with a redshift-dependent factor derived from a $\Lambda$CDM 
numerical simulation of Wechsler et al. (2002; see actual form of the factor in \S3.5.2). Systems with 
$M_{500}^{z=0} > 2\times 10^{14}$ $M_{\rm \odot}$ and with $R_{500}^{z=0}$ in the range of $1000-2000$ kpc
were selected. This reduced the number of candidates to 98. Second, to avoid
strong merging clusters, the positions of the X-ray brightness 
peak and the central dominant galaxy are required to coincide
with each other within 50 kpc, and the X-ray morphology should be
approximately symmetric. By examining the archival optical (DSS) and X-ray
({\it XMM-Newton} and {\it Chandra}) images of the 98 candidates, we
finally selected 34 clusters that also survive the second criterion. Basic
information for the 34 clusters are shown in Table 1. The sample was
further categorized into three subsamples by their redshifts, i.e., 
low-redshift subsample ($z \sim 0.11-0.22$, hereafter subsample L),
intermediate-redshift subsample ($z \sim 0.22-0.45$, hereafter
subsample M), and high-redshift subsample ($z\sim0.45-0.89$, hereafter
subsample H).     

The most important strategy of our study is to compare the X-ray and optical 
profiles of each cluster in the sample, to suppress the effects of intrinsic scatter 
in the cluster richness. As described in \S 2.2, the X-ray data for this purpose are
readily available in archives. A more difficult part is the optical information, because
we need to determine the galaxy membership in each cluster. For this purpose, however,  
existing archival optical databases (e.g., DSS, SDSS, 2MASS) either cannot fully
cover our sample, or are not deep enough for accurate galaxy light
measurements, especially for high redshift
clusters. Hence, we utilized the University of 
Hawaii 88-inch (UH88) telescope to observe the 34 clusters with deep optical
photometric data (PI: Inada). The UH88 telescope is
suited to our study, since its field of view, $7^{\prime}.5 \times
7^{\prime}.5$, can cover, even in the nearest objects in our sample, the central $\geq 0.5$ Mpc regions where the
galaxy-ICM interactions are mostly expected to take place. Details of the
observations are given in \S2.3.

\subsection{X-ray Data}

To achieve relatively homogeneous spatial resolutions for low redshift and high
redshift clusters, we use {\it XMM-Newton} data for $z<0.5$ and {\it
  Chandra} data for $z>0.5$ clusters, since {\it Chandra} has a
narrower point-spread-function than {\it XMM-Newton} by an
order of magnitude. As shown in Table 2, all the necessary X-ray data are available in the
archive. 

\subsubsection{\it Chandra}

We used the newest CIAO v4.4 and CALDB v4.4.7 for screening the {\it
  Chandra} data obtained with its advanced CCD imaging spectrometer
(ACIS). First, the bad pixels and columns were removed, as well as
events with \asca\ grades 1, 5, and 7. We then executed the gain, CTI,
and astrometry corrections. By
examining $0.3-10.0$ keV lightcurves extracted from source free
regions near the CCD edges (e.g., Gu et al. 2009), we removed the time
intervals contaminated by occasional background flares with count rate 
$>$$20$\% of the mean value. When available, ACIS-S1
data were also used to crosscheck the determination of the
contaminated intervals. As shown in Table 2, the exposure time was
reduced by $12-14$ ks for MS2053 and RXJ1334, and by $1-3$ ks for
the other clusters. In the ACIS spectral analysis, bright point
sources were detected and removed with the CIAO tool {\tt
  celldetect}. The spectral ancillary response files and
redistribution matrix files were calculated with the CIAO tools {\tt
  mkwarf} and {\tt mkacisrmf}, respectively.

Following Gu et al. (2012), the background was determined 
based on fitting a spectrum extracted from an off-center region, which is $\geq 2$ Mpc away
from the cluster center, with co-addition
of cosmic X-ray background (hereafter CXB), non-X-ray background, Galactic foreground,
and cluster emission components. Following, e.g., Markevitch et al. (2003), the non-X-ray background
spectra were obtained from stowed ACIS observations. After subtracting this component, 
we fitted the resulting spectrum with an absorbed power
law model (photon index $\Gamma = 1.4$) describing the CXB, an absorbed 
APEC model (temperature = 0.2 keV and abundance = 1 $Z\odot$) for the Galactic foreground,
and another absorbed APEC model for the cluster emission. The average $2.0-10.0$ keV CXB
flux was estimated to be $7.5 \times 10^{-8}$ ergs cm$^{-2}$ s$^{-1}$ sr$^{-1}$, which is consistent with 
the result of Kushino et al. (2002). The typical uncertainties on the CXB flux are $\sim 15$\% and $\sim 30$\%
for high-count and low-count datasets, respectively, which were accounted for in determining the error range of 
ICM mass profiles (see \S 3.2 for details).

\subsubsection{\it XMM-Newton}

Basic reduction and calibration of the {\it XMM-Newton} European
Photon Imaging Camera (EPIC) data were carried out with SAS
v11.0.1. In the screening process we set {\it FLAG} = 0,
and kept events with {\it PATTERNs} 0--12 for the MOS cameras and those
with {\it PATTERNs} 0--4
for the pn camera. By examining lightcurves extracted in
$10.0-14.0$ keV and $1.0-5.0$ keV from source free regions, we rejected time intervals
affected by hard- and soft-band flares, respectively, in
which the count rate exceeds a 2$\sigma$ limit above the quiescent
mean value (e.g., Katayama et al. 2004; Nevalainen et al. 2005).
The obtained MOS and pn exposures are shown in Table 2. Point sources
detected by a SAS tool {\tt edetect\_chain} were discarded in the spectral
analysis. The tool {\tt xissimarfgen} and {\tt xisrmfgen} were used to
calculate the ancillary responses and redistribution matrixes,
respectively. Like in the ACIS analysis, the EPIC background was
determined by fitting a CCD-edge region, which locates $\geq 1.5$ Mpc away from the cluster center, with combined CXB, 
non-X-ray background, Galactic foreground, and the cluster
emission components. The wheel-closed datasets were used as the
non-X-ray background. The average $2.0-10.0$ keV CXB flux was measured to 
be $6.3 \times 10^{-8}$ ergs cm$^{-2}$ s$^{-1}$ sr$^{-1}$, again 
in agreement with Kushino et al. (2002). The typical error of the CXB flux, $\sim 20$\%, 
was used for calculating the uncertainty on ICM mass profiles (see \S 3.2 for details).
Spectra of the three EPIC detectors, MOS1, MOS2, and pn, were fitted simultaneously with the same model, but leaving the model normalization free for each dataset.

\subsection{Optical Data}

As summarized in Table 3, optical images of our sample clusters were
obtained with the UH88 telescope on six nights 
in February, March, and August of 2010. The February and August
observations were carried out with the thinned Tek 2K CCD mounted at
the f/10 RC focus in $I$, $R$, and $B$, and in March the targets were observed
with the Wide Field Grism Spectrograph 2 (WFGS2) in
$i^{\prime}$, $r^{\prime}$, and $g^{\prime}$ imaging mode. Besides the central
pointing of each cluster as described above, we observed an offset
region in $I$ or $i^{\prime}$, which is about
$10^{\prime}-13^{\prime}$ ($1.5-5.1$ Mpc) away from the central
galaxy. The central and offset pointings were made with similar
observation depth. All images have a pixel scale of $0^{\prime\prime}.22$ $\rm
pixel^{-1}$. The exposure and limiting magnitude for each pointing are
given in Table 3.

The data were reduced with the IRAF software following the standard
procedure as described in Kodama et al. (2005). First, the images in
each band were debiased and flattened using a median of dome-flat
images in the same band taken before the cluster observation. For each
cluster the dithered images in the same band were shifted into
registration and combined after rejecting cosmic rays. As summarized in Table 3, the 
resulting images have a typical seeing of  $0^{\prime\prime}.5 -
1^{\prime\prime}.2$. Astrometry calibration
was performed using the USNO-A2.0 catalog (Monet 1998), and the
photometric zero points were calibrated based on the standard star
data of Landolt (1992) and Fukugita et al. (1995) in $B/R/I$ and
$g^{\prime}/r^{\prime}/i^{\prime}$ bands, respectively. According to
the transformation equations in
Lupton (2005), the $g^{\prime}$, $r^{\prime}$, and $i^{\prime}$ magnitudes were
converted into those in the $B$, $R$, and $I$ bands, respectively.

Object identifications and flux measurements were performed with the
Source Extractor algorithm (SExtractor; Bertin \& Arnouts 1996), and
sources with at least 10 connected pixels (comparable to the
spatial resolution of the observations $\approx 0^{\prime\prime}.7$),
of which the signal is higher than 3$\sigma$ above the contiguous background intensity, were
considered as a detection of an object. To separate stars and
galaxies, we constructed a star catalog based on the 
neural network method in SExtractor. All the objects with {\tt
  CLASS\_STAR} $>$ 0.6 measured by SExtractor were classified as stars
and were excluded from the subsequent analysis. The $I$-band images were
used in the star detection, and $B$- and $R$- band images were used for
crosscheck. The Kron-type total magnitudes and fluxes ({\tt MAG\_AUTO} and
{\tt FLUX\_AUTO}, respectively) were measured using automatic aperture
with the minimal value of $1^{\prime\prime}$ ($\sim 2-8$ kpc for different
redshifts). We inspected the aperture maps and confirmed the absence of strong
aperture overlap between neighboring objects. The galaxy color
was measured in the double image mode,
which ensures the same elliptical aperture between the different bands.

The calibrated apparent magnitude $m$ was transformed to absolute
magnitude $M$ by applying the relation,
\begin{equation}
M = m - 5 {\rm log_{10}} (D_{\rm L}/ 10 {\rm pc}) - A - K(z),
\end{equation}
where $D_{\rm L}$ is the luminosity distance, $A$ is the Galactic
extinction computed using the Galactic map of Schlegel et al. (1998),
and $K(z)$ is the K-correction. We used the band-dependent K-correction
by Fukugita et al. (1995) for early-type galaxies, which are assumed
to be the dominant population of the sample clusters.

\subsection{Determinations of $R_{500}$ and the Cluster Center}

To compensate for the differences in the cluster scale, we normalize all
the galaxy light and ICM mass profiles obtained in the subsequent
analysis to the scale of each cluster, which can be characterized by
$R_{500}$. To avoid possible systematic errors in the $R_{500}$
determinations among different authors (Table 1), we did not use the $R_{500}$
given in literature. Instead, $R_{500}$ was estimated based on an empirical relation
using the average ICM temperature $T_{\rm aver}$ as $R_{\rm 500} \approx 391 \times T_{\rm aver}^{0.63} \times E(z)^{-1} $ (Finoguenov et al. 2001), where $E(z) = H(z)/H_{0}$ is
the cosmological evolution factor. $T_{\rm aver}$ was calculated based on spectral fitting of 
the cluster central 500 kpc region X-ray data, but excluding the innermost
100 kpc to avoid the temperature affected by a cool ICM component that is often found 
therein. Specifically, the background-corrected EPIC/ACIS spectra
were fit with an absorbed, single-temperature APEC model in $0.7-8.0$
keV,  with the ICM temperature and abundance set free. The
obtained $T_{\rm aver}$ and $R_{\rm 500}$, listed in Table 4, agree
with the previous results quoted in Table 1 within 25\% and 20\%, respectively. 
We have tested another $R_{500}-T_{\rm aver}$ relation presented in Zhang et al. (2008),
which was obtained with a combined weak lensing and X-ray study. The $R_{500}$ estimated 
from the two relations are consistent within $\leq 8$\%.

Next, we determined the cluster center by comparing three types of
centers, i.e., optical peak of the brightest galaxy (Center I), X-ray brightness peak
(Center II), and centroid of the cluster X-ray surface brightness over a central
region (Center III). To calculate Center III, we  
applied a CIAO task {\tt dmstat} on a $r_1 = 5^{\prime}$ region centered on the X-ray peak. 
After the centroid was derived, we started an iteration based on the
new centroid. To approach the cluster center region, the new
examining radius was set to be $r_{i+1} = r_i - 1^{\prime}$. The
iterations were regarded as converged when the centroid
shifts by less than $1^{\prime\prime}$. As shown in Table 4, the
offsets among Center I, II, and III were smaller than 50 kpc ($\sim
0.05 R_{500}$).  
Since Center III is less affected by core disturbances (e.g., AGN
activity and cD galaxy sloshing) than the other two centers, we
employed it for the subsequent analysis. Our results remain virtually
the same even using Center I or Center II.

\section{Data Analysis and Results}

\subsection{Optical Light Profiles}
To calculate the radial profiles of the stellar light, we first need
to define the galaxy membership of each cluster.  
However, it is difficult to perform this directly,
since there is no spectroscopic data available for the cluster sample 
observed with UH88. Generally this problem may be overcome in two approaches. One is 
to subtract the foreground and background galaxy contributions,
based on the surface brightness of field galaxies in the offset
pointing data. The other is to set
color filter in the color-magnitude space, as the red sequence member
galaxies, which contribute a majority of cluster membership in the
central region, are expected to show a 
well defined relation on the color-magnitude diagram. The former background
subtraction (hereafter BS) method is mainly subject 
to uncertainties due to cosmic variance, arising from random projections of
large scale structures, while the latter color-magnitude filtering
(hereafter CMF) method may suffer biases by underestimating blue
member galaxies. 
To minimize the expected biases, we employ both the BS and CMF approaches in our analysis.

\subsubsection{Background Subtraction}

The BS method has been widely used to measure the cluster galaxy
luminosity (e.g., Abell 1958; Valotto et al. 1997; Goto et al. 2002;
Hansen et al. 2005). First, we divided the $I$-band images of the central pointings
into four radial bins, i.e., $0-0.3$ $R_{500}$, $0.3-0.5$ $R_{500}$, $0.5-0.7$ $R_{500}$,
and $0.7-0.9$ $R_{500}$. This binning was chosen because each bin contains a sufficient number 
of detected galaxies (at least 30, typically $>100$) before background subtraction. 
For the subsample-L clusters, the outermost radial bin was omitted in the actual measurement, 
because it was only partially covered by the central pointing. To exclude apparent foreground
galaxies, those with fluxes ({\tt FLUX\_AUTO}) higher than the
cluster central galaxy were discarded from the surface brightness
calculation. Then we estimated the background contribution based on
data of the offset regions about $1.5-5.1$ Mpc ($\sim 1-3$ $R_{200}$)
away from the cluster center, which were observed 
with similar depth and seeing as the cluster central regions ($\S2.3$;
Table 3). Bright foreground galaxies with fluxes higher than the central
galaxies were again removed. To discard any possible background clusters or voids in the
offset region, we divided it into 3$\times 3$ sub-regions, each with a size of $2^{\prime}.5 \times 
2^{\prime}.5$, and discarded those sub-regions which have a galaxy number deviated from the 
mean value by $>2\sigma$. To measure the surface brightness profile
of the remaining offset region, we re-divided it into three radial annuli
centered on the central galaxy, each with a width of $2^{\prime}.5$, and
calculated the galaxy flux per unit area for each bin. The obtained
central + offset surface brightness profiles for six representative clusters (two for each subsample) are shown in Figure
1 (see also Fig. A.1 for the other clusters).

Figure 2 shows surface brightness profiles from the offset regions, averaged over the three subsamples (L, M, and H). 
These profiles were calculated by summing up, over the relevant subsample, those galaxies which fall in a radial bin with
a width of $0.5 R_{500}$, and then dividing by the sky area. Thus, the results still show a mild
outward decline, suggesting that the cluster galaxies should
distribute up to $\geq 3R_{500}$. Considering this, we fit the combined central + offset 
$I$-band surface brightness profiles with a King model plus constant, $S_{I} (r) = S_{0} [1 + (r/r_{\rm c})^2]^{-3/2} + S_{\rm bkg}$ (King 1962), 
where $S_{0}$ is a normalization, $r$ is projected radius, $r_{\rm c}$ is core radius, and
$S_{\rm bkg}$ represents the background component. As plotted in Figure 1, this model 
gave reasonable fits to the combined (central plus offset) surface brightness profiles of the six
representative clusters, as well as to the rest, all with $\chi^{2}/\nu \lesssim 2.0$ (Table 5). As shown in, e.g., Budzynski et al. (2012),
the cluster galaxy distributions can also be described by a Navarro-Frenk-White model (Navarro, Frenk, \& White 1996, hereafter
NFW model), which was first introduced to describe dark matter distributions. We fit the surface brightness profiles 
by $S_{I} (r) = S_{0} x^{-1}(1+x)^{-2}+ S_{\rm bkg}$, with $x \equiv r/r_{\rm s}$, where $r_{\rm s}$ is the projected
scale radius. The results of this fit are also given in Table 5. Thus, the two 
models yield essentially similar cluster galaxy components for the offset region,
so that the background components obtained with the two fittings are
consistent with each other within errors (Table 5). Therefore, we selected $S_{\rm bkg}$ from the fit
whichever gave a smaller $\chi^{2}/\nu$. Then, by subtracting this $S_{\rm bkg}$ 
from each radial bin in the central region, we obtained the
member galaxy surface brightness profiles as shown in Figure
3 (see also Fig. A.2).


The error of the obtained surface brightness profiles is dominated by
statistical uncertainties of the data, as well as systematic uncertainties
of the background which originates from the cosmic variance,
i.e., the galaxy luminosity variations due to the spatial distributions
of large-scale structures. To quantify this effect, we adopted a
theoretical result of Trenti and Stiavelli (2008), who provide
analytic estimation of galaxy number fluctuation as a function of
region size and the target redshift. Their calculation is
based on a two point correlation function of galaxy halos generated
with Press-Schechter formalism. In this way, the fractional
background uncertainties in our analysis were estimated to be $\sim
40\%$ and $\sim 20\%$ for low-redshift and high-redshift ends,
respectively. The final error bars shown in Figure 3 and Figure A.2 were obtained by
combining in quadrature the systematic background uncertainty and the
statistical flux measurement error. 

The background-subtracted surface brightness enclosed in radius $r$ was integrated 
to calculate the integral light profile, $L_{\rm opt, BS}(r)$. Figure 4 and Figure A.3 show
the result for each cluster, in the form of a quasi-continuous integrated light profile.
This was obtained by adding up all galaxies (including background) detected in I-band, and
then subtracting the background contribution which scales as the radius squared.
At the same time, based on the surface brightness uncertainties
obtained above, we evaluated error ranges at five representative radii, i.e., $r=0.25 R_{500}$, $0.35 R_{500}$, $0.45 R_{500}$,
$0.55 R_{500}$, and $0.65 R_{500}$. The typical errors are $\sim 5$\% and $15$\% for 
$r=0.25 R_{500}$ and $0.65 R_{500}$, respectively.

\subsubsection{Color-Magnitude Filtering}

To cross-check the background subtraction result, we conducted the member
galaxy selection based on the standard color-magnitude relation, that the old
passively evolving cluster galaxies form a tight horizontal relation in the
color-magnitude space (Bower et al. 1992; Stanford et al. 1998; Kodama
et al. 1998). The colors $m_{\rm R}-m_{\rm I}$ were acquired from the Kron-type $I$-
and $R$- band magnitudes calculated with the SExtractor, and corrected
for photometric zero points as described in \S2.3. As shown in Figure 5 (also seen in Fig. A.4),  the
bright red sequence galaxies form an apparent ridge-line with weak
negative slopes. This line-up creates a peak on the projected number-color
histogram shown in the right side panels of Figure 5. First, a Gaussian fitting to the number-color
histogram was used to determine the approximate color range which contains the color-magnitude line.  
The fitting gave an approximate central color ($C_0$), as well as the
associated color dispersion ($\sigma_{\rm C}$).  For the high
redshift ones ($z \geq 0.5$), we set $C_0$ to the color of the central
dominant galaxy and $\sigma_{\rm C} = 0.2$, as the relatively high background level caused poor
determinations of the central color. Then, to further determine the slope of
the relation, we performed linear fitting, $m_{\rm R}-m_{\rm I}$= $A$
+ $B \times m_{\rm I}$,
where $A$ and $B$ are free parameters, to the color-magnitude
diagram over a color range $C_0 - 3 \sigma_{\rm C} < m_{\rm
  R}-m_{\rm I} < C_0 + 3
\sigma_{\rm C}$ and a magnitude range $14 < m_{\rm I} < 24$. The fitting was
optimized with a robust biweight linear least square method (Beers,
Flynn \& Gebhardt 1990), which is insensitive to data points much deviated
from the relation, e.g., foreground/background galaxies. The fitting was repeated
for a few times, as the data points outside the 3$\sigma$ range
of the fitted relation were rejected in the next iteration. Typically,
the color-magnitude relation fitting converged within less than five
iterations. The derived relations, characterized by the best-fit
parameters $A$ and $B$ listed in Table 5, are consistent with the
predicted redshift-dependent relations reported in Kodama et
al. (1998). Then the red-sequence members were selected as data within
the fitting uncertainty range of the derived relations, typically
$\delta(m_{\rm R}-m_{\rm I}) \sim 0.1-0.2$ mag (Figure 5, Table
5). The integral light profiles, $L_{\rm opt, CMF}(r)$, were obtained
based on the fluxes of the selected galaxies. 

According to e.g., Kodama \& Bower (2001), the faint end of
the color-magnitude relation is strongly contaminated by background
non-members that by chance fall in the same range as the members. To
remove such faint-end contaminants, candidates with apparent $m_{\rm I}>23$
were discarded. To account for the background contaminants on the
brighter part of the red sequence relation, we followed Hao et
al. (2009) and fit the color diagram of non-selected $m_{\rm I}<23$ galaxies
with a Gaussian model. By integrating the best-fit model over the color
range of red sequence relation, we estimated the
number of non-member contaminants to selected galaxies. In
this way, the systematic uncertainty of the CMF method 
was estimated to be $10-30$\%, which is comparable to those shown in
Kodama \& Bower (2001). Assuming a uniform spatial distribution of the
non-members, the error was propagated to each radial bin to define
the systematic uncertainty of the integral galaxy light
profile. 

Another source of uncertainty is the fitting error of the
color-magnitude relation, which was assessed with Monte-Carlo
simulations. Following, e.g., Hilton et al. (2009), the color of each
galaxy has been replaced with a random variable within its measurement
error, and the randomized color-magnitude diagram was fitted with the robust
regression method. By 1000 realizations for each cluster, we
determined the errors on the slope and intercept of the relation, and
hence the uncertainty of the galaxy flux caused by fitting. In Figure
3 and Figure A.2, the systematic and fitting errors are combined in 
quadrature and used as the final error range of the $L_{\rm opt, CMF}(r)$
profiles.






The member selection results obtained with the BS and the CMF methods were
cross-examined for consistency in two ways, the radial surface
brightness profile and the luminosity function. In Figure 3 (see also
Fig. A.2),
we compare the $I$ surface brightness profiles obtained with the two
methods. In most bins the two profiles agree with each other within
error ranges. Their differences, mainly caused either by local background
variations in the BS method or the omission of blue populations in
the CMF, are thus found to be on average $\leq 10$\% to 20\% from
innermost to outermost bins, respectively. In addition, the luminosity function 
of the entire member galaxies, detected in our 34 sample clusters, was 
constructed using the BS and CMF methods separately. For
the BS method, this was performed by subtracting
another luminosity function created from the offset regions of the 34 clusters. We excluded the
central dominant galaxies, which are known to be outliers in this
respect. As shown in Figure 6, the two methods again give consistent
luminosity functions.

To compare our luminosity functions with previous works, two
best-fit functions in Rudnick et al. (2009) are also plotted in Figure 6. Their curves
were given in the Schechter form (Schechter 1976) as
\begin{equation}
\phi(m)dm= 0.4 {\rm ln}(10) N_{\rm clu} \phi^{\ast} 10^{-0.4(m-m^{\ast})(\alpha+1)} \times {\rm exp}\left(-10^{-0.4(m-m^{\ast})} \right) {\rm d}m,
\end{equation} 
where $N_{\rm clu}$ is the galaxy number, $\phi^{\ast}$ is normalization,
while $m^{\ast}$ and $\alpha$ are parameters defining the shape of the
function. The two curves from Rudnick et al. (2009) are defined by shape parameters ($m^{\ast}_{\rm AB}, \alpha$) = (-21.46, -0.75) and
(-21.83, -0.58), which were obtained for local clusters with $z<0.06$ and high redshift ones 
with $0.4<z<0.6$, respectively. Our sample-averaged BS and CMF results are consistent
with both curves within errors. The agreement between our results and their local cluster curve can be represented 
by $\chi^{2}/\nu$ = 2.8/5 and 2.5/4 for the BS and CMF curves, respectively, which is slightly 
better than that with the high redshift one, $\chi^{2}/\nu$ = 4.1/5 and 5.5/4 for the BS and CMF 
curves, respectively. This is probably because most (two-thirds) of our sample clusters have
redshifts below 0.4. Hence, we have proved the agreement between the BS
and CMF radial light profiles,
which are also well in line with those reported in previous optical studies.



\subsection{X-ray Mass Profiles}

To derive cluster gas mass profiles, the gas density profiles were calculated based on 
deprojection spectral analysis with the {\it XMM-Newton} EPIC and {\it Chandra} ACIS data after point source removal and background correction (\S2.2).
First we extracted spectra from several concentric annulus regions for each cluster. The radial
boundaries of each annulus, $r_{\rm out}$ and $r_{\rm in}$, were chosen by $r_{\rm out}/ r_{\rm in} = 1.1$ and 1.25 for high-flux clusters 
and low-flux ones, respectively. The extracted spectra were fit in $0.7-8.0$ keV with an absorbed, single-temperature APEC model in {\tt XSPEC}. All annuli were linked by a PROJCT model, which evaluates 
a two-dimensional (hereafter 2-D) projected model from intrinsic three-dimensional
(hereafter 3-D) one; in this way all the APEC parameters are obtained
in 3-D.  For each 3-D annulus, the gas temperature and metal abundance
were set free. When the model parameters cannot be well constrained in
some annuli due to relatively low statistics, we tied them to the values of their adjacent regions. The column density of neutral absorber was fixed to the Galactic value given in Kalberla et al. (2005).
All the fits were acceptable, with reduced
chi-square $\approx 0.9 - 1.1$ for a typical degree of freedom of
$\sim 500-1000$. Then, the 3-D gas density profiles were calculated
from the best-fit model normalization of each annulus; the ICM density is proportional to
the square root of the model normalization. The results are
presented in Figure 7 (see also Fig. A.5), where the density
error bars were obtained by taking into account
both statistical and systematic uncertainties. The former was
estimated by scanning over the parameter space with an {\tt XSPEC} tool
{\em steppar} iteratively, while the latter was assessed by
re-normalizing the level of CXB component by 20\% (\S 2.2) for each annulus.

As shown in Figure 7 and Figure A.5, the gas density profiles were fit with a canonical $\beta$ model, $n_{\rm g}(R) = n_{\rm g,0} [1+(R/R_{\rm c})^2]^{-3\beta/2}$, where $R_{\rm c}$ is the core radius and $\beta$ is the slope. 
In several clusters (i.e., A963, A1835, A68, A1576, RBS 797, MS0451, and
CL0016), the observed profiles show significant central excess over
the $\beta$ model. Such an excess often indicates a hierarchical
potential structure associated with the cD galaxy, sometimes augmented
by the central cool component (see Makishima et
al. 2001 for a review). In these clusters the ICM density
profiles are empirically better described with a double-$\beta$ model, $n_{\rm g}(R) =
\{n_{\rm g1,0}^2 [1+(R/R_{\rm c1})^2]^{-3\beta_1}+n_{\rm g2,0}^2
   [1+(R/R_{\rm c2})^2]^{-3\beta_2}\}^{1/2} $, where suffixes 1 and
   2 indicate the compact and extended components,
   respectively. Actually in previous works, this model successfully described the ICM emissivity
   profiles of some representative cool-core clusters (e.g., Ikebe et al. 1999; Xu et al. 1998). To
   reduce degeneracy among the double-$\beta$ parameters, we fixed
   $\beta_1 = \beta_2$ in the fittings. Allowing them free does not change the resulting model profile
   significantly. For the seven clusters showing the central excess, the
   double-$\beta$ model yields significantly better fit than the
   $\beta$ model with reduced
   chi-square $< 1.3$. The best-fit parameters for the
   $\beta$/double-$\beta$ models are listed in Table 4. Like in the optical case,
we calculated the 2-D ICM mass profile, $M_{\rm X, 2D}(r)$, by projecting the best-fit
$\beta$/double-$\beta$ density profile along the line of sight, and integrating it over the 2-D radius. 
The obtained quasi-continuous 2-D ICM mass profiles are shown in Figure 4 and Figure A.3,
together with the errors estimated at $r=0.25 R_{500}$, $0.35 R_{500}$, $0.45 R_{500}$, $0.55 R_{500}$,
and $0.65 R_{500}$. The typical uncertainties at $r=0.55 R_{500}$ are $\sim 4$\%
and $\sim 8$\% for high flux and low flux clusters, respectively.


\subsection{Galaxy Light vs ICM Mass Ratio Profiles}

As revealed in Figure 4 and Figure A.3, the $L_{\rm opt}(r)$ profiles of L-subsample clusters 
are significantly flatter than their $M_{\rm X, 2D}(r)$ profiles. In
contrast, the optical and X-ray profiles of H-subsample clusters show similar
gradients. To enable direct comparison between their slopes, we
normalized each of them to
its central values within $r=0.25 R_{500}$, and calculated the ratios
between the normalized profiles. This gives ``galaxy light vs. ICM mass ratio'' 
(hereafter GLIMR) at each radius, together with the uncertainties
at the representative points selected in Figure 4; $r=0.35 R_{500}$, $0.45 R_{500}$, $0.55 R_{500}$, and $0.65 R_{500}$. In 
Figure 8, we present the subsample-averaged GLIMR profiles obtained with the 
BS and CMF methods. Uncertainties of the averaged profiles were calculated by combining
in quadrature the errors of all clusters in the subsample. The BS-GLIMR profiles
are consistent with the CMF-GLIMR ones within 1$\sigma$ error range.

As shown in Figure 8, the average GLIMR profiles of the subsample L clusters drop
steeply towards outer regions, reconfirming the inference of Figure
4 and Figure A.3. The GLIMR profiles of the
subsample H clusters, in contrast, show significantly flatter distributions
than the low redshift counterparts, and the profiles of the subsample M
clusters appear to have intermediate gradients. The normalized
BS-GLIMR at $r=0.65$ $R_{500}$ are $0.38^{+0.06}_{-0.05}$, $0.54^{+0.05}_{-0.04}$, and
$0.87^{+0.13}_{-0.06}$ for the subsample L, M, and H, respectively.    
The CMF method gave consistent
values, $0.45^{+0.04}_{-0.04}$, $0.58^{+0.06}_{-0.05}$, and
$0.80^{+0.11}_{-0.12}$ for the subsample L, M, and H,
respectively. The galaxy and ICM components thus followed a similar
distribution in high-redshift ($z>0.5$) clusters, while their optical
size relative to that of ICM evolved to become considerably smaller within central $0.65$ $R_{500}$ in nearby ($z \sim 0.1$) systems.

To confirm that Figure 8 reflects systematic differences among the three subsamples, rather than, e.g., biased by some extreme outliers, we carried out
Kolmogorov-Smirnov (K-S) test among the three subsamples. In Figure 9
we show the cumulative fraction probability distributions as a
function of the GLIMR values of individual objects at $r=0.45$ $R_{500}$ and $0.65$ $R_{500}$. The K-S probability curves are clearly separated among the three subsamples. Given the D-statistics shown in Figure 9, the hypothesis that the BS-GLIMRs follow the same distribution at $0.65$ $R_{500}$ can be rejected at $94\%$ and $98\%$ confidence levels between subsamples L and M, and between subsamples M and H, respectively. The CMF-GLIMR profiles show similar significance, that the hypothesis can be rejected at $97\%$ and $98\%$ levels between subsamples L and M, and between M and H, respectively. To account for the measurement uncertainties, we performed 1000 Monte-Carlo realizations, in which the GLIMR profiles of the individual clusters were randomly varied within the error range. The D-statistics were not found to vary significantly. 

To examine whether the gradient of GLIMR profile evolves with redshift
continuously or not, we plot in Figure 10 the logarithmic slopes of the GLIMR
profiles of the 34 clusters, defined as indices of the power-law models which best fit the five radial GLIMR 
points of each cluster, against cluster redshifts. The figure shows a
smooth trend, that the logarithmic slope changes gradually from $\sim
-0.7$ for the nearest objects to $\sim 0$ for the farthest ones. By fitting the
distribution with a constant model, we obtained $\chi^2 = 530$ (BS) and $154$ (CMF) for $\nu = 33$. This means that the evolution of the
GLIMR slope is significant at $>99\%$ confidence level.
Thus, the implication of Figure 10 agrees with those of Figure 8 and Figure 9.

\subsection{GLIMR Profiles with the SDSS Data and Spectroscopic Measurements}

To crosscheck the GLIMR profiles obtained with the UH88 data, we
carried out an independent photometric analysis of SDSS DR6 (data release 6)
public data (Adelman-McCarthy et al. 2008). As shown in Figure 11, a
total of nine clusters in our sample were found available in the current SDSS DR6
catalog. From the catalog we obtained $i^{\prime}$-band magnitudes for
all galaxy-type objects ({\tt CLASS} = 3) within $0.65$ $R_{500}$, and their
magnitudes were converted to fluxes to calculate surface brightness
profiles. Galaxies with fluxes higher than that of the central dominant
galaxy were excluded. For the remaining galaxies we applied the BS
method (\S 3.1.1), wherein the background value was obtained by the
average galaxy flux in an annulus with $r=10-15$ Mpc centered on the
cluster center. To remove possible contamination from other clusters,
the background region was divided equally into 20 azimuthal sectors, and
those sectors of which the flux is deviated from the annulus-average
value by 2$\sigma$ were discarded. After subtracting the background
value from the surface brightness profile, the integral light profile,
and hence the GLIMR profile was obtained in the same way as described
in \S 3.3. The SDSS profiles have smaller uncertainties than the UH88 profiles, 
because the SDSS data have much better sky coverage for both source and background regions. 
As shown in Figure 11, the typical differences between the
SDSS BS-GLIMR profiles and UH88 BS-GLIMR profiles are within 8\% at
all radii, which are smaller than the estimated error bars.


Multi-object spectroscopy gives the most reliable determination of
galaxy members. Previous spectroscopic studies have provided
membership information for five clusters in our sample, i.e., A959 (Boschin et
al. 2009), RXJ2308 (Braglia et al. 2009), CL0024 (Moran et al. 2007),
MS0451 (Moran et al. 2007), and CL0016 (Dressler \& Gunn 1992), each
with $50-130$ detected members in $r<0.65$ $R_{500}$. By calculating
the $I$-band fluxes with the UH88 data for the spectroscopically confirmed member galaxies, we
obtained the GLIMR profiles as shown in Figure 11 in blue. The spectroscopic
GLIMR profiles are consistent with our photometric (either BS or CMF) GLIMR profiles
within errors. The average fractional differences between the spectroscopic- and the BS-
(CMF-) GLIMR are about 6\% (10\%), which agree with the systematic
errors estimated in \S 3.1.1 and \S 3.1.2 for the BS and CMF methods,
respectively. 

For a further comparison, we examined the GLIMR profile
of a typical relaxed cluster in local universe, Abell 1650 (hereafter A1650;
$z=0.084$; $T_{\rm aver} \approx 6$ keV; Donahue et al. 2005). While this cluster satisfies 
the criteria used in our sample definition, it is located closer than those in 
subsample L. Furthermore, its spectroscopic membership is available in Pimbblet et
al. (2006). By deriving the converted $I$-band profile of the spectroscopic members
using the SDSS photometric data, and combining it with the ICM density profile measured in Gu et
al. (2009; see their Fig. 2), we calculated the GLIMR profile of A1650 and compared it
with our cluster sample. As shown in Figure 8, A1650 has an even more
centrally-concentrated GLIMR distribution than the average profile of
subsample L. This reinforces the GLIMR evolution detected in \S 3.3.


\subsection{Systematics Errors and Biases}

So far, we have detected significant evolution in the galaxy light-to-ICM
mass ratio profiles among subsamples L, M and H. However, the result
might be subject to various systematic errors and biases. In Figure 12, 
new five-point GLIMR profiles are presented to account for these effects one by one.

\subsubsection{Possible Biases in High Redshift Clusters}

First of all, the GLIMR evolution could be an artifact caused by the
inconstant galaxy selection completeness, which obviously becomes lower at
higher redshifts. To assess this redshift-dependent bias, we set a
uniform absolute magnitude limit to the CMF method which corresponds to the
observation depth at $m_{I} < 25.4$ for the highest redshift
cluster WJ1226. This limit was converted for each cluster to $I$-band
magnitude, and is shown as
dashed lines in Figure 5 and Figure A.4. When galaxies fainter than this limit are
discarded, the selection completeness 
applied to each cluster should be approximately the same. As shown in
Figure 12{\it a}, the average CMF-GLIMR profile with the new limiting
magnitude is consistent, within errors, with the previous one for each
subsample; the difference caused by varying the limiting magnitude is
about 5\%. The BS method yielded similar results by applying the new
limiting magnitude. Hence, the result is robust against the galaxy selection completeness.

As shown in, e.g., Cucciati et al. (2012), galaxies at high redshift tend to have more dust than 
those in local universe. The mean dust attenuation in far ultraviolet band increases by about 1 mag from $z = 0$
to $z \sim 1$. This introduces a systematic uncertainty to the GLIMR profiles, especially for subsample H.
This effect can be addressed by considering several observational facts. As shown in e.g., Garn \& Best (2010), the dust attenuation has
a positive correlation with the star forming rate, while such activity is known to become more enhanced 
towards outskirt regions of clusters with $z \leq 1$ (e.g., Kodama et al. 2004; Koyama et al. 2008). Hence, the dust
would more strongly attenuate the light emitted from cluster outer regions, and the effect should be stronger in higher redshift.    
However, this is opposite to the observed GLIMR evolution. Therefore, it is unlikely that the dust effect, if any,
has significantly affected our results.

Since the cluster-average X-ray source-to-background ratio is generally smaller in 
more distant objects, it is possible that the ICM density profiles at $0.65 R_{500}$ are suppressed
due to background over-subtraction in high redshift clusters, and the GLIMR profiles are
biased to be flatter. To examine this effect, we calculated the X-ray source-to-background
ratio at $0.65 R_{500}$ ($S/B_{0.65}$ hereafter) for each cluster as
shown in Table 4. The mean ratios
are 1.7, 1.2, and 1.5 for subsample L, M, and H, respectively. These are larger than the limiting source-to-background ratio
in previous studies using {\it Chandra} and {\it XMM-Newton} data
(e.g., $S/B_{\rm limit} = 0.6$; Leccardi \& Molendi 2008), and are not strongly subsample dependent. Then we divided the 34 clusters
into two groups by the mean source-to-background ratio ($S/B_{0.65} = 1.4$). As shown
in Figure 12{\it b}, the average GLIMR profiles of the two groups are consistent with each other
within errors. This indicates that the shape of the GLIMR profile is nearly independent of 
the X-ray source-to-background ratio, and the observed evolution cannot be explained 
by background over-subtraction of the X-ray data of high redshift clusters.

\subsubsection{Possible Scale- and Richness- Dependences}

To examine the uncertainty of the $R_{500}$ determinations based on the
$T_{\rm aver}-R_{500}$ scaling relation (\S2.4), we calculated a new set of
$R_{500}$ directly from the hydrostatic mass estimates using the X-ray
data. Based on the best-fit 3-D gas temperature profiles $T_{\rm
  X}(R)$ and density profiles $n_{\rm g}(R)$ obtained with the
deprojected analysis (\S 3.2), and assuming spherical symmetry and a
hydrostatic equilibrium, the total gravitating mass within a 3-D radius $R$
can be calculated generally as
\begin{equation}
M(R) = \frac{-kT_{\rm X}(R) R}{G \mu m_{\rm p}}\left( \frac{d {\rm ln}
  n_{\rm g}(R)}{d {\rm ln} R} + \frac{d {\rm ln} T_{\rm X}(R) }{d {\rm
    ln} R}\right),
\end{equation}
where $G$ is the gravitational constant, $\mu = 0.609$ is the average
molecular weight, and $m_{\rm p}$ is the proton mass. Then we
calculated $R_{500}$ according to its original definition, i.e., the radius
within which the average mass density equals to 500 times the critical
density of the Universe. In Table 6 the new $R_{500}$ (re-named as
$R_{500}^{\rm HM}$ for clarify) are shown, together with the total
enclosed gravitating mass $M_{500}$. The obtained
$R_{500}^{\rm HM}$ and $M_{500}$ are consistent within 1$\sigma$ uncertainties with the previous
measurements using both X-ray (e.g., Arnaud et al. 2007 and Zhang et al. 2008) and weak lensing
(e.g., Dahle 2006) techniques. By
adopting $R_{500}^{\rm HM}$, the BS-GLIMR profiles were updated as
shown in Figure 12{\it c}. The new profiles are consistent with the previous
ones within 1$\sigma$ errors, and the 
different measurements of $R_{500}$ introduce insignificant ($<8$\%)
bias on the GLIMR profile.

Since our sample clusters have a non-negligible scatter in the system richness,
it is important to examine whether or not the gradient of GLIMR is related to the
cluster mass. Since some ICM heating processes, e.g., AGN
outbursts (see McNamara \& Nulsen 2007 for a review), are expected to
have stronger effects of expelling hot gas in systems with shallower
potentials (Sun et al. 2009), the gas mass profiles might be more
extended in such systems. Since the stellar component would not feel
such heating processes, the GLIMR profiles could be steeper in
low-mass systems than in high-mass ones. In addition, there can be other unknown 
richness-dependent effects that may affect the GLIMR profiles. Because the higher-redshift
subsamples would tend to pick up relatively richer clusters, the
possible GLIMR dependence on cluster mass might mimic the observed evolution
shown in Figure 8. We therefore divided each subsample equally into two subgroups
according to the obtained $M_{500}$ as listed in Table 6, and compared
the subgroup-average BS-GLIMR profiles. As shown in Figure 12{\it d}, the
higher-mass subgroups do have slightly more extended GLIMR profiles as expected. However, the
differences between them are within 6\%,
smaller than the $1\sigma$ errors. Both high-mass and low-mass
profiles show essentially the same evolution. Hence, the
systematic errors due to cluster mass differences do not significantly
affect our result, and we can exclude the possibility that the GLIMR evolution is an artifact produced
by the selection effect on cluster mass.


Since the clusters are expected to grow via long-term matter
accretion from cluster outskirts, the cluster scale (i.e., mass and radius) should increase
over several Gyrs. Therefore, a young
cluster may have been growing by taking in infalling new materials,
and eventually constitute a central region of an old 
cluster. In another word, the $R_{500}$ region of a subsample  
L cluster might correspond to a $>$$R_{500}$ region in a subsample
H cluster. To compensate for the underlying difference in cluster evolution stage,
we again considered the evolved scale, $R_{500}^{z=0}$, which was defined 
in \S2.1 as the radius that each $R_{500}$ will evolve to at $z=0$.
Based on the empirical cosmic
growth function derived from $\Lambda$CDM numerical simulations, e.g.,
Wechsler et al. (2002), the cluster mass is expected to increase by a factor of
$\sim e^{1.33z_{0}}$ from $z=z_{0}$ to $z=0$, so that the
$R_{500}^{z=0}$ can be estimated as $R_{500}^{z=0} \approx R_{500}^{z=z_{0}} \times
e^{0.56 z_{0}} E(z_{0})^{0.58}$, where $E(z_{0})$ is
the cosmological evolution factor defined in \S2.4. By adopting the new radial scale, the
BS-GLIMR profiles were calculated in the same way as in \S3.1.1. As
shown in Figure 12{\it e}, the new BS-GLIMR profiles are consistent with
our original results within error ranges, and the systematics caused by the cosmic growth in cluster outer regions should be $\leq 10$\%.


\subsubsection{Possible Artifact by Galaxy Color and Luminosity Evolution}

Another major concern is possible redshift-dependent deviation biased
by radial color gradient of member galaxies, and galaxy luminosity
variations, coupled with the obvious differences in the rest-frame
colors among the three subsamples. As shown in e.g., 
Boselli \& Gavazzi (2006), the spatial distribution of blue galaxies are 
considerably flatter than that of the red ones. Since
higher-redshift clusters are observed in bluer rest-frame light, the
apparent GLIMR evolution in Figure 8 could simply because our optical data of higher-z 
clusters are contributed more by bluer galaxies which have flatter spatial
distributions. The K-correction which we adopted (\S 2.3)
cannot remove such a bias. Furthermore, the galaxy
luminosity might have varied with time by, e.g., star forming
events, which are considered to have been proceeding during the 
observed redshift range. As shown in, e.g., Hayashi et al. (2009) and Tadaki et al. (2011),
the star formation activity took place in a larger radius in higher-redshift
clusters, while it was reduced rapidly in cluster central regions with
z$\leq 0.5$. Hence, the gradient of the GLIMR profiles might be further
biased by such radius- and color- dependent luminosity variations. To
examine these effects, we re-calculated the galaxy light profiles 
$L_{\rm opt}(r)$ based on a same rest-frame band for different
redshifts. Specifically, the $B$-band, $R$-band, and $I$-band data were used
for $z=0.119-0.235$, $z=0.235-0.600$, and $z=0.600-0.890$ clusters,
respectively, so that the fluxes are all measured in approximately
rest-frame $B$-band. Since the $B$/$R$ data are not
available for the offset regions, we can obtain light profiles only via
the CMF method. As shown in Figure 12{\it f}, the subsample-averaged CMF-GLIMR profiles
derived in the same rest-frame band agree, within $7$\%, with
the one derived with $I$ data alone.

As another approach to more thoroughly eliminate 
the effects of color gradient and galaxy luminosity changes, we adopted
galaxy number counts instead of the optical luminosity, and derived
the galaxy number vs. ICM mass ratio profile (hereafter GNIMR profile) for each cluster. Both the BS
and the CMF methods were employed. For a crosscheck, we compared the number density
profiles with the one presented in Budzynski et al. (2012), which was
calculated by averaging over 20000 groups and clusters with $z=0.15-0.4$ using
the SDSS data. As shown in Figure 13{\it a}, our subsample-averaged profiles
are in rough agreement with the SDSS result within a scatter of $\sim
15$\%. Then we calculated the GNIMR profiles in a similar
way to the GLIMR profiles, and show the results in Figure 13{\it b}. The BS and CMF methods 
yield consistent profiles within 1$\sigma$ error ranges. The shapes of the subsample-averaged 
GNIMR profiles exhibit significant dependence on redshift, and
the evolution is consistent with that implied by the GLIMR profiles. Hence, from the above two examinations, i.e., 
the GLIMR profiles obtained in the same rest-frame band and the GNIMR profiles, we can
exclude the possibility that the detected GLIMR evolution is caused artificially by
the radius-/redshift- dependent galaxy color/luminosity variations.


We conclude that none of the above factors, i.e., redshift-dependent
galaxy selection and X-ray background subtraction, $R_{500}$ uncertainty, mass-dependent effects
of central ICM heating, matter accretion in outer region, 
galaxy color gradient, and luminosity variation, can induce significant
bias to the detected evolution of GLIMR profiles. The data favor the
scenario that cluster member galaxies have concentrated towards the
cluster center more than the hot gas component from $z=0.9$ to $0.1$.

\subsection{Galaxy Light vs. Total Mass Ratio}

To explore one step further, we studied the galaxy light
vs. total mass ratio (hereafter GLTMR) profiles. Based on the total
gravitating mass profiles $M(R)$ obtained by Eq.3 in \S3.5.2, the 3-D total mass
density profiles were calculated as $\rho(R) = (4 \pi R^{2})
{\rm d}M(R)/{\rm d}R$. Since the mass density distribution of rich clusters 
is known to be well described by the NFW model (e.g., Vikhlinin et al. 2006), it is
better to calculate the 2-D total mass profiles from continuous NFW curves.
Then, we fit the $\rho(R)$ profiles with the model used in \S 3.1.1 as
\begin{equation}
\rho (R) = \frac{\delta_{\rm c} \rho_{\rm crit,z}}{\frac{R}{R_{\rm s}}\left(1+\frac{R}{R_{\rm s}}\right)^2},
\end{equation} 
where $R_{\rm s}$ is the 3-D scale radius, $\rho_{\rm crit,z}$ is the critical density of the Universe at redshift $z$, and $\delta_{\rm c}$ is the density contrast which can be expressed in terms of the concentration parameter $c$ as 
\begin{equation}
\delta_{\rm c} = \frac{200}{3} \frac{c^3}{[ln(1+c)-c/(1+c)]}.
\end{equation}
The fits were acceptable with reduced chi-square $< 1.2$,
and yielded the best-fit parameters $c$ and $R_{\rm s}$ as listed in Table
6. The two parameters are consistent with previous
reportings in, e.g., Arnaud et al. (2007) and Vikhlinin et al. (2006).
To obtain the integral 2-D total mass profiles, $M_{\rm 2D}(r)$,
the best-fit NFW mass density profiles was projected along the line-of-sight
and integrated. By replacing $M_{\rm X, 2D}(r)$ with $M_{\rm 2D}(r)$, we
 calculated the GLTMR profiles, and derived the average profile for
 each subsample as shown in Figure 14. Compared to the BS-GLIMR
 profiles, the BS-GLTMR profiles exhibit relatively flatter
 gradients. This agrees with the consensus that the density of the
 dark matter follows a $\sim R^{-3}$ distribution outside the core
 region, which is steeper than the ICM component which roughly follows
 $\sim R^{-2}$ (Navarro, Frenk, \& White 1996). Nevertheless, just like the case of GLIMR, the
 GLTMR profiles show significant evolution in their gradient; the average
 profile of subsample L is $\sim 60$\% of that of subsample H at the
 $r=0.65$ $R_{500}$ bin, while the latter agrees with unity throughout
 the central $0.65$ $R_{500}$ region. Thus, while the three
 mass components are likely to have followed a similar distribution in
 high-redshift ($z>0.5$) clusters, the stellar component concentrated
 relative to the ICM and dark matter components, and have evolved to
 be nearly twice more compact in the central $0.65$ $R_{500}$ at local
 universe.   


\section{Discussion}

\subsection{Summary of the Results}

By analyzing the optical photometric data obtained with the UH88
telescope, together with the X-ray data from 
{\it XMM-Newton} and {\it Chandra}, we studied the galaxy light
vs. ICM mass ratio profiles for a sample of 34 X-ray bright clusters with
$z=0.1-0.9$, which are selected to be relaxed systems with similar
richness. To enhance reliability of the member galaxy selection, we 
employed two independent methods, i.e., offset background subtraction
and color-magnitude filtering. The GLIMR
profiles of nearby clusters were found to drop more steeply, by a factor of $\sim
2$ at $\sim 0.65$ $R_{500}$, than their higher redshift
counterparts. According to a K-S test, the evolution is significant on
$\geq 94$\%. 

The GLIMR profiles derived here are still subject to measurement
uncertainties caused by, e.g., cosmic variance in optical background
and blue/background galaxies contaminations in color-magnitude
relation. Furthermore, the GLIMR evolution indicated by our data may suffer from a range of
radius-/redshift- dependent systematic errors and biases from, e.g.,
magnitude limits on galaxy selection, uncertainties in X-ray
background and $R_{500}$ estimation,
difference in cluster richness, growth of cluster outer regions, cluster color 
gradient, and galaxy luminosity variations. However, by assessing each of these
errors and biases, none of them was found to affect the observed
GLIMR profiles significantly by $>15$\%, so that the evolution of the
GLIMR profiles remains intact. The galaxy-to-total mass profiles also
show significant evolution of galaxy concentration. We hence conclude that the
galaxies in cluster have actually evolved, from $z\sim 0.9$ to $z\sim 0.1$, to
become more concentrated relative to the ICM and dark matter.

\subsection{Concentration of the GLIMR Profiles}

The concentration of the galaxy distributions, relative to the ICM and dark matter, can be
explained by four competing scenarios. First, the galaxy central
excess in nearby clusters might be produced by local star
formation. However, this scenario is opposite to the current
understanding of star-formation-rate evolution, which is thought to drop exponentially from
$z \sim 1$ to $z \sim 0$ (e.g., Tresse et al. 2007). Although the
luminosities of central galaxies might increase with time due to
strong star formation with a rate of a 
few tens of $M_{\odot}$yr$^{-1}$ in the core (e.g., Smith et al. 1997), it still fails to
explain the detected evolution on the GNIMR profiles
shown in Figure 13{\it b}. Second, the GLIMR evolution can be caused by
gradual expansion of the ICM halo. However, this is again contrary to
the observed evolution of the ICM surface brightness distributions as
reported by, e.g., Santos et al. (2008). Furthermore, as mentioned in \S 1, the effects of 
gradual accretion of metal-poor gas is considered insignificant. The third scenario is based
on a redshift-dependent accretion rate of galaxies. As suggested in, e.g.,
Ellingson et al. (2001), more field galaxies are likely to have accreted into clusters in
higher redshift ($z>0.5$) than in lower redshift. Such a field galaxy
component might not be fully virialized right after the accretion (e.g., Balogh et
al. 2000), and is expected to exhibit a more extended spatial distribution
than the red sequence galaxies. Thus, the GLIMR evolution might be
caused by a decline with time in the field-galaxy accretion
rate. However, as shown in Figure 8 (see also Fig. 12{\it a},
Fig. 12{\it f}, and Fig. 13), the GLIMR profiles of red sequence cluster members,
which entered the cluster region before $z\sim 2$ (e.g., Kriek et al. 2008),
also show significant evolution from $z=0.9$ to $z=0.1$. Furthermore, as shown in 
Figure 13{\it a}, the galaxy density profiles concentrated more significantly in the central
0.2 $R_{500}$ region. This indicates
that the redshift-dependent accretion rate alone cannot fully explain the
observed phenomena. Hence, it is natural to
propose a fourth scenario, that the cluster galaxies have actually been falling,
relative to the ICM and dark matter, towards the cluster center from $z \sim 0.9$
to $z \sim 0.1$.

\subsection{Origin of the Galaxy Infall}

The galaxy infall relative to the ICM and dark matter, as concluded above, indicates interactions taking place between
the cluster galaxies and other two components,
by which the galaxies lose a significant portion of their kinetic
energy. In this subsection, we examine possible origins of the GLIMR evolution in detail.

\subsubsection{Dynamical Friction}

One obvious candidate process to cause the galaxies to fall to the cluster center is dynamical friction,
which occurs when the potential of a moving galaxy interacts with a gravitational
wake created behind it, leading to gravitational energy exchange with the galaxy and
the surrounding media including stars, ICM and dark matter (Dokuchaev
1964; Rephaeli \& Salpeter 1980; Miller 1986). As shown in, e.g., El-Zant et al. (2004), the dynamical friction in a typical rich cluster may convert a level of $10^{44}$ erg s$^{-1}$ from galaxy kinetic energy to the ICM thermal energy, while the galaxy will fall to the cluster center significantly. To examine whether such a process can reproduce the observed GLIMR evolution or not, we investigate quantitatively the orbital evolution of a model galaxy that is subject to the dynamical friction.



We begin with a model galaxy cluster with its gravitating mass distribution
following the NFW density profile defined in Eq.4 and Eq.5. The
total mass enclosed in $R$ is then given as 
\begin{equation}
M(R) = 4 \pi  \delta_{\rm c} \rho_{\rm crit,z} R_{\rm s}^3
\left[{\rm ln} \left( \frac{R_{\rm s}+R}{R_{\rm s}} \right)-\frac{R}{R_{\rm s}+R}\right].
\end{equation} 
The sample-averaged values of $c=3.7$ and $R_{\rm s}=502$ kpc
(Table 6) are employed to calculate the $M(R)$ profile. For reference, when adopting
the value of $\rho_{\rm crit,z}$ at $z=0.3$, the model cluster has $M(R_{\rm vir}) \approx 2.9 \times
10^{15}$ $M_{\rm \odot}$ within its virial radius $R_{\rm vir} = 1857$
kpc.
 


Consider a perturber galaxy with a mass $m$, on a circular orbit with a radius of $R$ and an orbit velocity of $v=\sqrt{{\rm G} M(R)/R}$. Since the galaxy typically has a transonic velocity ($v \sim 500-1500$ km $\rm s^{-1}$), the gravitational interaction between the galaxy and cluster matter can be expressed by the dynamical friction force given as 
\begin{equation}
F_{\rm DF}(R) \approx \frac{4 \pi \rho(R) ({\rm G} m)^2 }{v^2}
\end{equation}  
(e.g., Ostriker 1999; Nath 2008), where $\rho(R)$ is the total mass density estimated as the derivative of $M(R)$ (\S 3.6). The angular momentum of the galaxy, $L \sim m v R$, decreases with time by ${\rm d}L/{\rm d}t \sim F_{\rm DF}(R) \times R$, so that the galaxy moves in a spiral trajectory with the radius changing by 
\begin{equation}
\frac{{\rm d}R}{{\rm d}t} \approx \frac{4 \pi \rho(R) {\rm G}^2 m R}{(k+1)v^3}, 
\end{equation}
where $k \equiv {\rm d} {\rm ln} v / {\rm d} {\rm ln} R$ is the logarithmic velocity gradient.  
By numerically solving Eq.8, we calculate how the circular orbit of a
massive and a less-massive galaxy decays, and show  
the results in Figure 15. The effect thus depends clearly on the infalling
galaxy mass. In about 5 Gyrs, a massive galaxy ($m=1
\times 10^{12}$ $M_{\rm \odot}$) spirals inwards significantly by
$\approx 8$\%, $19$\%, and $>80$\% from the initial radius of $R_{0} =
700$ kpc, $500$ kpc, and $300$ kpc, respectively. In contrast, the orbit of a less
massive one ($m=1 \times 10^{11}$ $M_{\rm \odot}$)
is much less affected, with a decay of $<1$\%, $2$\%, and $4$\%
for $R_{0} = 700$ kpc, $500$ kpc, and $300$ kpc, respectively. This
result agrees with e.g., Nath (2008); the dynamical friction induces 
a spatial mass segregation in galaxy clusters in such a way that
massive galaxies become concentrated towards the center more quickly,
while less massive ones remain widely scattered.



To assess the effect of dynamical friction, it is hence crucial to
examine whether or not the GLIMR evolution depends strongly on the 
galaxy mass. Based on above calculation, a dividing mass may be set 
at $m_{\rm limit} = 1 \times 10^{11}$ $M_{\rm \odot}$, below which the 
dynamical friction becomes negligible. By
adopting the observed mass-to-light vs. I-band luminosity 
relation given in Cappellari et al. (2006; Eq.9 therein), we converted the
I-band luminosity to mass $m$, and re-calculated the BS-GLIMR profiles 
for $m<m_{\rm limit}$ galaxies. As shown in Figure 16, the
GLIMR profiles for the low-mass galaxies still exhibit significant evolution among subsamples H, M, and L.
Since this feature cannot be explained by dynamical friction, we conclude that the 
gravitational effect alone is insufficient to account for the observed GLIMR evolution.

\subsubsection{Galaxy-ICM Interaction}

Observed in dozens of clusters and groups, e.g., the Virgo
cluster (Randall et al. 2008), Abell 2125 (Wang et al. 2004), and
Abell 3627 (Sun et al. 2006), interactions between the moving galaxies
and the ambient hot plasmas via, e.g., ram pressure, also exert drag
force on the galaxies. In such a way free energy is transferred from
galaxies to the ICM, which may lead to the galaxy infall, and possible ICM heating. The ram pressure force on a single galaxy is written as 
\begin{equation}
F_{\rm RP}(R) \approx \pi R_{\rm int}^2 \rho_{\rm ICM}(R) v^2 
\end{equation}
(Sarazin 1988), where $R_{\rm int}$ is the effective interaction
radius of the moving galaxy, and $\rho_{\rm ICM}(R)$ is the ICM mass
density distribution. Considering the observations mentioned just
above, we may assume $R_{\rm int} \approx R_{\rm D}$ (see also Makishima et
al. 2001), where $R_{\rm D}$ is the radius of the galaxy disk. In our
calculation, let us employ $R_{\rm D}$ that is determined from its scaling with the galaxy stellar mass, based on the SDSS result
presented in Figure 15 of Fathi et al. (2010). For reference, this implies that galaxies
with stellar mass of $10^{10}$ $M_{\rm \odot}$ and $10^{11}$ $M_{\rm
  \odot}$ have on average $R_{\rm D} \approx 2$ kpc and 6 kpc,
respectively. The $\rho_{\rm ICM}(R)$ profile was modeled
by a $\beta$ profile (\S3.2), by the sample-averaged parameters of
$\beta=0.44$ and $R_{\rm c} =145$ kpc (Table 4).

The ram pressure primarily affects the galactic interstellar medium,
which has a considerably large collision cross section than the stellar
component. The gas disk will start to be removed when the ram pressure
force reaches the gravitational force, $F_{\rm G} = 2 \pi
G \sigma_{\rm D} M_{\rm ISM}$, where $\sigma_{\rm D}$ is the surface
density of the disk, and $M_{\rm ISM}$ is the mass of interstellar
medium in the disk (Gunn \& Gott 1972). Instantaneous gas
stripping continues for a rather short timescale $\sim 10^7 - 10^8$ yrs
(Quilis et al. 2000), and the remaining disk shrinks to satisfy
$F_{\rm RP} = F_{\rm G}$. The
$R_{\rm D}$ evolution may be predicted by employing the empirical relation between
$R_{\rm D}$ and ram pressure, $\rho_{\rm ICM}(R) v^2$, given by a
numerical simulation of Roediger \& Hensler (2004; Eq.24 therein). For
reference, this relation predicts that $R_{\rm D}$ decreases by 22\%
and 47\% from the original value when $\rho_{\rm ICM}(R) v^2$ increases by one and two orders of magnitude, respectively.

By including the ram pressure, the galaxy orbit decay equation (Eq.8) can be updated as
 \begin{equation}
\frac{{\rm d}R}{{\rm d}t} \approx \frac{4 \pi \rho(R) {\rm G}^2 m R} {(k+1) v^3} +
\frac{\pi R_{\rm int}^2 \rho_{\rm ICM}(R) v R} {(k+1) m}. 
\end{equation}
We employ the same model cluster as in \S 4.3.1 to
calculate the orbit evolution for galaxies with different masses. By
numerically solving Eq.10, the orbit evolution has been modified as plotted in red
in Figure 15. In contrast to the profiles with dynamical
friction alone, which affects massive galaxies mostly, the new
profiles predict a similar infall for the low-mass and high-mass model
galaxies. In $\sim 5$ Gyrs, a $10^{12}$ $M_{\rm
  \odot}$($10^{11}$ $M_{\rm \odot}$) galaxy drops by $\approx
10$\%(12\%), 21\%(20\%), and $>85$\%(35\%) from the initial radius
$r_{0} = 700$ kpc, $500$ kpc, and $300$ kpc, respectively. This
indicates that, while the dynamical friction may account for the
infall of most massive galaxies, the galaxy-ICM interaction (e.g., ram
pressure) should be responsible for the sink of intermediate- and low-
mass galaxies towards cluster center. As proposed in, e.g., Krick et al. (2006),
the history of such galaxy-environment interaction might be recorded by
the intracluster light which has been observed extensively
in cluster field.


The energy loss rate during the galaxy infall can be represented by $L
= (F_{\rm DF} + F_{\rm RP}) \times v$. As shown in Figure 15, the
model galaxy with $m = 1 \times 10^{11}$ $M_{\rm \odot}$ suffers an
average loss rate of $2.9 \times 10^{42}$ erg $\rm s^{-1}$, $4.0
.\times 10^{42}$ erg $\rm s^{-1}$, and $5.6 \times 10^{42}$ erg $\rm
s^{-1}$ when falling from 700 kpc, 500 kpc, and 300 kpc,
respectively. The $m = 1 \times 10^{12}$ $M_{\rm \odot}$ galaxy loses
energy at a higher rate, $\sim 3.3 \times 10^{43}$ erg $\rm s^{-1}$,
$7.4 \times 10^{43}$ erg $\rm s^{-1}$, and $1.1 \times 10^{44}$ erg
$\rm s^{-1}$ for the three starting radii. If we assume that 10\% of
the output energy is converted to the ICM thermal energy, the
heating luminosity of a few hundreds of galaxies in the central 500 kpc
region will reach a few times $10^{44}$ to $10^{45}$ erg $\rm s^{-1}$, which
is sufficient to compensate for the radiative 
cooling of the ICM ($\sim 10^{44}-10^{45}$ erg $\rm s^{-1}$, Table
1). Such a heating by galaxy infall has been successfully reproduced
with numerical simulations (e.g., Asai et al. 2007).
Furthermore, as suggested by more recent numerical works of, e.g., Ruszkowski et al. (2011), 
Ruszkowski \& Oh (2011), and Parrish et al. (2012), gas motion driven by the infall might reshape the magnetic
field in the ICM and enhance inward conductive heating down to the cluster core.


\section{Conclusion}

Based on optical and X-ray data for a sample of 34 relaxed rich clusters with $z=0.1-0.9$, we studied the relative spatial distributions of
the two major baryon components, the cluster galaxies and the ICM, over
the central $0.65$ $R_{500}$ regions. To determine the galaxy light
profiles, we employed two independent methods, i.e., the background
subtraction and the color-magnitude filtering. The ICM mass
profiles was derived from a spatially-resolved spectral analysis using {\it XMM-Newton} and {\it Chandra} data. When binned into three redshift 
subsamples of $z=0.11-0.22$, $z=0.22-0.45$, and $z=0.45-0.89$, the normalized galaxy light vs. ICM mass ratio profiles exhibit steeper drop towards outside
in lower redshift subsamples. A K-S test confirmed that the evolution in the galaxy light vs. ICM mass ratio profile is significant 
at $\geq 94$\% confidence level. A range of measurement uncertainties and radial-/redshift- dependent biases have been assessed, but none of 
them was found significant against the observed galaxy-to-ICM
evolution. Besides, the galaxy light vs. total mass ratio profiles was discovered also to
exhibit concentration towards low redshifts. Our result indicates that
the cluster galaxies have been falling, over $> 6$ Gyr, towards the
center relative to the ICM (and also dark matter component). Such galaxy infall is likely to be caused by
the drag exerted from the interaction between the moving galaxy and
the ICM (e.g., ram pressure), even though the dynamical friction could
enhance the infall of 
the most massive galaxies.

\section*{Acknowledgments}
We thank Yutaka Fujita, Yosuke Matsumoto, Jimmy A. Irwin, Raymond E. White III,
Yuanyuan Su, and Daisuke Nagai for helpful suggestions and comments.
This work was supported by the MOST 973 Programs (gr ant Nos. 2009CB824900,
2009CB824904, and 2013CB837900), and the Grant-in-Aid for Scientific
Research (S) of Japan, No. 18104004, titled "Study of Interactions
between Galaxies and Intra-Cluster Plasmas". L. G. was supported by
the Grand-in-Aid for JSPS fellows, through the JSPS Postdoctoral
Fellowship program for Foreign Researchers. H. X. was supported by the NSFC
(Grant Nos. 10973010, 11125313, and 11261140641).

\begin{deluxetable}{lccccccccc}
\centering
\tabletypesize{\tiny} \tablecaption{Sample clusters
\label{tbl:Genlog} }\tablewidth{0pt} \tablecolumns{9} 
\tablehead{
\colhead{Cluster name} & \colhead{RA} & \colhead{Dec} &
\colhead{z} & \colhead{subsample} & \colhead{BCG name} &
\colhead{$L_{\rm x, bol}$} & \colhead{$T_{\rm aver}^{\rm lite}$} & \colhead{$R_{\rm
    500}^{\rm lite}$}\tablenotemark{a} \\
  & \colhead{(h m s; J2000)} & \colhead{(d m s; J2000)} &  \colhead{} &  \colhead{} &  \colhead{} &\colhead{($10^{44}  {\rm erg} {\rm s^{-1}}$)} & \colhead{(keV)}  & \colhead{(kpc)} & }

\startdata

RX J1141.4-1216 & 11 41 24.0 & -12 16 00 & 0.1195 & L & 2MASX
J11412420-1216386 & 3.8 & 3.3 & $885 $\tablenotemark{1} \\
RX J1044.5-0704 & 10 44 37.1 & -07 04 47 & 0.1323 & L & 2MASX J10443287-0704074 & 7.4 & 3.4 & 932\tablenotemark{1} \\ 
Abell 2204 & 16 32 45.7 & +05 34 43 & 0.1522 & L & TXS 1630+056 & 38.0 & 6.8 & 1370\tablenotemark{2} \\
RX J0958.3-1103 & 09 58 21.1 & -11 03 22 & 0.1527 & L & 2MASX J09582201-1103500 & 10.4 & 5.3 & 1120\tablenotemark{2}\\
Abell 665 & 08 30 45.2 & +65 52 55 & 0.1819 & L & 2MASX J08305736+6550299 & 21.0 & 7.5 & 1370\tablenotemark{2}\\
Abell 383 & 02 48 06.9 & -03 29 32 & 0.1871 & L & 2MASX J02480342-0331447 & 7.6 & 4.5 & 980$^{3}$\\
Abell 963 & 10 17 13.9 & +39 01 31 & 0.2055 & L & 2MASX J10170363+3902500 & 11.6 & 6.4 & 1140$^{3}$\\
RX J1504.1-0248 & 15 04 07.5 & -02 48 16 & 0.2153 & L & 2MASX J15040752-0248161 & 57.6 & 7.9 & 1298$^{4}$\\
MS 0735.6+7421 & 07 41 50.2 & +74 14 51 & 0.2160 & L &  2MASX
J07414444+7414395 & 8.7 & 4.7 & 941$^{2}$ \\		
Abell 2261 & 17 22 28.3 & +32 09 13 & 0.2240 & M & 2MASX J17222717+3207571 & 26.6 & 7.2 & 1310$^{2}$ \\
Abell 267 & 01 52 52.2 & +01 02 46 & 0.2310 & M & 2MASX
J01524199+0100257 & 11.1 & 4.9 & 1040$^{2}$ \\
RX J2129.3+0005 & 21 29 37.9 & +00 05 39 & 0.2350 &  M & 2MASX J21293995+0005207 & 20.2 & 5.6 & 1200$^{2}$\\
MS 1231.3+1542 & 	12 33 55.3 & +15 25 58 & 0.2380 & M & 2MASX
J12335533+1525593 & 4.2 & 4.5  & 906$^{2}$\\
Abell 1835 & 14 01 02.0 & +02 51 32 & 0.2532 & M & 2MASX J14010204+0252423 & 51.8 & 7.2 & 1220$^{3}$ \\
Abell 68 & 00 36 59.4 & +09 08 30 & 0.2550 & M & 2MASX J00370686+0909236 & 16.7 & 8.1 & 1190$^{2}$ \\
CIZA J1938.3+5409 & 19 38 18.6 & +54 09 33 & 0.2600 & M & 2MASX
J19381810+5409402 & 20.3 & 6.8  & 1165$^{2}$\\
Abell 1576 & 12 36 49.1 & +63 11 30 & 0.2790 & M & 2MASX J12365866+6311145 & 38.4 & 7.6 & 1300$^{5}$ \\  
RX J0437.1+0043 & 04 37 09.8 & +00 43 37 & 0.2850 & M & 2MASX J04370955+0043533 & 6.2 & 5.1 & 1170$^{6}$ \\
Abell 959 & 10 17 34.3 & +59 33 39 & 0.2883 & M & 2MASX J10173435+5933390 & 29.4& 7.5 & 1177$^{5}$  \\
ZwCl 3146 & 10 23 39.6 & +04 11 10 & 0.2906 & M & 2MASX J10233960+0411116 & 44.4 & 6.4 & 1300$^{2}$  \\
RX J2308.3-0211 & 23 08 16.4 & -02 10 44 & 0.2950 & M & 2MASX J23082221-0211315 & 12.0 & 7.9 & 1240$^{6}$  \\
MS 1241.5+1710 & 12 43 59.9 & +16 53 46 & 0.3120 & M  & SDSS J124357.98+165354.1 & 22.6 & 6.1 & 1055$^{2}$ \\
RBS 797 & 09 47 12.9 & +76 23 13 & 0.3540 & M & RHS 30 & 45.9 & 6.0 & 1130$^{2}$\\
Cl 0024.0+1652 & 00 26 36.0 & +17 08 36 & 0.3900 & M & 2MASXi J0026356+170943 & 4.0 & 3.5 & 740$^{7}$  \\
MS 0302.7+1658 & 03 05 33.9 & +17 10 06 & 0.4240 & M & GMF2004 J46.3821+17.1674 & 6.4 & 3.8 & 899$^{8}$ \\
RX J0030.5+2618 & 00 30 33.2 & +26 18 19 & 0.5000 & H & BWE97 J003034.0+261808 & 5.4 & 5.7 & 906$^{8}$  \\
MS 0451.6-0305 & 04 54 10.9 & -03 01 07 & 0.5500 & H & PKB2009 06 & 50.9  & 8.0 & 1263$^{8}$ \\
 Cl 0016+16 & 00 18 33.3 & +16 26 36 & 0.5410 & H & SES2008 J001833.68+162615.1 & 50.8 & 8.9 & 1190$^{7}$ \\
 MS 2053.7-0449 & 20 56 22.4 & -04 37 43 & 0.5830 & H & CXOU J205621.2-043749 & 5.40 & 5.5 & 931$^{8}$ \\
 RX J1120.1+4318 & 11 20 07.2 & +43 18 12 & 0.6000 & H & SDSS J112005.61+431809.0 & 13.0 & 4.9 & 940$^{7}$  \\
 RX J1334.3+5030 & 13 34 20.0 & +50 30 54 & 0.6200 & H & SDSS J133420.56+503103.5 & 7.5 & 4.6 & 780$^{7}$ \\
 MS 1137.5+6624 & 11 40 23.3 & +66 08 41 & 0.7820 & H & MS 1137.5+6625:DVS99 09 & 15.3 & 6.9 & 964$^{8}$  \\
 RX J1716.4+6708 & 17 16 49.6 & +67 08 30 & 0.8130 & H & RX J1716.6+6708:HGM97 08 & 13.9 & 6.8  & 896$^{8}$\\
 WARP J1226.9+3332 & 12 26 58.0 & +33 32 54 & 0.8900 & H & SDSS J122653.11+333330.8 & 54.6 & 11.2 & 1134$^{8}$ \\								

\enddata
\scriptsize
\tablenotetext{a}{X-ray luminosity, ICM temperature, and $R_{500}^{\rm
    lite}$
  cited from (1) Pratt et al. (2009); (2) Maughan et al. (2008); (3) Zhang
et al. (2007); (4) Santos et al. (2010); (5) Novicki et al. (2002); (6)
Zhang et al. (2006); (7) Kotov \& Vikhlinin (2005); and (8) Ettori et al. (2004).}

\end{deluxetable}

\begin{deluxetable}{lcccccc}
\tabletypesize{\scriptsize} \tablecaption{Summary of X-ray Observations
\label{tbl:XObsLog}} \tablewidth{0pt} \tablecolumns{10} \tablehead{
\colhead{Cluster name} & \colhead{Date} & \colhead{Detector} & \colhead{ObsID} &
\colhead{Mode}\tablenotemark{a} & 
\colhead{Raw/Clean Exposure}\tablenotemark{a}\\
& \colhead{dd mm yyyy} & \colhead{}
& & & \colhead{(ks)} } \
\startdata

RX J1141.4-1216 & 09/07/2004 & {\it XMM-Newton} EPIC  & 0201901601 & FF(EFF) & 32.3/27.6(32.8/--) \\
RX J1044.5-0704 & 23/12/2004 & {\it XMM-Newton} EPIC & 0201901501 & FF(EFF) & 29.2/23.4(25.2/17.5) \\
Abell 2204 & 08/02/2006 &  {\it XMM-Newton} EPIC  & 0306490201 & FF(EFF) & 22.9/14.0(21.2/4.7) \\
RX J0958.3-1103 & 27/05/2006 & {\it XMM-Newton} EPIC & 0404910601 & FF(EFF) & 14.7/6.8(11.3/4.1) \\
Abell 665  & 04/04/2001 &   {\it XMM-Newton} EPIC & 0109890401 & FF(EFF) & 40.4/19.7(--/--) \\
Abell 383 & 16/08/2002 & {\it XMM-Newton} EPIC & 0084230501 & FF(EFF) & 33.4/28.1(30.0/14.7) \\
Abell 963 & 02/11/2001 &  {\it XMM-Newton} EPIC & 0084230701 & FF(EFF) & 26.9/24.0(22.5/14.6) \\
RX J1504.1-0248 & 22/01/2007 & {\it XMM-Newton} EPIC & 0401040101 & FF(EFF) & 38.8/35.6(34.9/25.9) \\
MS 0735.6+7421 & 01/04/2005 & {\it XMM-Newton} EPIC & 0303950101 & FF(FF) & 71.6/46.9(69.9/35.1) \\
Abell 2261 & 15/08/2003 &  {\it XMM-Newton} EPIC & 0093030301 & FF(EFF) & 6.4/2.3(7.8/0.9) \\
Abell 267 & 02/01/2002 &  {\it XMM-Newton} EPIC & 0084230401 & FF(EFF) & 18.6/16.5(25.2/11.1) \\
RX J2129.3+0005 & 29/10/2002 & {\it XMM-Newton} EPIC &  0093030201 & FF(EFF) & 58.7/28.7(55.3/14.5) \\
MS 1231.3+1542 & 19/12/2006 & {\it XMM-Newton} EPIC & 0404120101 & FF(FF) & 31.6/30.7(30.0/25.4) \\
Abell 1835 &  25/07/2008 & {\it XMM-Newton} EPIC  & 0551830201 & FF(EFF) & --/--(100.3/21.5) \\
Abell 68 & 14/12/2001 & {\it XMM-Newton} EPIC & 0084230201 & FF(EFF) & 29.6/27.4(25.2/15.2) \\
CIZA J1938.3+5409 & 28/03/2007 & {\it XMM-Newton} EPIC & 0404120301 & FF(FF) & 7.7/1.1(4.1/0.9) \\
Abell 1576 & 25/10/2006 & {\it XMM-Newton} EPIC & 0402250101 & FF(EFF) & 20.6/6.9(17.1/1.8) \\
RX J0437.1+0043 & 05/03/2004 & {\it XMM-Newton} EPIC & 0205330201 & FF(FF) & 10.4/4.0(8.7/2.6) \\
Abell 959 & 12/04/2007 & {\it XMM-Newton} EPIC & 0406630201 & FF(EFF) & 23.7/7.9(25.4/2.1) \\
ZwCl 3146 & 13/12/2009 &  {\it XMM-Newton} EPIC & 0605540201 & FF(EFF) & 121.9/99.7(118.5/72.0) \\
RX J2308.3-0211 & 10/06/2001 & {\it XMM-Newton} EPIC & 0042341201 & FF(EFF) & 11.7/8.0(10.0/6.8) \\
MS 1241.5+1710 &  22/06/2005 & {\it XMM-Newton} EPIC & 0302581501 & FF(EFF) & 29.3/28.9(23.5/20.1) \\
RBS 797 & 15/04/2008 & {\it XMM-Newton} EPIC &  0502940301 & FF(FF) & 31.6/14.5(28.3/10.3) \\
Cl 0024.0+1652 & 06/01/2001  & {\it XMM-Newton} EPIC & 0050140201 & FF(EFF) & 52.1/47.0(48.2/37.0) \\
MS 0302.7+1658 & 23/08/2002 & {\it XMM-Newton} EPIC & 0112190101 & FF(EFF) & 13.4/11.4(10.0/5.7) \\
RX J0030.5+2618 & 17/08/1999 & {\it Chandra} ACIS-S & 1190 & VFaint & 18.1/15.2 \\
MS 0451.6-0305 & 08/10/2000 &  {\it Chandra} ACIS-S & 902 & Faint &  44.8/41.2 \\
Cl 0016+16 & 18/08/2000 &   {\it Chandra} ACIS-I & 520 & VFaint & 68.3/67.4 \\
MS 2053.7-0449 & 07/10/2001 & {\it Chandra} ACIS-I & 1667 & VFaint & 45.1/33.2 \\
RX J1120.1+4318 & 11/01/2005 & {\it Chandra} ACIS-I  & 5771 & VFaint & 20.1/19.3 \\
RX J1334.3+5030 & 05/08/2005 & {\it Chandra} ACIS-I & 5772 & VFaint & 19.8/5.9 \\
MS 1137.5+6624 & 30/09/1999 &  {\it Chandra} ACIS-I &  536 & VFaint & 119.2/117.3  \\
RX J1716.4+6708 & 27/02/2000 & {\it Chandra} ACIS-I & 548 & Faint & 52.4/51.7 \\
WARP J1226.9+3332 & 07/08/2004 & {\it Chandra} ACIS-I  & 5014 & VFaint & 33.2/32.7 \\

\enddata
\scriptsize
\tablenotetext{a}{Mode and exposure time of the {\it XMM-Newton} pn and MOS data
  are shown within and without brackets, respectively.}

\end{deluxetable}

\begin{deluxetable}{lcccccc}
\centering
\tabletypesize{\tiny} \tablecaption{Summary of Optical Observations
\label{tbl:OObsLog}} \tablewidth{0pt} \tablecolumns{10} \tablehead{
\colhead{Cluster name} & \colhead{Date} & \colhead{Passband} & \colhead{Exposure} &
\colhead{Limited Magnitude} & 
\colhead{Seeing}\tablenotemark{a}\\
& \colhead{dd mm yyyy} & \colhead{}
& \colhead{(s)} & & (arcsec)   } \
\startdata

RX J1141.4-1216 & 21/03/2010 & $g^{\prime}$  & 500 & 25.0 & $0.8$ \\
 			     & 21/03/2010 & $r^{\prime}$ &  500 & 24.7 & $0.8$ \\
			     & 21/03/2010 & $i^{\prime}$ & 500 & 24.6 & $0.8$ \\
			     & 21/03/2010 & $i^{\prime}$ offset & 500 & 24.4 & $0.8$ \\
RX J1044.5-0704 & 16/02/2010 & $B$ & 1000 & 24.2 & $1.2$ \\
			     & 16/02/2010 & $R$ & 500 & 24.2  & $1.2$ \\
			     & 16/02/2010 & $I$ & 500 & 23.8 & $1.2$ \\
			     & 16/02/2010 & $I$ offset & 500 & 24.1 & $1.2$ \\
Abell 2204 & 20/03/2010 & $g^{\prime}$ & 500 & 24.9 & $0.8$\\
		  & 20/03/2010 & $r^{\prime}$ & 500 & 25.0 & $0.8$\\
		  & 20/03/2010 & $i^{\prime}$ & 500 & 24.8 & $0.8$\\
		  & 20/03/2010 & $i^{\prime}$ offset & 500 & 24.6 & $0.8$\\
RX J0958.3-1103 & 16/02/2010 & $B$ & 1000 & 25.0 & $1.0$ \\
			     & 16/02/2010 & $R$ & 500 & 24.2 & $1.0$ \\
			     & 16/02/2010 & $I$ & 500 & 23.8 & $1.0$ \\
			     & 16/02/2010 & $I$ offset & 500 & 24.0 & $1.0$ \\
Abell 665  & 20/03/2010 & $g^{\prime}$  & 500 & 25.5 & $0.8$ \\
		 & 20/03/2010 & $r^{\prime}$ & 500 & 25.3 &  $0.8$ \\
		 & 20/03/2010 & $i^{\prime}$ & 500 & 24.7 &  $0.8$ \\
		 & 20/03/2010 &  $i^{\prime}$ offset & 500 & 24.8 & $0.8$ \\  
Abell 383 & 16/08/2010 & $B$ & 300 & 24.2 &  $0.5$ \\
		& 16/08/2010 & $I$ & 300 & 24.0 & $0.5$ \\
		& 16/08/2010 & $I$ offset & 300 & 23.7 & $0.5$ \\
Abell 963 & 15/02/2010 & $B$ & 1000 & 24.8 &  $0.8$ \\
	         & 15/02/2010 & $R$ & 500 & 24.2 & $0.8$ \\
	         & 15/02/2010 & $I$ & 500 & 24.0 &  $0.8$ \\
	         & 15/02/2010 & $I$ offset & 500 & 23.9 &  $0.8$ \\
RX J1504.1-0248 & 20/03/2010 & $g^{\prime}$ & 500 & 25.3 &  $0.8$ \\
	  		     & 20/03/2010 & $r^{\prime}$ & 500 & 25.4 & $0.8$ \\
			     & 20/03/2010 & $i^{\prime}$ & 500 & 25.0 & $0.8$ \\
			     & 20/03/2010 & $i^{\prime}$ offset & 500 & 24.8 & $0.8$ \\
MS 0735.6+7421 & 15/02/2010 & $B$ & 500 & 25.0 & $0.5$ \\
			    & 15/02/2010 & $R$ & 500 & 24.5 & $0.5$ \\
			    & 15/02/2010 & $I$ & 500 & 23.8 &  $0.5$ \\
			    & 15/02/2010 & $I$ offset & 500 & 24.2 &  $0.5$ \\ 
Abell 2261 & 16/08/2010 & $B$ & 300 & 23.8 & $0.5$ \\
		  & 16/08/2010 & $R$ & 300 & 24.5 & $0.5$ \\
		  & 16/08/2010 & $I$ & 300 & 23.9 & $0.5$ \\
		  & 16/08/2010 & $I$ offset & 300 & 24.1 & $0.5$ \\
Abell 267 	  & 16/08/2010 & $B$ & 300 & 24.1 & $0.5$ \\
		  & 16/08/2010 & $R$ & 300 & 24.0 & $0.5$ \\
		  & 16/08/2010 & $I$ & 300 & 24.0 & $0.5$ \\
		  & 16/08/2010 & $I$ offset & 300 & 23.9 & $0.5$ \\
RX J2129.3+0005 & 13/08/2010 & $B$ & 300 & 24.2 & $0.5$ \\
			      & 13/08/2010 & $R$ & 300 & 24.4 & $0.5$ \\
			      & 13/08/2010 & $I$ & 300 & 23.6 & $0.5$ \\
			      & 13/08/2010 & $I$ offset & 300 & 24.0 & $0.5$ \\
MS 1231.3+1542 & 21/03/2010 &  $g^{\prime}$ & 500 & 25.4 &  $0.8$ \\
			    & 21/03/2010 &  $r^{\prime}$ & 500 & 25.1 &  $0.8$ \\
			    & 21/03/2010 &  $r^{\prime}$ & 500 & 25.2 &  $0.8$ \\
			    & 21/03/2010 &  $r^{\prime}$ offset & 500 & 25.1 &  $0.8$ \\
Abell 1835 & 15/02/2010 & $B$ & 1000 & 24.6 & $0.6$ \\
		  & 15/02/2010 & $R$ & 400 & 23.7 & $0.6$ \\
		  & 15/02/2010 & $I$ & 500 & 24.4 & $0.6$ \\
Abell 68 & 13/08/2010 & $B$ & 400 & 25.0 & $0.7$ \\
	       & 13/08/2010 & $R$ & 400 & 23.9 & $0.7$ \\
	       & 13/08/2010 & $I$ & 400 & 23.7 & $0.7$ \\
	       & 13/08/2010 & $I$ offset & 400 & 23.7 & $0.7$ \\
CIZA J1938.3+5409 & 13/08/2010 & $B$ & 400 & 24.4 & $0.5$ \\
	      			& 13/08/2010 & $R$ & 400 & 24.3 & $0.5$ \\
				& 13/08/2010 & $I$ & 400 & 24.3 & $0.5$ \\
				& 13/08/2010 & $I$ offset & 400 & 23.9 & $0.5$ \\
Abell 1576 & 21/03/2010 & $g^{\prime}$ & 500 & 25.3 &  $0.8$ \\
		  & 21/03/2010 & $r^{\prime}$ & 500 & 25.4 &  $0.8$ \\
		  & 21/03/2010 & $i^{\prime}$ & 500 & 25.2 &  $0.8$ \\
		  & 21/03/2010 & $r^{\prime}$ offset & 500 & 25.0 &  $0.8$ \\
RX J0437.1+0043 & 15/02/2010 & $B$ & 1000 & 25.4 & $0.5$ \\
			      & 15/02/2010 & $R$ & 500 & 24.8 & $0.5$ \\
			      & 15/02/2010 & $I$ & 500 & 24.7 & $0.5$ \\
			      & 15/02/2010 & $I$ offset & 500 & 24.6 & $0.5$ \\
Abell 959 & 16/02/2010 & $B$ & 500 & 24.1 &  $1.2$ \\
		& 16/02/2010 & $R$ & 500 & 24.7 &  $1.2$ \\
		& 16/02/2010 & $I$ & 500 & 24.6 &  $1.2$ \\
		& 16/02/2010 & $I$ offset & 500 & 24.5 &  $1.2$ \\
ZwCl 3146 & 15/02/2010 & $B$ & 1000 & 24.9 & $0.9$ \\ 
		  & 15/02/2010 & $R$ & 500 & 24.7 & $0.9$ \\
		  & 15/02/2010 & $I$ & 500 & 24.9 & $0.9$ \\  
		  & 15/02/2010 & $I$ offset & 500 & 24.7 & $0.9$ \\  
RX J2308.3-0211 & 16/08/2010 & $B$ & 400 & 25.0 & $0.5$ \\
			     & 16/08/2010 & $R$ & 400 & 24.6 & $0.5$ \\
			     & 16/08/2010 & $I$ & 400 & 24.8 & $0.5$ \\ 
			     & 16/08/2010 & $I$ offset & 400 & 24.8 & $0.5$ \\
MS 1241.5+1710 &  21/03/2010 & $g^{\prime}$ & 500 & 25.7 &  $0.9$ \\
			    &  21/03/2010 & $r^{\prime}$ & 500 & 24.9 &  $0.9$ \\
			    &  21/03/2010 & $i^{\prime}$ & 500 & 24.7 &  $0.9$ \\
			    &  21/03/2010 & $i^{\prime}$ offset & 500 & 25.0 &  $0.9$ \\
RBS 797 & 15/02/2010 & $B$ & 1000 & 24.7 & $0.8$ \\
	        & 15/02/2010 & $R$ & 500 & 24.6 & $0.8$ \\
	        & 15/02/2010 & $I$ & 500 & 23.7 & $0.8$ \\
	        & 15/02/2010 & $I$ offset & 500 & 23.7 & $0.8$ \\
Cl 0024.0+1652 & 13/08/2010 & $B$ & 800 & 24.6 & $0.7$ \\
			  & 13/08/2010 & $R$ & 800 & 23.9 & $0.7$ \\ 
			  & 13/08/2010 & $I$ & 800 & 23.7 & $0.7$ \\
			  & 13/08/2010 & $I$ offset & 800 & 24.0 & $0.7$ \\
MS 0302.7+1658 & 13/08/2010 & $B$ & 800 & 24.3 & $0.7$ \\
			    & 13/08/2010 & $R$ & 800 & 23.8 & $0.7$ \\ 
			    & 13/08/2010 & $I$ & 800 & 23.8 & $0.7$ \\
			    & 13/08/2010 & $I$ offset & 800 & 23.7 & $0.7$ \\
RX J0030.5+2618 & 16/08/2010 & $B$ & 1200 & 25.1 & $0.5$ \\
			      & 16/08/2010 & $R$ & 1200 & 24.7 & $0.5$ \\
			      & 16/08/2010 & $I$ & 1200 & 24.3 & $0.5$ \\ 
			      & 16/08/2010 & $I$ offset & 1200 & 24.4 & $0.5$ \\
MS 0451.6-0305 & 16/02/2010 & $B$ & 1600 & 24.5 & $1.2$ \\
			   & 16/02/2010 & $R$ & 800 & 23.6 & $1.2$ \\
			   & 16/02/2010 & $I$ & 800 & 23.5 & $1.2$ \\
			   & 16/02/2010 & $I$ offset & 800 & 23.4 & $1.2$ \\
Cl 0016+16 & 16/08/2010 & $B$ & 800 & 24.6 & $0.5$ \\
		    & 16/08/2010 & $R$ & 800 & 24.6 & $0.5$ \\
		    & 16/08/2010 & $I$ & 1200 & 24.3 & $0.5$ \\
		    & 16/08/2010 & $I$ offset & 1200 & 24.3 & $0.5$ \\
MS 2053.7-0449 & 16/08/2010 & $B$ & 800 & 24.1 & $0.7$ \\
			   & 16/08/2010 & $R$ & 800 & 24.5 & $0.7$ \\
			   & 16/08/2010 & $I$ & 5000 & 25.3 & $0.7$ \\
RX J1120.1+4318 & 15/02/2010 & $B$ & 1600 & 25.3 & $0.6$ \\
			      & 15/02/2010 & $R$ & 800 & 24.5 & $0.6$ \\
			      & 15/02/2010 & $I$ & 800 & 24.2 & $0.6$ \\
			      & 15/02/2010 & $I$ offset & 800 & 24.2 & $0.6$ \\
RX J1334.3+5030 & 21/03/2010 & $g^{\prime}$ & 800 & 25.5 &  $0.8$ \\
			      & 21/03/2010 & $r^{\prime}$ & 1200 & 25.7 &  $0.8$ \\
			      & 21/03/2010 & $i^{\prime}$ & 800 & 25.1 &  $0.8$ \\
			      & 21/03/2010 & $i^{\prime}$ offset & 800 & 25.0 &  $0.8$ \\
MS 1137.5+6624 & 20/03/2010 & $g^{\prime}$ & 800 & 25.5 &  $0.8$ \\
			    & 20/03/2010 & $r^{\prime}$ & 1200 & 25.8 &  $0.8$ \\
			    & 20/03/2010 & $i^{\prime}$ & 800 & 25.5 &  $0.8$ \\
			    & 20/03/2010 & $i^{\prime}$ offset & 800 & 25.5 &  $0.8$ \\
RX J1716.4+6708 & 13/08/2010 & $R$ & 800 & 24.2 &  $0.7$ \\
			      & 13/08/2010 & $I$ & 7200 & 25.3 &  $0.7$ \\
WARP J1226.9+3332 & 20/03/2010 & $g^{\prime}$ & 800 & 25.6 &  $0.8$ \\
				   & 20/03/2010 & $r^{\prime}$ & 1200 & 25.5 &  $0.8$ \\
				   & 20/03/2010 & $i^{\prime}$ & 800 & 25.4 &  $0.8$ \\
				   & 20/03/2010 & $i^{\prime}$ offset &
800 & 25.3 &  $0.8$ \\
\hline

\enddata
\scriptsize
\tablenotetext{a}{Seeings were estimated from the combined frames.}

\end{deluxetable}

\begin{deluxetable}{cccccccccc@{}}

  \tabletypesize{\tiny} \tablecaption{Summary of X-ray fittings}
 \tablehead{            
& \multicolumn{2}{c}{Center Offset} & & & \multicolumn{3}{c}{ICM Density Model} \\
\cline{2-3}
\cline{6-8}
    Cluster name &  center I-III  &
    center II-III & $T_{\rm aver}$ & $R_{500}$  &
    $\beta$\tablenotemark{a}& $R_{\rm c}$ or $R_{\rm c,1}$, $R_{\rm c,2}$ \tablenotemark{a} &
    $\chi^2/\nu$ & $S/B_{0.65}$\\
 & (kpc) & (kpc) & (keV) & (kpc) & & (kpc) & & &}
\startdata
RX J1141.4-1216 &  11.2 & 11.6 & $3.4\pm0.1$ & 804 & $0.39 \pm 0.01$ & $54.9 \pm 5.0$ &  6.8/5 & 2.0 \\

RX J1044.5-0704 & 3.3 & 4.4 & $4.3\pm0.1$ & 927  & $0.42\pm 0.01$ & $63.9 \pm 6.2$ & 6.5/5 & 1.1 \\

Abell 2204 & 9.9 & 9.1 & $6.9\pm0.2$ & 1239 &  $0.42 \pm 0.01$ & $71.6 \pm 4.2$ & 12.1/8 & 0.8 \\

RX J0958.3-1103 & 0.6 & 6.0 & $5.5 \pm 0.2$ & 1063 & $0.42\pm 0.02$ & $95.5 \pm 6.3$ &  3.2/5 & 1.4 \\
                            
 Abell 665 & 46.8 & 38.8 &$7.4 \pm 0.2$ & 1272 & $0.44 \pm 0.01$ & $223.1 \pm 10.0$ & 10.8/9 & 3.5 \\

Abell 383 & 13.0 & 16.5 &$4.1 \pm 0.1$ & 877  & $0.45\pm 0.01$ & $22.1 \pm 1.2$ & 2.9/8 & 2.2 \\

Abell 963 & 19.9 & 14.2 & $6.6 \pm 0.1$ & 1167 &  $0.53 \pm 0.01$ & $101.1 \pm 7.0$, $383.3 \pm 25.0$ & 12.4/10 & 0.7 \\

RX J1504.1-0248 & 7.2 & 8.7 & $6.8 \pm 0.1$ & 1185  & $0.46\pm 0.02$ & $84.9 \pm 3.2$  & 12.5/8 & 1.0 \\

MS 0735.6+7421 & 4.1 & 2.3 & $4.7 \pm 0.1$& 941 & $0.46\pm 0.01$ & $143.3 \pm 3.5$ & 6.9/5 & 2.6 \\

Abell 2261 & 21.1 & 18.4  & $5.8 \pm 0.3$ & 1065  & $0.43\pm 0.01$ & $160.1 \pm 9.0$ & 1.1/3 & 1.7 \\
                            
 Abell 267 & 15.2 & 12.4 &$5.1 \pm 0.1$ & 987  & $0.48 \pm 0.01$ & $217.9 \pm 11.0$ & 7.8/10 & 1.5 \\

RX J2129.3+0005 & 21.8 & 20.1 & $5.3 \pm 0.1$  & 1001 &  $0.43\pm 0.01$ & $108.7 \pm 2.4$ & 10.9/7 & 1.0 \\

MS 1231.3+1542  & 42.1 & 38.1 & $4.5 \pm 0.1$ & 906 & $0.39\pm 0.01$ & $193.1 \pm 11.6$ & 9.5/7 & 1.2 \\

Abell 1835 & 29.0 &  29,8  & $7.7 \pm 0.1 $ & 1257  &  $0.60 \pm 0.01 $  & $75.2 \pm 8.4$, $887.4 \pm 214.2$ & 9.6/7 & 1.0 \\

Abell 68 & 21.4 & 19.6  & $6.9 \pm 0.2$ & 1179  &  $0.54 \pm 0.01$ & $11.5 \pm 5.4$, $359.2 \pm 102.2$  & 8.4/10 & 1.2 \\

CIZA J1938.3+5409 & 20.0 & 7.9 & $6.8 \pm 0.8$ & 1165 & $0.47\pm 0.02$ & $218.4 \pm 23.0$ & 2.0/3 & 1.0 \\

Abell 1576 & 30.4 & 35.5 & $6.3 \pm 0.3$ & 1099  & $0.51 \pm 0.04$  & $7.6 \pm 7.2$, $324.9 \pm 22.1$ & 12.3/10 & 1.0 \\
                            
RX J0437.1+0043 & 28.4 & 10.0 & $6.7 \pm 0.3 $ & 1142  &  $0.43 \pm 0.01$ & $121.6 \pm 5.2$ & 12.4/8 & 0.9 \\

Abell 959  & 16.5 & 12.6 & $8.1 \pm 0.8$ & 1280  & $0.49\pm 0.01$ & $108.7 \pm 2.4$ & 6.5/8 & 0.7 \\

Zwcl 3146 &12.4 & 18.3 & $6.5 \pm 0.1$ & 1110  & $0.46\pm 0.01$ & $116.5 \pm 1.0$ & 13.3/9 & 1.5 \\

RX J2308.3-0211 & 31.0 &  28.7 &$ 6.5 \pm 0.3$ & 1110 & $0.41 \pm 0.01$ & $141.3 \pm 6.9$ & 9.7/8 & 0.7 \\

MS 1241.5+1710 & 24.2 & 26.2 & $4.3 \pm 0.3$ & 849 & $0.43\pm 0.01$ & $72.4 \pm 3.1$ & 10.4/8 & 0.6 \\

RBS 797 & 14.2 & 14.7 & $ 6.4 \pm 0.2$ & 1061 & $0.50\pm 0.01$ & $7.8 \pm 6.5$, $167.9 \pm 65.3$ & 12.8/10 & 2.3 \\

Cl 0024.0+1652 & 24.7 & 24.7 & $3.8 \pm 0.1$& 756  &  $0.44\pm 0.01$ & $154.8 \pm 8.0$ & 6.7/6 & 1.3 \\
                            
MS 0302.7+1658 & 13.4 & 30.1 & $4.4 \pm 0.4$& 805  & $0.42 \pm 0.02$ & $182.4\pm 27.6$ & 4.0/3 & 1.3 \\

RX J0030.5+2618 & 32.2 & 35.2 &$ 5.7 \pm 0.3$ & 916 & $0.40\pm 0.01$ & $179.7 \pm 20.1$ & 11.3/8 & 0.7 \\

MS 0451.6-0305 & 16.9 & 25.7 & $8.6 \pm 0.2$ & 1151  & $0.59\pm 0.02$ & $10.8 \pm 7.7$, $425.5 \pm 124.3$ & 12.2/9 & 2.8 \\

Cl 0016+16  & 9.2 & 14.5 & $8.9\pm 0.3 $ & 1189  &  $0.53\pm 0.02$ & $8.8 \pm 6.7$, $449.5 \pm 45.4$ &  11.2/8 & 3.0\\
                            
MS 2053.7-0449  & 8.1 & 43.8 & $4.8 \pm 0.6 $& 792 & $0.66 \pm 0.04$ & $299.2\pm 60.2$ & 2.5/3 & 0.7 \\

RX J1120.1+4318 & 6.2 & 6.0 &$ 4.7 \pm 0.2$ & 764 &  $0.42\pm 0.01$ & $211.8 \pm 31.3$ & 8.9/7 & 2.3 \\
                      
RX J1334.3+5030 & 19.5 & 21.0 & $4.2 \pm 0.2$ & 710  &  $0.35\pm 0.02$ & $144.5 \pm 24.0$ &  6.5/5 & 1.2 \\

MS 1137.5+6624 & 5.9  & 23.1  & $6.8 \pm 0.9$ & 869  & $0.46\pm 0.02$ & $174.9\pm 19.7$ & 7.6/8 & 1.5 \\
                            
RX J1716.4+6708 & 23.0 & 30.0 & $5.8 \pm 0.6$ & 781  & $0.43 \pm 0.02$ & $167.4\pm 27.5$ & 3.3/5 & 0.9 \\

WARP J1226.9+3332 & 26.9 & 32.6 &  $10.1\pm 0.4$ & 1049  & $0.46\pm 0.01$ & $192.3 \pm 8.7$& 7.9/8 & 0.7 \\

\hline

 \enddata
\scriptsize
\tablenotetext{a}{Best-fit double-$\beta$ model parameters are shown
  for A963, A1835, A68, A1576, RBS 797, MS0451, and CL0016, and
  $\beta$ parameters are shown for the other clusters. }

\end{deluxetable}

\begin{deluxetable}{cccccccccc@{}}

  \tabletypesize{\tiny} \tablecaption{Summary of optical fittings}

 \tablehead{            
& \multicolumn{3}{c}{King Model} &  \multicolumn{3}{c}{NFW
    Model} & \multicolumn{3}{c}{Color-Magnitude Fitting} \\
\cline{2-4}
\cline{5-7}
\cline{8-10}
    Cluster name &  $r_{\rm c}$ \tablenotemark{a} &
    $S_{\rm bkg}$ \tablenotemark{a} &$\chi^2/\nu$& $r_{\rm s}$
    \tablenotemark{a} & $S_{\rm bkg}$\tablenotemark{a} &
    $\chi^2/\nu$&  A \tablenotemark{b}&B \tablenotemark{b}&$\delta
    (m_{\rm R}-m_{\rm I})$ \tablenotemark{b} \\
 & (kpc) & (mag/$\rm arcmin^{2}$) & & (kpc) & (mag/$\rm arcmin^{2}$) & & & & (mag)
 }
\startdata
RX J1141.4-1216  & $16.7 \pm 13.2$ & $ 17.72\pm 0.07$
                            & $3.7/5$ & $59.4 \pm 20.1$ &  $17.65 \pm 0.09$ & 2.0/5 & 0.023 & -0.0040 & 0.12\\

RX J1044.5-0704  &  $79.1 \pm 44.0$   & $18.10 \pm 0.09$ & $5.9/5$
                            & $66.4 \pm 29.9$ & $18.08 \pm 0.09$ & 5.2/5 & 0.59 & -0.030 & 0.11\\

Abell 2204  &  $60.4 \pm 31.8$ & $17.32 \pm 0.15$ & $1.3/5$
                            & $320.3 \pm 122.0$ & $17.32 \pm 0.14$ & 1.8/5 & 0.038 & -0.0025 & 0.19\\

RX J0958.3-1103   &  $ 20.3\pm 18.6$ & $16.96 \pm 0.09$ & $9.2/6$
                            & $ 77.5\pm 32.8$ & $16.95 \pm 0.08$ & 7.2/6  & 0.40 & -0.0046 & 0.11\\
                            
 Abell 665  & $ 25.5\pm 20.9$  & $17.89 \pm 0.16$ & $8.6/6$
                            & $ 62.8\pm 21.0$ & $17.84 \pm 0.18$ & 7.0/6 & 0.36 & -0.013 & 0.13\\

Abell 383  & $39.7 \pm 17.5$  & $19.23 \pm 0.09$ & $9.5/6$
                            & $55.3 \pm 9.9$ & $19.25 \pm 0.08$ & 15.5/6 & 3.99 & -0.051 & 0.23\\

Abell 963   & $295.4\pm 43.7$& $18.02 \pm 0.11$ & 2.3/5
                            & $407.6 \pm 190.9$ & $18.04 \pm 0.13$ & 2.6/5 & 0.67 & -0.025 & 0.15\\

RX J1504.1-0248  & $18.5 \pm 41.3$ & $17.62 \pm 0.08$ & 10.2/6
                            & $59.4 \pm 20.9$ & $17.63 \pm 0.09$ & 9.6/6 &  0.59 &  -0.013 & 0.19\\

MS 0735.6+7421  &  $130.3 \pm 84.3$& $17.74 \pm 0.09$ & 6.6/5
                            & $56.8 \pm 28.2$ & $17.73 \pm 0.09$ & 6.6/5 & 0.77 & -0.021 & 0.15\\

Abell 2261  &  $262.5 \pm 98.9$ & $17.45 \pm 0.08$ & 9.0/5
                            & $800.3 \pm 214.4$ & $17.49 \pm 0.25$ & 7.8/5 & 0.23 & -0.010 & 0.15\\
                            
 Abell 267  &  $381.8\pm 47.3$& $18.44 \pm 0.09$ & $8.7/5$
                            & $257.6 \pm 35.2$ & $18.58 \pm 0.15$ & 4.0/5 & 0.038 & -0.0015 & 0.13\\

RX J2129.3+0005  &  $603.2 \pm 456.2$ & $17.47 \pm 0.10$ & 2.6/5
                            & $329.6 \pm 158.1$ & $17.44 \pm 0.11$ & 3.8/5 & 0.067 & -0.0093 & 0.17 \\

MS 1231.3+1542   & $170.8 \pm 50.8$ & $18.09 \pm 0.10$ & $3.9/5$
                            & $52.1 \pm 18.0$ & $18.09 \pm 0.13$ & 3.8/5 & 0.20 & -0.0013 & 0.15\\

Abell 1835  & $121.9 \pm 93.2 $ & $17.03 \pm 0.69$ & $0.1/3$
                            & $136.9 \pm 28.6$ & $16.97 \pm 0.64$ & 0.1/3 & 0.70 & -0.025 & 0.12\\

Abell 68  & $185.4 \pm 38.0$ & $18.50 \pm 0.09$ & $10.0/6$
                            & $ 90.3\pm 29.2$ & $18.49 \pm 0.10$ & 9.9/6 & 0.25 & -0.0018 & 0.11\\

CIZA J1938.3+5409  & $312.1 \pm 167.7$ & $16.87 \pm 0.10$ & $1.3/5$
                            & $731.9 \pm 408.8$ & $16.87 \pm 0.12$ & 1.3/5 & 0.093 & -0.0089 & 0.07\\

Abell 1576  & $265.0 \pm 46.7$ & $17.91 \pm 0.10$ & $ 8.6/6$
                            & $171.9 \pm 19.0$ & $17.94 \pm 0.11$ & 7.4/6 &  0.17 & -0.0043 & 0.15\\
                            
RX J0437.1+0043  & $78.1 \pm 13.7$ & $17.45 \pm 0.15$ & $2.0/5$
                            & $137.4 \pm 38.8$ & $17.46 \pm 0.10$ & 2.0/5  & 0.20 & -0.0046 & 0.12\\

Abell 959 & $350.4 \pm 44.9$  & $18.73 \pm 0.12$ & $11.3/6$
                            & $498.5 \pm 154.4$ & $18.78 \pm 0.12$ & 9.7/6  & 0.41 & -0.0090 & 0.13\\

Zwcl 3146  &  $304.8 \pm 46.4$ & $18.14 \pm 0.13$ & $10.7/6$
                            & $567.6 \pm 193.2$ & $18.17 \pm 0.15$ & 8.7/6 & 0.56 & -0.012 & 0.10\\

RX J2308.3-0211 & $307.0 \pm 46.8$ & $17.88 \pm 0.13$ & $2.8/5$
                            & $756.1 \pm 288.1$ & $17.89 \pm 0.14$ & 3.4/5 &  0.17 & -0.00033 & 0.10 \\

MS 1241.5+1710 & $301.3 \pm 47.9$ & $18.15 \pm 0.13$ & 14.4/6
                            & $392.5 \pm 63.6$ & $18.16 \pm 0.12$ & 8.3/6 & 0.77 & -0.012 & 0.13 \\

RBS 797  & $323.3 \pm 97.3$ & $18.18 \pm 0.13$ & $7.6/5$
                            & $285.9 \pm 131.3$ & $18.19 \pm 0.14$ & 7.7/5 & 0.72 & -0.018 & 0.22\\

Cl 0024.0+1652   & $356.4 \pm 21.6 $ & $17.26 \pm 0.14$ & 7.6/5
                            & $ 775.8\pm 256.1$ & $17.31 \pm 0.13$ & 26.0/5 &  0.23 & -0.00013 & 0.18\\
                            
MS 0302.7+1658   & $658.3 \pm 190.7$ & $18.15 \pm 0.14$ & 9.9/5
                            & $416.6 \pm 198.6$ & $18.09 \pm 0.18$ & 12.8/5 & 0.52 & -0.0088 & 0.14\\

RX J0030.5+2618 & $423.2 \pm 31.8$ & $17.64 \pm 0.14$ & 2.5/5
                            & $702.6 \pm 188.7$ & $17.58 \pm 0.11$ & 12.1/5  & 1.68 & -0.051 & 0.09 \\

MS 0451.6-0305  & $442.8 \pm 64.3$ & $17.86 \pm 0.14$ & $8.3/6$
                            & $926.7 \pm 355.7$ & $17.89 \pm 0.18$ & 8.8/6 & 1.18 & -0.0017 & 0.18\\

Cl 0016+16   & $222.2 \pm 29.5$  & $18.51 \pm 0.10$ & 13.9/6
                            & $588.2 \pm 190.4$ & $18.56 \pm 0.12$ & 10.5/6  & 0.73 & -0.0014 & 0.12\\
                            
MS 2053.7-0449   & $289.6 \pm 290.9$ & $16.96 \pm 0.33$ & 0.6/4
                            & $387.3 \pm 132.6$ & $16.97 \pm 0.32$ & 0.6/4 & 1.60 &  -0.0015 & 0.14 \\

RX J1120.1+4318  & $298.0 \pm 68.4$  & $17.79 \pm 0.13$ & 11.3/6
                            & $401.2 \pm 102.4$ & $17.78 \pm 0.12$ & 9.6/6 &  0.80  &  -0.013 & 0.18\\
                      
RX J1334.3+5030  & $396.2 \pm 75.7$ & $18.66 \pm 0.15$ & $2.2/5$
                            & $214.8 \pm 149.2$ & $18.60 \pm 0.09$ & 6.4/5 &  1.45 & -0.035 & 0.18\\

MS 1137.5+6624   & $138.8 \pm 96.1$ & $18.80 \pm 0.13$ & $0.7/5$
                            & $249.3 \pm 164.1$ & $18.80 \pm 0.13$ & 0.7/5 & 1.98 & -0.022 & 0.16\\
                            
RX J1716.4+6708   &  $423.4 \pm 230.4$ & $18.66 \pm 0.23$ & 1.7/4
                            & $364.4 \pm 150.2$ & $18.73 \pm 0.34$ & 1.6/4 & 2.66 & -0.034 & 0.12\\

WARP J1226.9+3332  & $490.2 \pm 451.3$ & $18.96 \pm 0.16$ & $1.5/5$
                            & $685.2 \pm 302.5$ & $18.91 \pm 0.15$ & 1.5/5 &  2.35 & -0.067 & 0.12\\

\hline

\enddata

\scriptsize
\tablenotetext{a}{Best-fit King model and NFW model
  parameters used in the BS method (\S 3.1.1).}
\tablenotetext{b}{Best-fit color-magnitude relations for red sequence
  galaxies (\S 3.1.2).}

\end{deluxetable}

\begin{deluxetable}{cccccc}
\centering
\tabletypesize{\scriptsize}
\tablewidth{0pt}
 \tablecaption{NFW model fitting}

\tablehead{            

    Cluster name & $R_{\rm 500}^{\rm HM}$\tablenotemark{a} &
    $M_{500}$\tablenotemark{a} & $c$& $R_{\rm s}$ &
    $\chi^2/\nu$ \\
 & (kpc) & ($10^{14} M_{\rm \odot}$) & & (kpc) & }

\startdata
RX J1141.4-1216 &   879 &  $2.1\pm 0.1$  & $3.6 \pm 0.5$ & $368 \pm 52$ &  3.3/6\\

RX J1044.5-0704 &  930 &  $2.0 \pm 0.1$ & $4.6\pm 0.7$ & $278 \pm 39$ & 1.7/6 \\

Abell 2204 &  1269 & $8.6 \pm 0.4$ &  $4.1 \pm 0.7$ & $500 \pm 90$ & 7.0/9 \\

RX J0958.3-1103 &1078 & $3.9 \pm 1.0$ & $4.1 \pm 0.5$ & $456 \pm 99$ &  5.4/8\\
                            
 Abell 665  & 1283 & $6.9 \pm 0.2$ & $1.9 \pm 0.2$ & $1216 \pm 574$ & 2.7/9 \\

Abell 383 & 1014 & $3.5 \pm 0.4$ & $3.8\pm 0.3$ & $447 \pm 134$ & 2.4/11 \\

Abell 963 & 1170 &  $7.2\pm0.6$ &  $3.1 \pm 0.2$ & $710 \pm 283$ & 2.1/7 \\

RX J1504.1-0248 &  1236 & $6.6 \pm 0.2$ & $5.3\pm 0.7$ & $340 \pm 48$  & 3.6/9 \\

MS 0735.6+7421 & 1002 & $4.3\pm 0.3$ & $3.0\pm 0.3$ & $587 \pm 124$ & 1.5/7 \\

Abell 2261 & 1158 & $5.1 \pm 0.8$ & $6.8\pm 1.9$ & $256 \pm 82$ & 2.4/4 \\
                            
 Abell 267 &  1065 & $4.2 \pm 0.3$ & $1.9 \pm 0.2$ & $999 \pm 191$ & 5.9/9 \\

RX J2129.3+0005 & 1081 & $3.4\pm0.2$ &  $3.3\pm 0.5$ & $480 \pm 67$ & 5/12 \\

MS 1231.3+1542  & 967 & $3.1\pm2.1$ & $3.7\pm 0.4$ & $378 \pm 84$ & 1.9/6 \\

Abell 1835 & 1315 & $10.9 \pm 0.4$ &  $3.5 \pm 0.5 $  & $480 \pm 68$ & 8.2/8  \\

Abell 68 &  1289 & $8.0\pm0.6$ &  $2.9 \pm 0.5$ & $698 \pm 301$ & 2.5/9 \\

CIZA J1938.3+5409 & 1182 & $7.1\pm2.2$ & $8.1\pm 2.9$ & $232 \pm 83$ & 1.1/3 \\

Abell 1576 & 1107 & $4.9\pm0.8$ & $2.9 \pm 0.5$  & $584 \pm 253$ & 2.5/10 \\
                            
RX J0437.1+0043 & 1219 & $5.8\pm0.5$ &  $4.0 \pm 0.8$ & $418 \pm 93$ & 2.1/7 \\

Abell 959  &1112 & $7.9\pm0.9$ & $2.8\pm 0.3$ & $857 \pm 581$ & 1.4/3  \\

Zwcl 3146 & 1135 & $5.5\pm0.2$ & $3.9\pm 0.6$ & $442 \pm 62$ & 6.9/9 \\

RX J2308.3-0211 & 1174 & $4.5\pm0.5$ & $3.0 \pm 0.8$ & $542 \pm 153$ & 5.5/10 \\

MS 1241.5+1710 &  846  & $2.5\pm0.3$ & $4.3\pm 0.9$ & $280 \pm 63$ & 1.4/4 \\

RBS 797 & 1146 & $6.6\pm0.6$ & $4.1\pm 0.6$ & $432 \pm 61$ & 6.8/10\\

Cl 0024.0+1652 & 723 & $1.8\pm0.2$ &  $4.3\pm 0.5$ & $260 \pm 48$ & 1.5/5\\
                            
MS 0302.7+1658 & 888 & $2.2\pm0.3$ & $3.7 \pm 1.0$ & $310\pm 88$ & 1.8/4\\

RX J0030.5+2618 &  943 & $2.4\pm0.3$ & $2.0\pm 0.3$ & $637 \pm 360$ & 4.8/7 \\

MS 0451.6-0305 &1346 & $5.9\pm0.3$ & $2.0\pm 0.2$ & $1009 \pm 285$ & 5.4/7 \\

Cl 0016+16  & 1279 & $6.1\pm0.4$ &  $2.0\pm 0.3$ & $846 \pm 252$ &  2.7/8\\
                            
MS 2053.7-0449  &  975 & $2.9\pm0.2$  & $3.5 \pm 0.9$ & $362\pm 102$ & 2.1/7 \\

RX J1120.1+4318 & 900 & $2.3\pm0.3$ &  $2.4\pm 0.3$ & $486 \pm 186$ & 9.0/8 \\
                      
RX J1334.3+5030 & 835 & $2.5\pm0.5$ &  $4.5\pm 1.9$ & $268 \pm 113$ &  0.8/2\\

MS 1137.5+6624 & 968 & $5.1\pm0.8$ & $1.6\pm 0.2$ & $1125 \pm 648$ & 1.3/2 \\
                            
RX J1716.4+6708 & 904 & $4.1\pm0.2$ & $6.4 \pm 2.7$ & $196\pm 83$ & 1.3/2 \\

WARP J1226.9+3332 &  1109 & $5.5\pm0.8$ & $3.3\pm 1.4$ & $418 \pm 177$& 3.9/10 \\

\hline

 \enddata
\scriptsize
\tablenotetext{a}{$R_{500}^{\rm HM}$ and $M_{500}$ were calculated based on
  hydrostatic mass estimates described in \S 3.5.2.}
\end{deluxetable}

\begin{figure}
\begin{center}
\includegraphics[angle=-0,scale=.8]{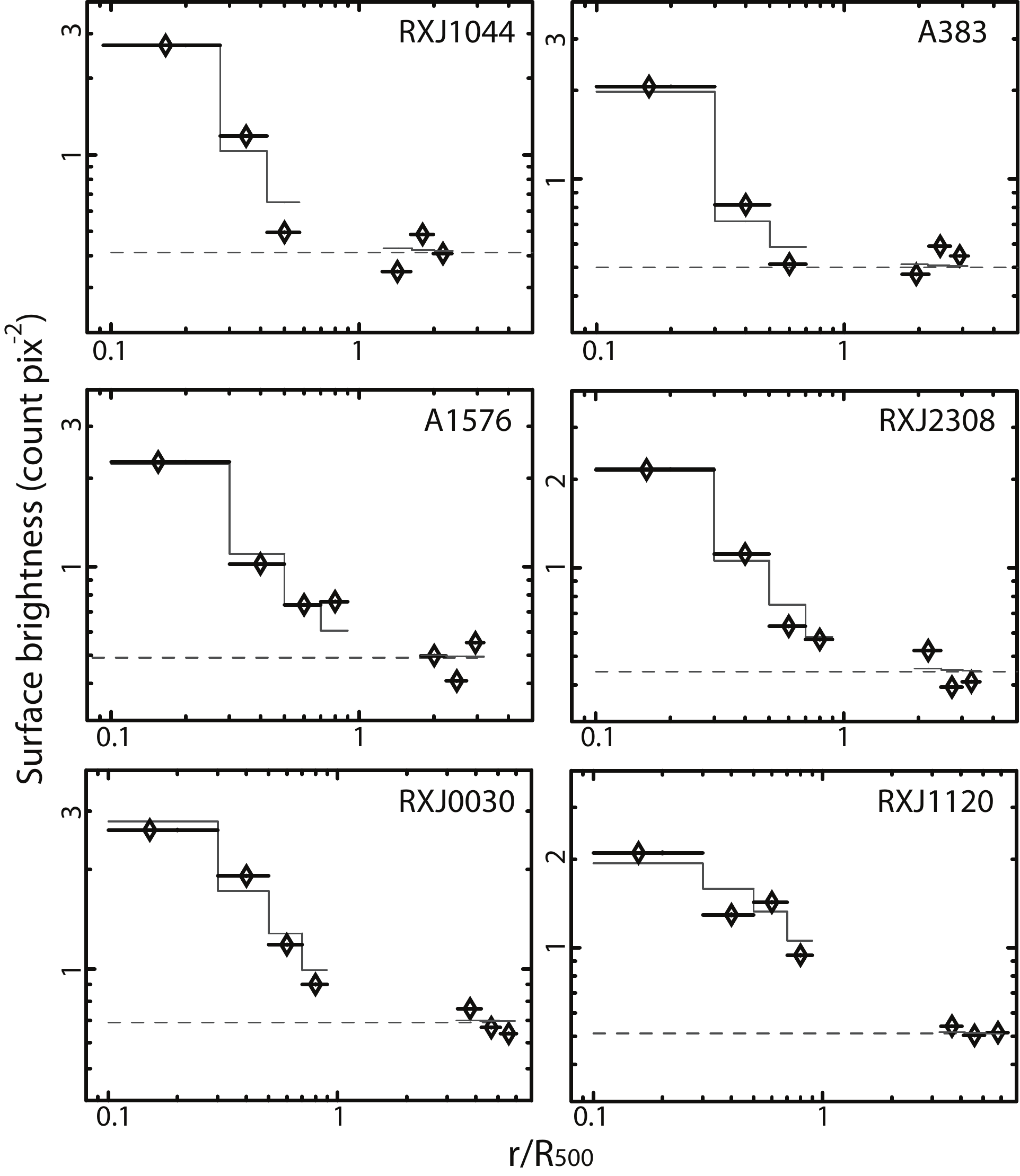}
\caption{Background-inclusive $I$- band surface brightness profiles of six representative
  clusters, fitted with the king model plus constant background (solid
  line). The outermost three data points are from the offset pointings, while the rest are from the central 
pointing onto each cluster. The background values are shown in dashed line.  }
\end{center}
\end{figure}

\begin{figure}
\begin{center}
\includegraphics[angle=-0,scale=.5]{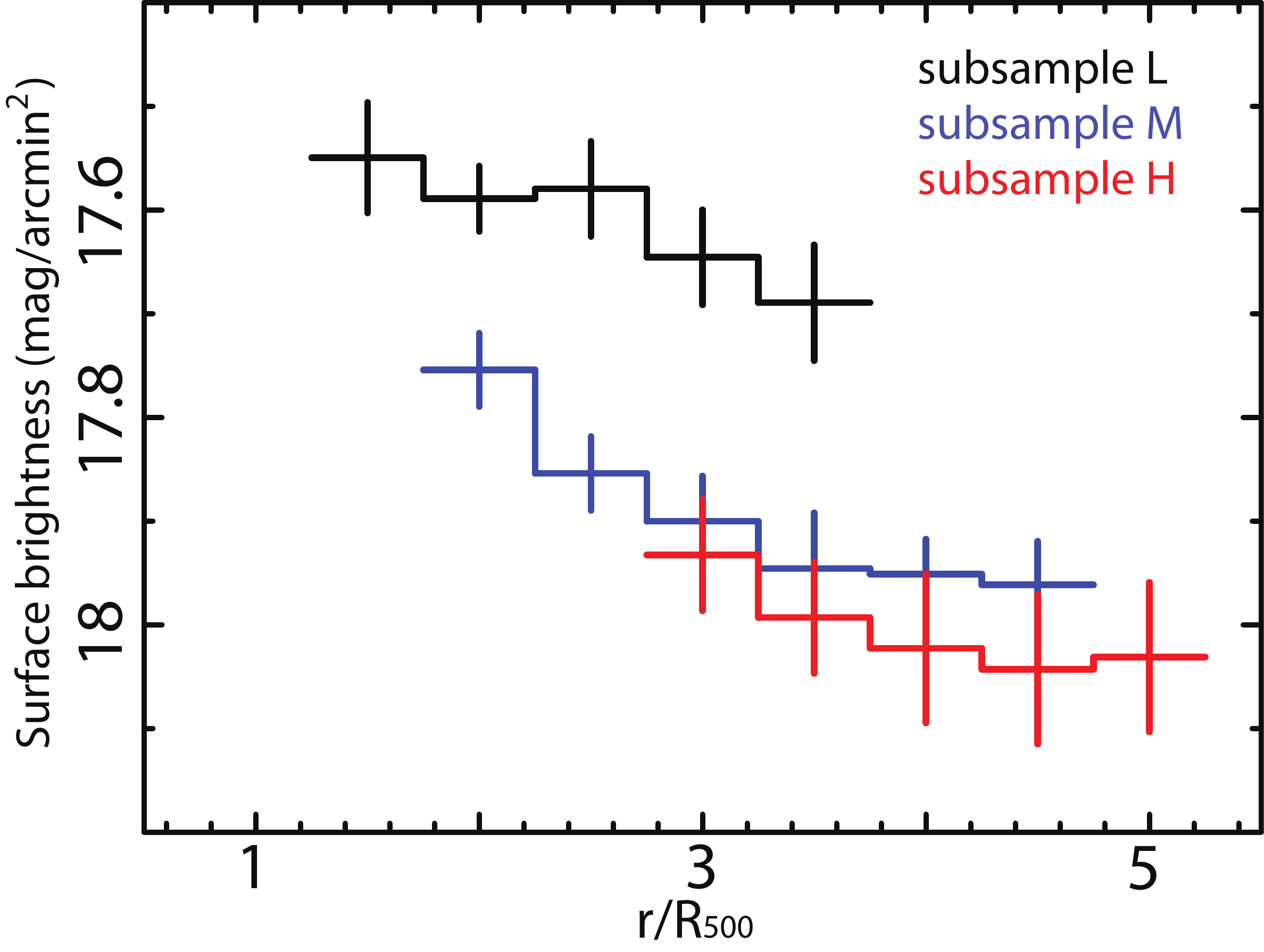}
\caption{$I$- band surface brightness measured for the offset
  region. Galaxies with fluxes higher than that of the central dominant galaxies
  were discarded in the calculation. The average profiles (see text) for subsample L, M, and H are shown in
  black, blue, and red, respectively.}
\end{center}
\end{figure}

\begin{figure}
\begin{center}
\includegraphics[angle=-0,scale=.8]{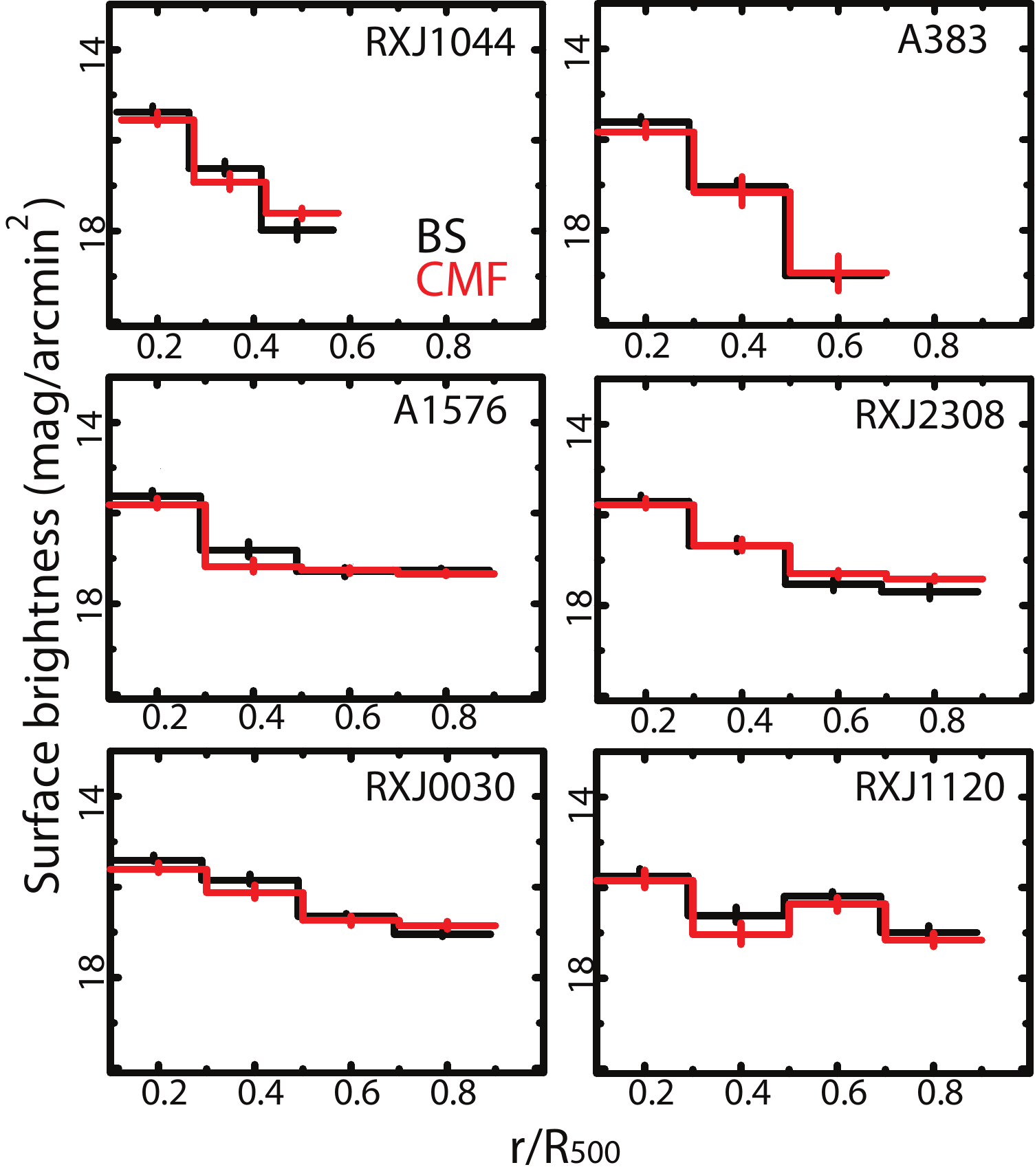}
\caption{$I$-band surface brightness profiles of member galaxies for the six
  representative clusters,
  obtained with
  the BS (black) and CMF (red) methods. The constant background is subtracted in the BS results.}
\end{center}
\end{figure}

\begin{figure}
\begin{center}
\includegraphics[angle=-0,scale=.6]{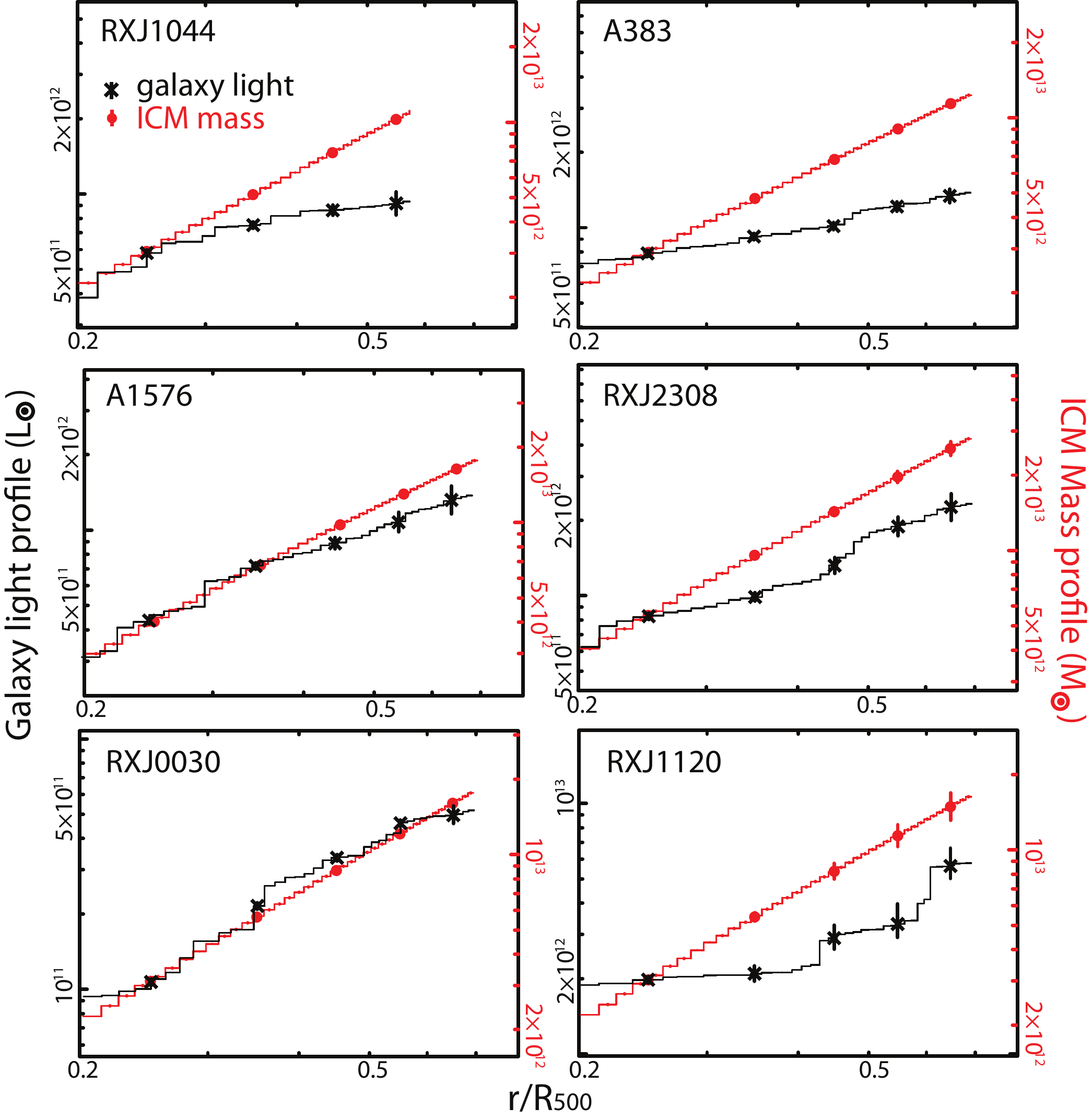}
\caption{Optical light profiles obtained with the BS method (black; see text) of the six representative
clusters, compared with their ICM mass profiles (red). Both are radially integrated over two dimensions.
Five representative radial points are also shown, together with the estimated errors therein (see text). }
\end{center}
\end{figure}

\begin{figure}
\begin{center}
\includegraphics[angle=-0,scale=0.8]{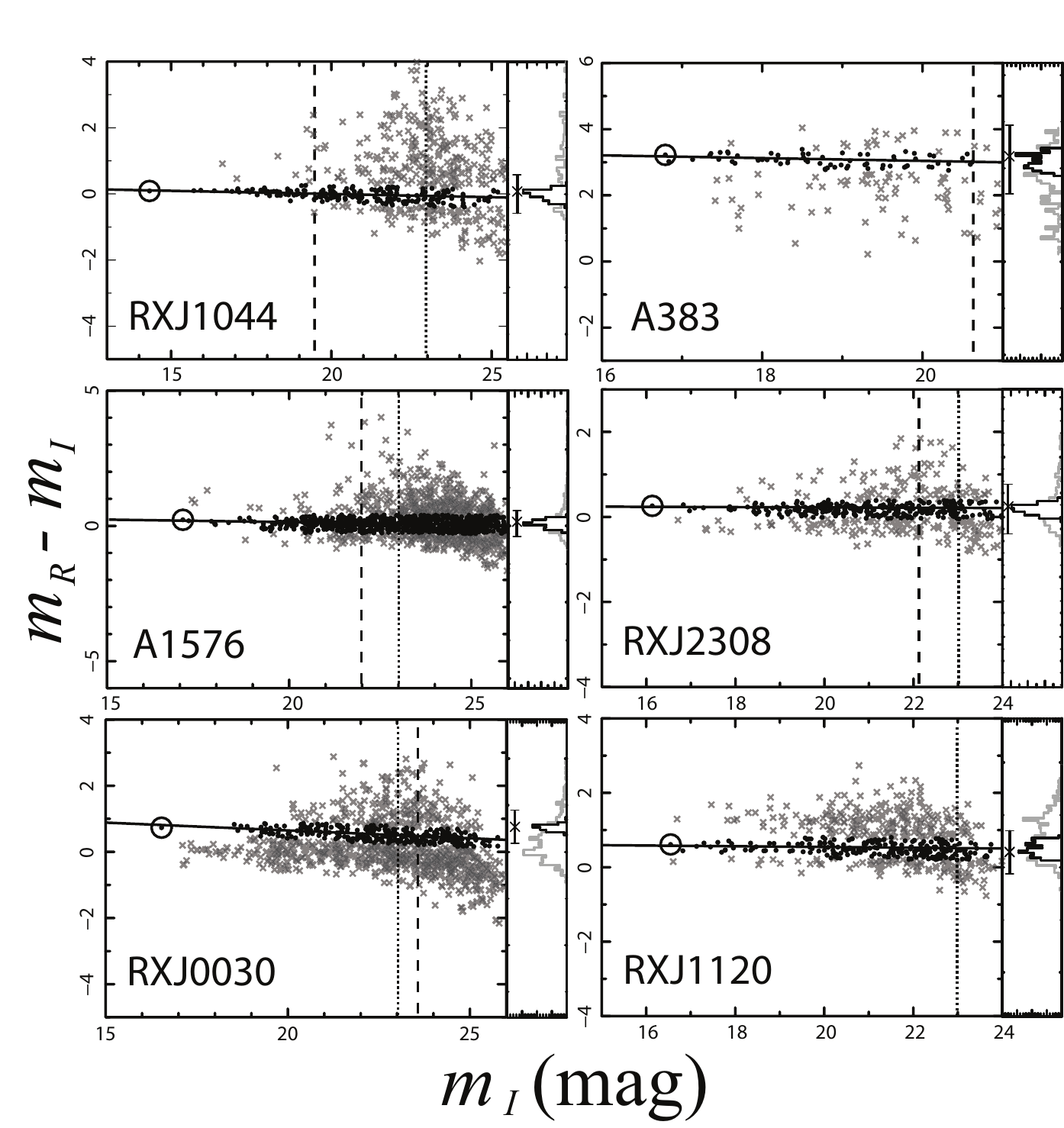}

\caption{Color-magnitude diagrams of galaxies obtained within a $\sim
  4^{\prime}$ region of the six representative clusters. The color $m_{\rm R} - m_{\rm I}$ is plotted,
  except for A383, for which $m_{\rm B} - m_{\rm I}$ is used due to the absence of
  $R$-band observation. Solid lines show the location of the
best-fit red sequence curves. The selected red sequence galaxies are
shown by filled black points, while the others are in gray
crosses. The central dominant galaxies are highlighted with
circles. The vertical dotted and dashed lines
indicate the two types of limiting magnitude described in \S 3.1.2 and \S
3.5.1. In the side panel, the number-color diagrams of selected galaxies and all galaxies
are plotted in black and gray, respectively, where the central color ($C_{0}$) and
color range (3$\sigma_{\rm c}$) used for the robust curve
fitting (\S 3.1.2) are shown by an error bar. }

\end{center}
\end{figure}

\begin{figure}
\begin{center}
\includegraphics[angle=-0,scale=.5]{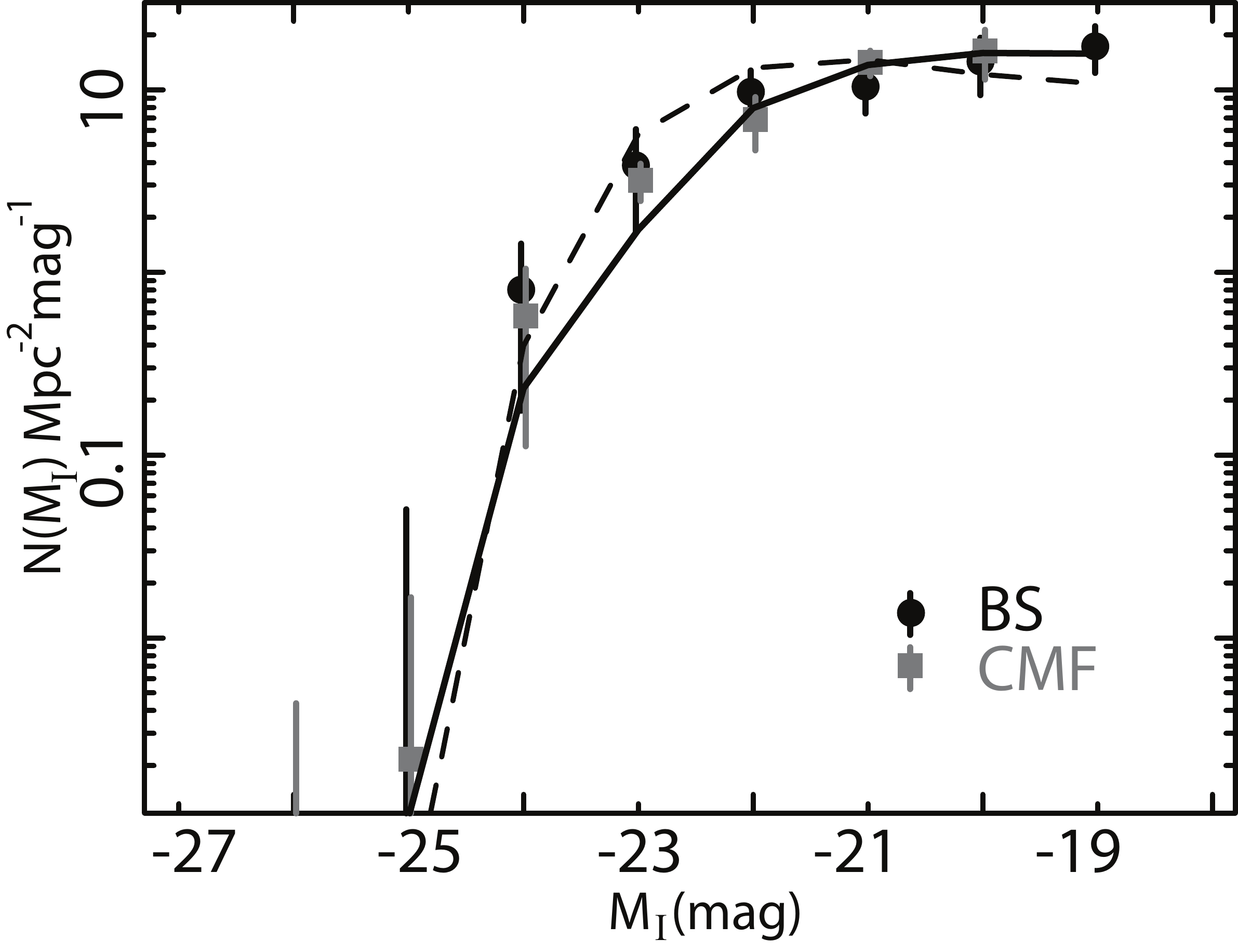}
\caption{The number of member galaxies detected in all our sample clusters, plotted as a function of magnitude in $I$
  band. The BS and the CMF results are plotted in black circles and
  gray boxes, respectively. The lines show the Schechter luminosity function with
  parameters fixed to those obtained in Rudnick et al. (2009). The solid and dashed lines show 
their results for nearby clusters and distant clusters, respectively. See text for details. }
\end{center}
\end{figure}

\begin{figure}
\begin{center}
\includegraphics[angle=-0,scale=1.1]{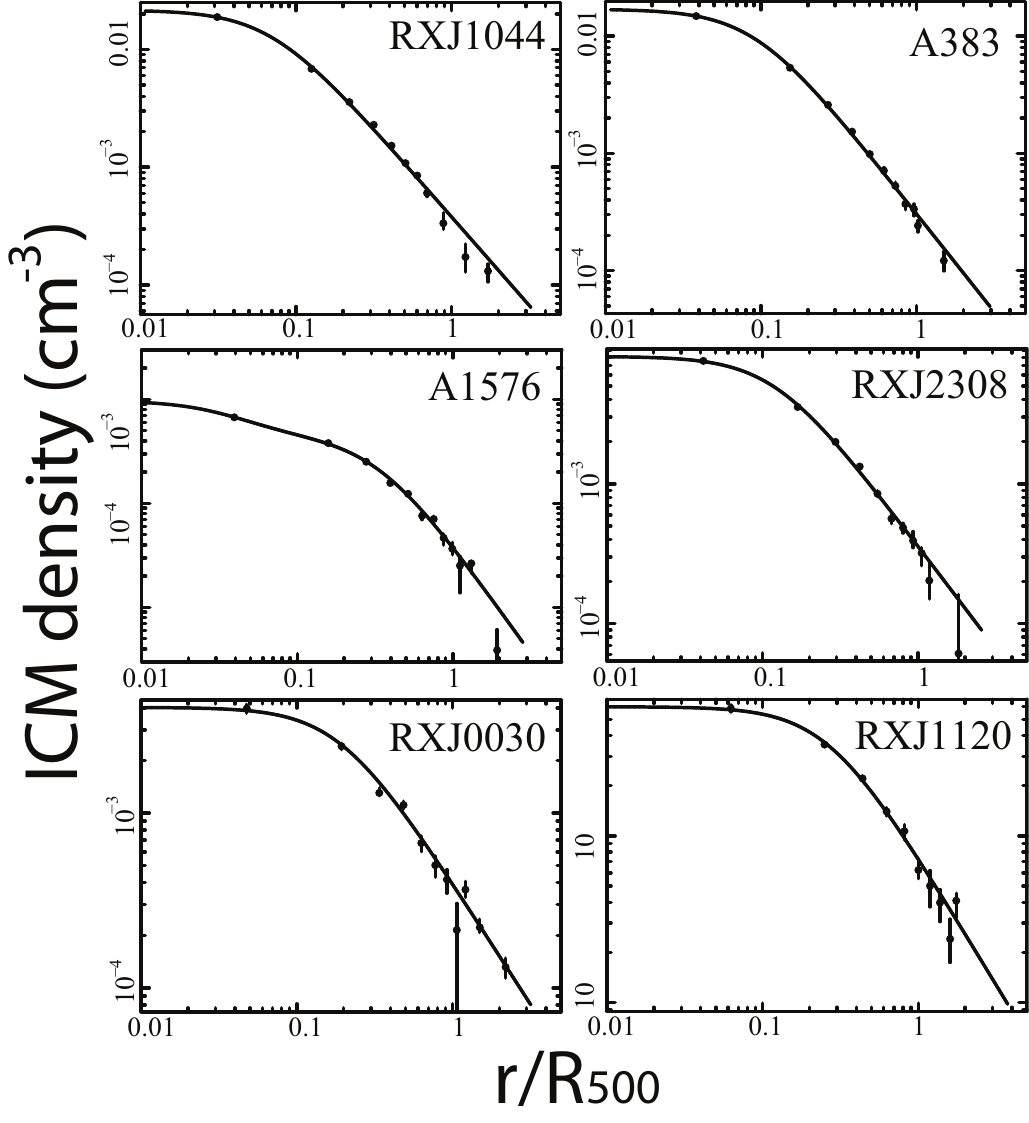}
\caption{ICM density profiles of the six representative clusters obtained with the {\it XMM-Newton} (for RXJ1044, A383, A1576, and RXJ2308) and {\it
    Chandra} (for RXJ0030 and RXJ1120) data. The best-fit $\beta$/double-$\beta$ models (\S3.2) are
  shown in black curves.}
\end{center}
\end{figure}



 


\begin{figure}
\begin{center}
\includegraphics[angle=-0,scale=.4]{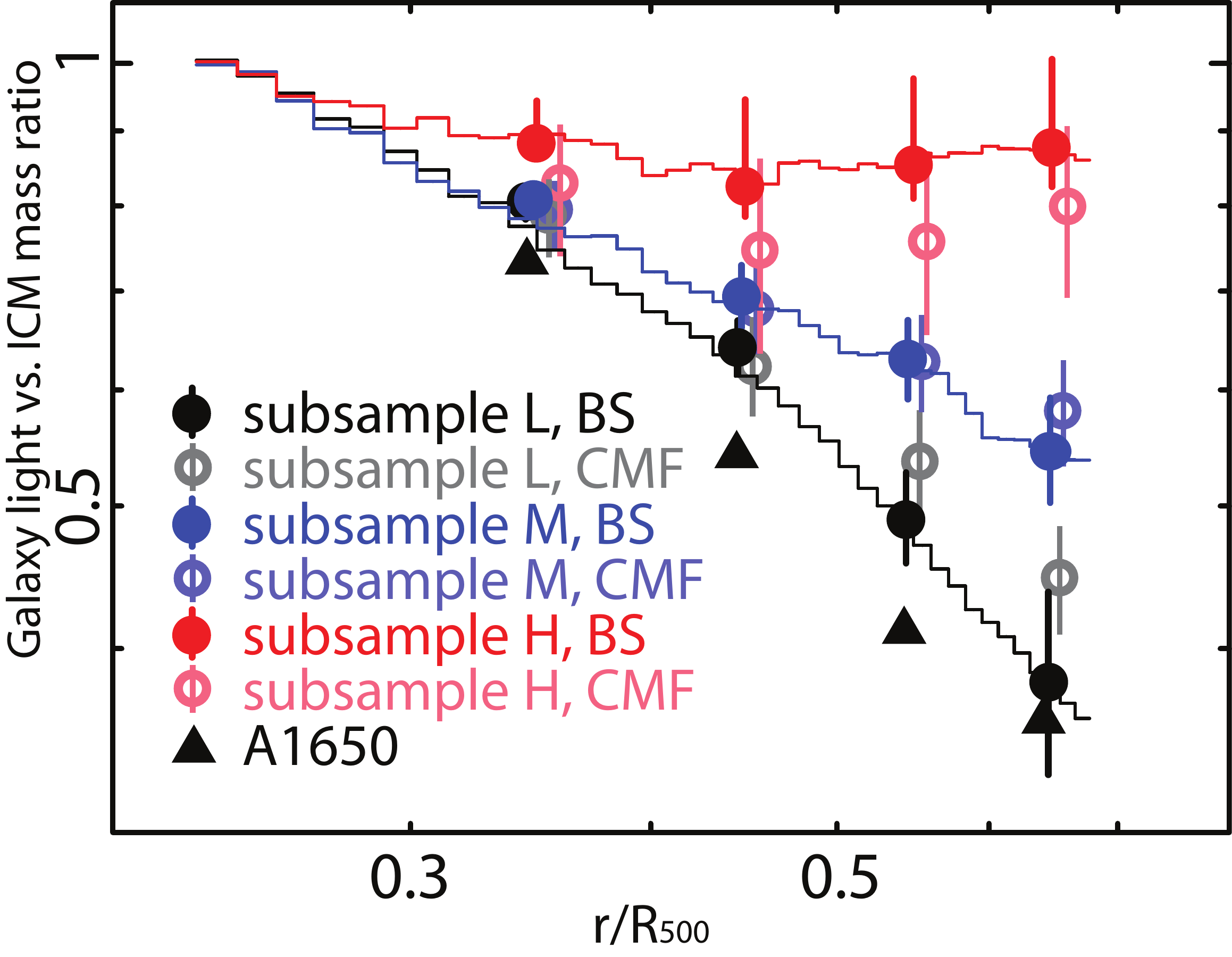}
\caption{GLIMR profiles averaged over the subsample L, M, and H, 
  shown in black,
  blue, and red, respectively. The quasi-continuous curves are from the BS method, with its 
errors given in filled symbols at the five representative radii. The results from the 
CMF method are given, at the five radii, in open symbols. For reference, the GLIMR
profile of a nearby cluster, A1650 (\S 3.4), is shown in black triangles. }
\end{center}
\end{figure}

\begin{figure}
\begin{center}
\includegraphics[angle=-0,scale=.6]{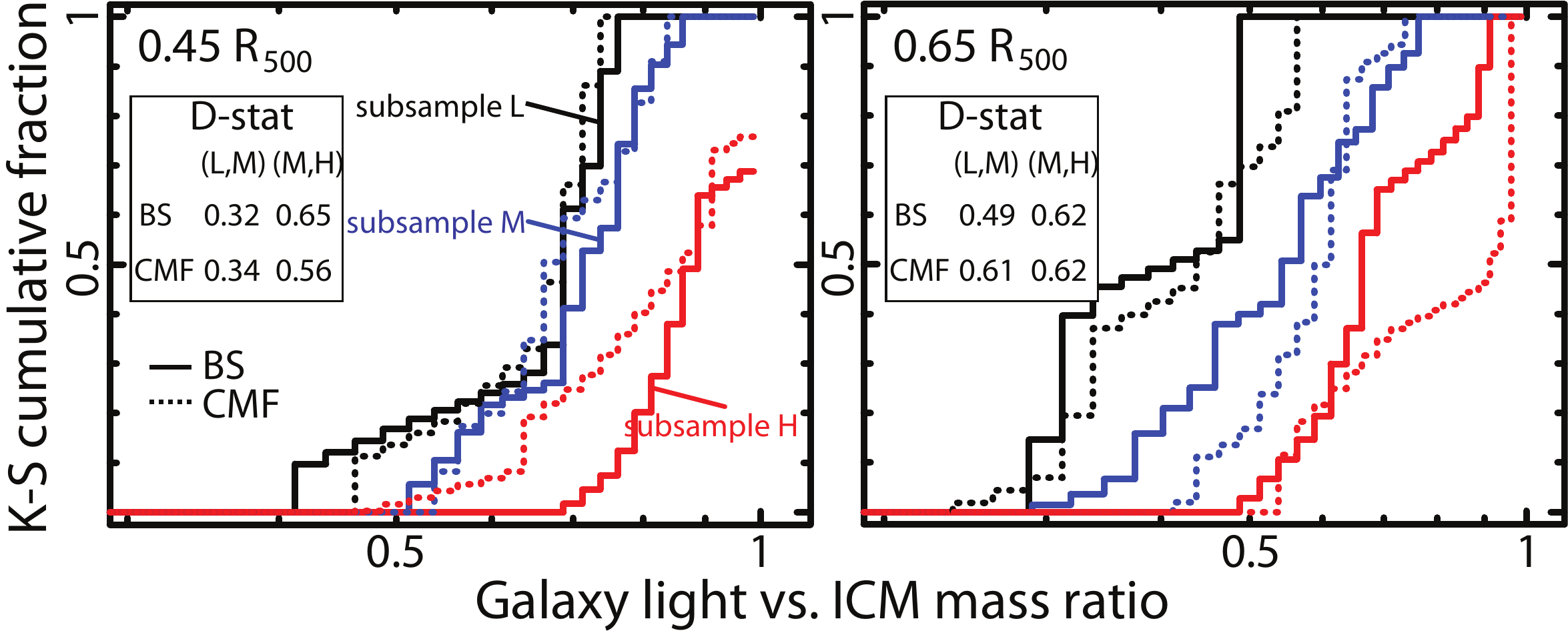}
\caption{Fractional K-S cumulative distributions of GLIMR values at the
$r=0.45$ $R_{500}$ (left) and $r=0.65$ $R_{500}$ bins (right). The
  subsample L, M, and H results with the BS (solid) and the CMF (dotted)
  methods are shown in black, blue, and red lines, respectively. The
  D-statistics calculated from the differences in the cumulative
  distributions between subsamples L and M, and between
  M and H, are shown in the figures. }
\end{center}
\end{figure}

\begin{figure}
\begin{center}
\includegraphics[angle=-0,scale=.4]{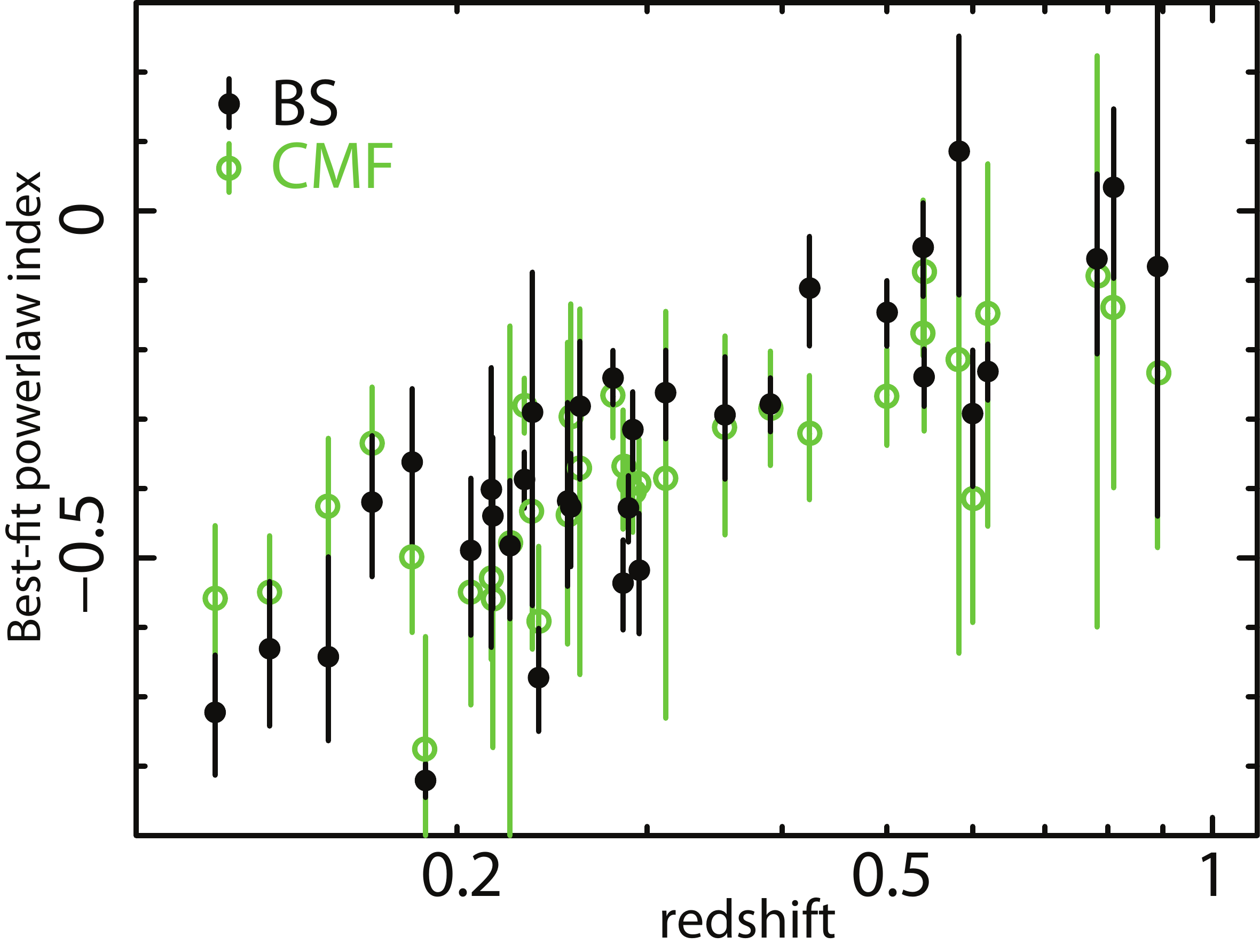}
\caption{Logarithmic slopes of the GLIMR profiles of the 34 clusters,
  plotted as a function of redshift. The BS and CMF
  results are marked with filled and empty circles, respectively.}
\end{center}
\end{figure}

\begin{figure}
\begin{center}
\includegraphics[angle=-0,scale=.7]{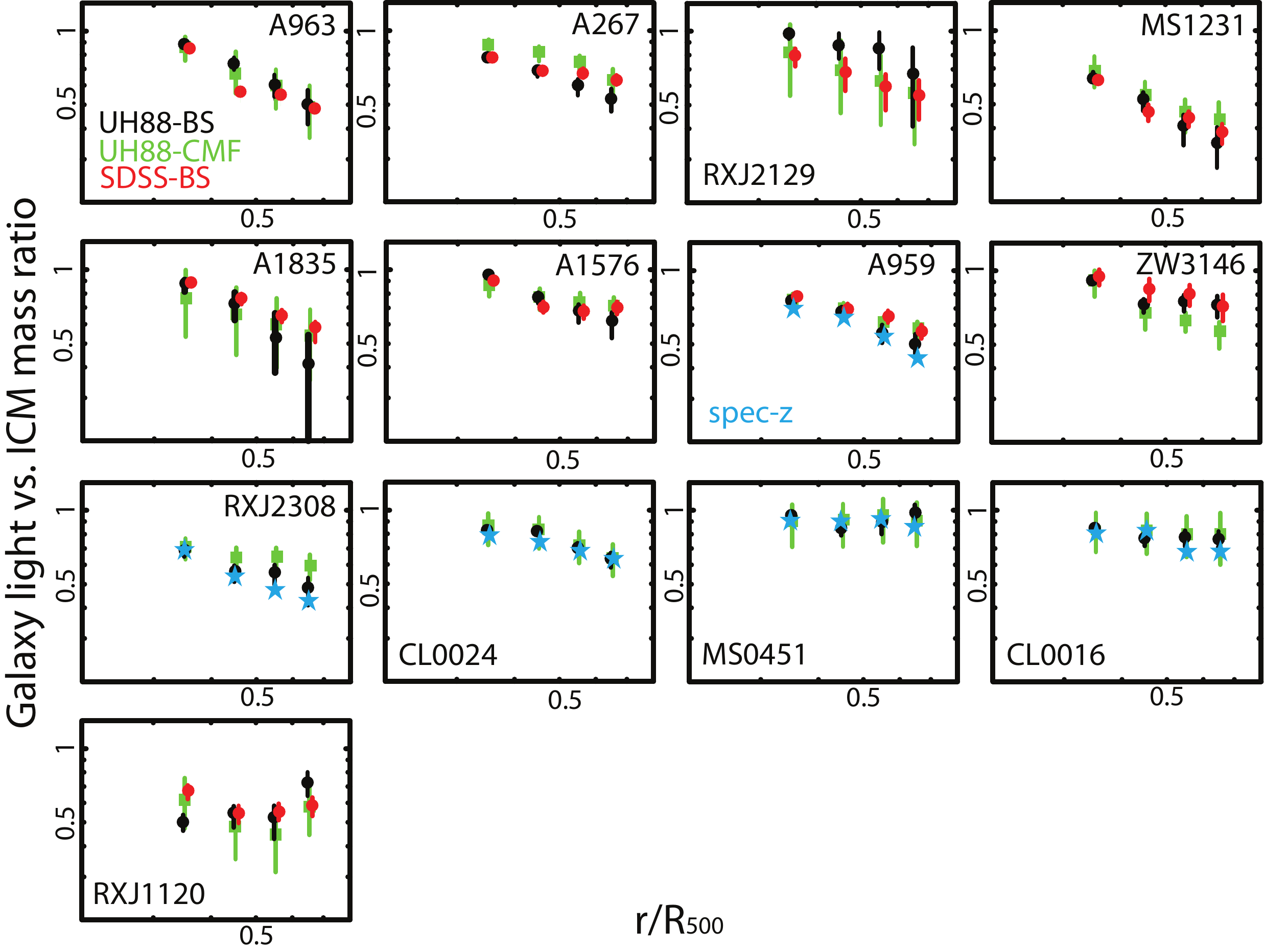}
\caption{GLIMR profiles obtained with the SDSS DR6 photometric data
  (red) and previous spectroscopic membership determinations (blue; see \S3.4 
 for reference), compared with our
  UH88 results (the same as in Fig. 8) using the BS method (black) 
and the CMF method (green). }
\end{center}
\end{figure}

\begin{figure}
\begin{center}
\includegraphics[angle=-0,scale=.5]{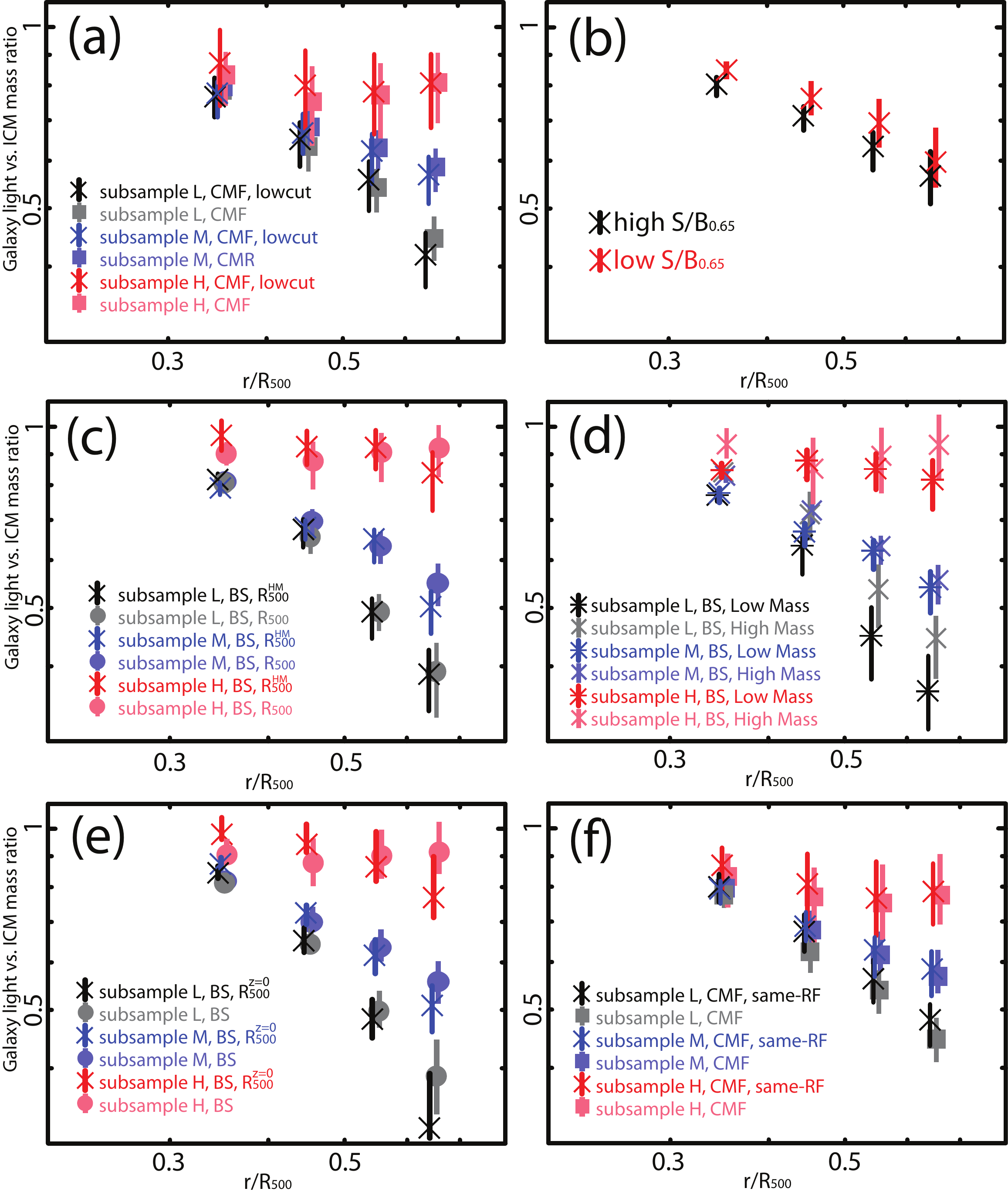}
\caption{Examinations of various systematic errors and observational biases. ({\it a}) Subsample-averaged CMF-GLIMR 
profiles (crosses) obtained by applying the redshift-dependent
  limiting magnitude (an approximately redshift-independent
  absolute-magnitude) shown as dashed line in Figure 5 and Figure A.4, compared with
the original CMF results (squares). The color
  for each subsample is the same as in Figure 8. ({\it b}) Average
  BS-GLIMR profiles of clusters with high (black) and low (red) X-ray signal-to-background ($S/B_{0.65}$) 
  ratios. ({\it c})
  Subsample-averaged BS-GLIMR profiles based on $R_{500}^{\rm HM}$ obtained
  in hydrostatic mass measurement (crosses), plotted together with the original BS results
(filled circles). ({\it d}) BS-GLIMR profiles obtained from higher (lighter stars) and
lower (darker crosses) mass objects in each subsample. ({\it e})
Subsample-averaged BS-GLIMR profiles based on the predicted evolutionary scale 
$R_{500}^{z=0}$ (crosses), plotted together with the original BS results
(filled circles). ({\it f}) Subsample-averaged CMF-GLIMR profiles obtained in approximately the same
  rest-frame bands (crosses), compared with the original profiles with
  the CMF method (boxes).}
\end{center}
\end{figure}

\begin{figure}
\begin{center}
\includegraphics[angle=-0,scale=.3]{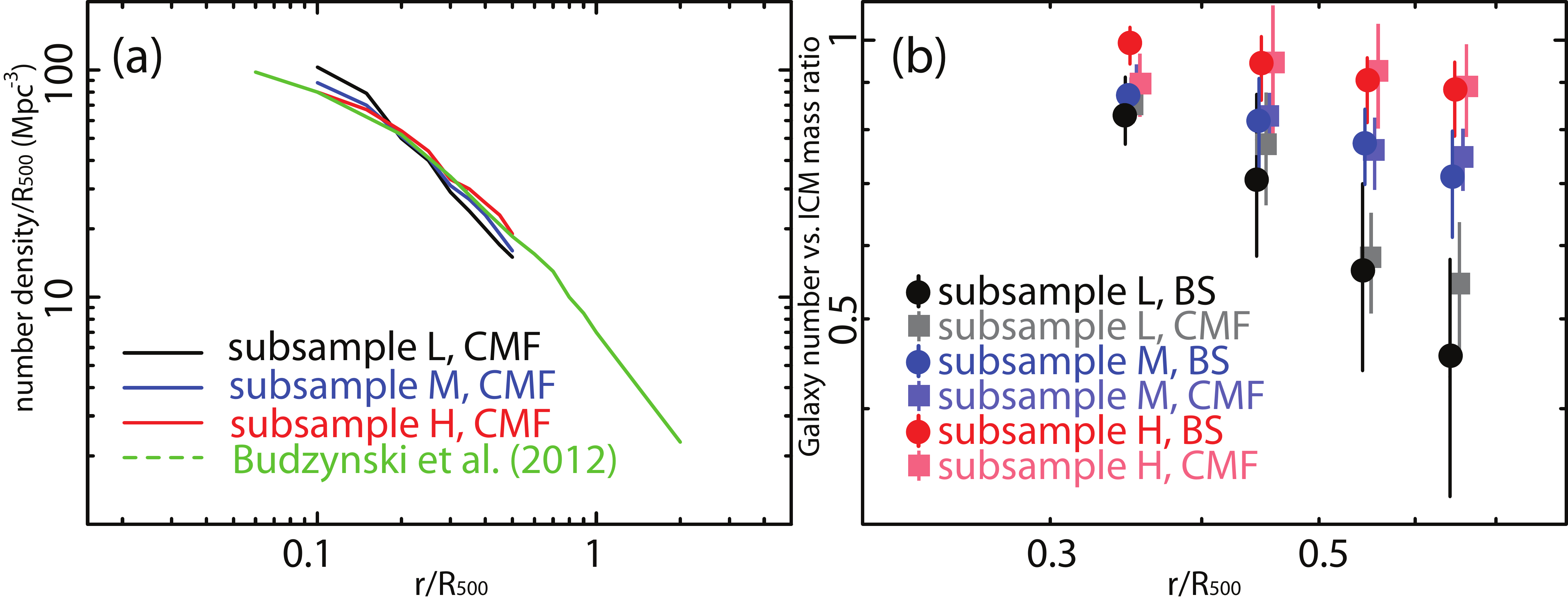}
\caption{({\it a}) Subsample-averaged galaxy number density profiles (in differential form)
  obtained with the CMF method, compared with the mean number density
  profile shown in Budzynski et al. (2012). The limiting
  magnitude adopted in Budzynski et al. (2012), $m_{\rm r, AB} <
  21.5$, was applied to calculate the density profiles of our
  sample. ({\it b}) The same as Figure 8, but using the galaxy number density
profiles instead of their luminosity profiles.}
\end{center}
\end{figure}

\begin{figure}
\begin{center}
\includegraphics[angle=-0,scale=.4]{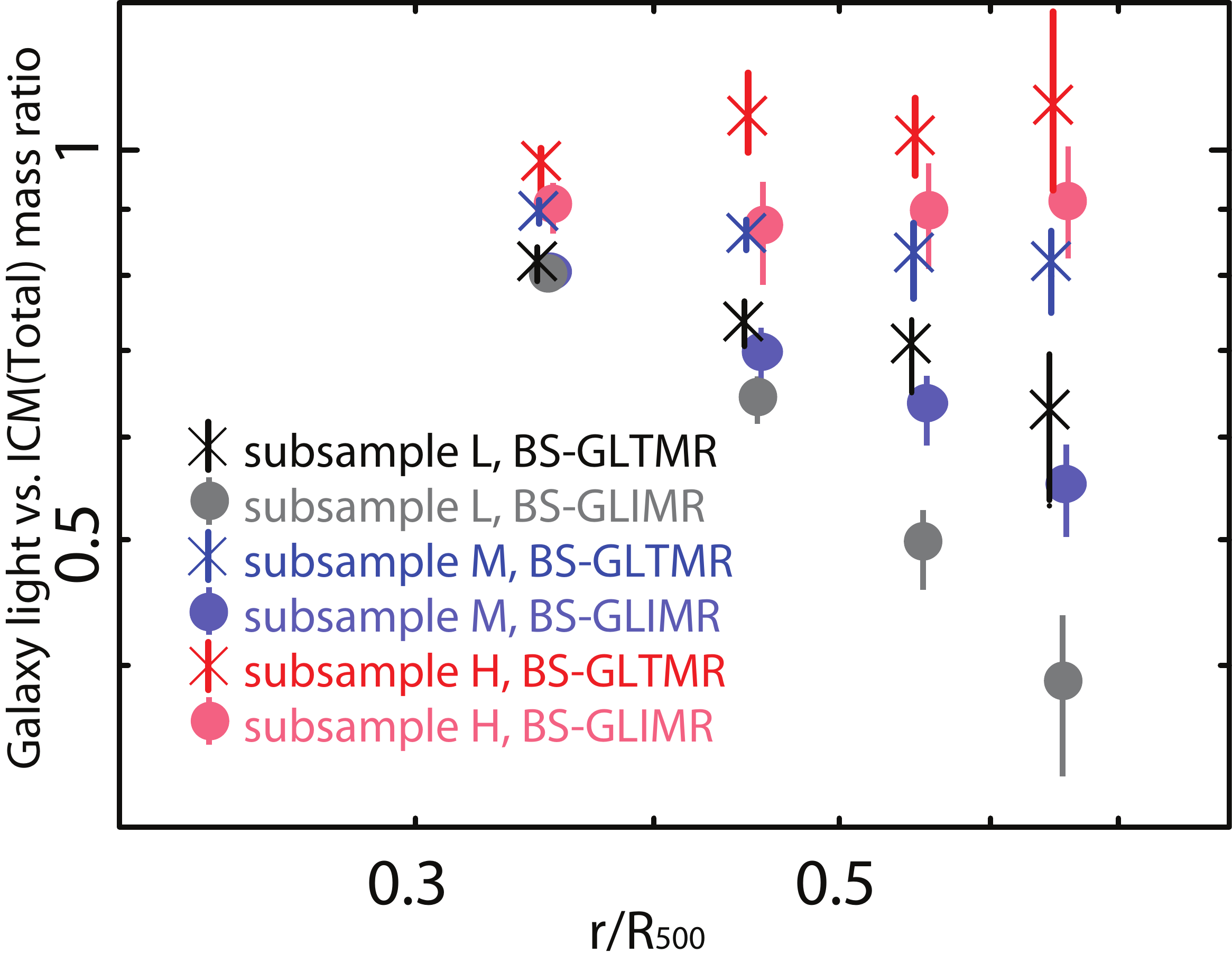}
\caption{Subsample-averaged GLTMR profiles (crosses) compared with the GLIMR
  profiles using the BS method (filled circles; the same as in Fig. 8). The color style
  for each subsample is the same as in Figure 8.}
\end{center}
\end{figure}

\clearpage

\begin{figure}
\begin{center}
\includegraphics[angle=-0,scale=.6]{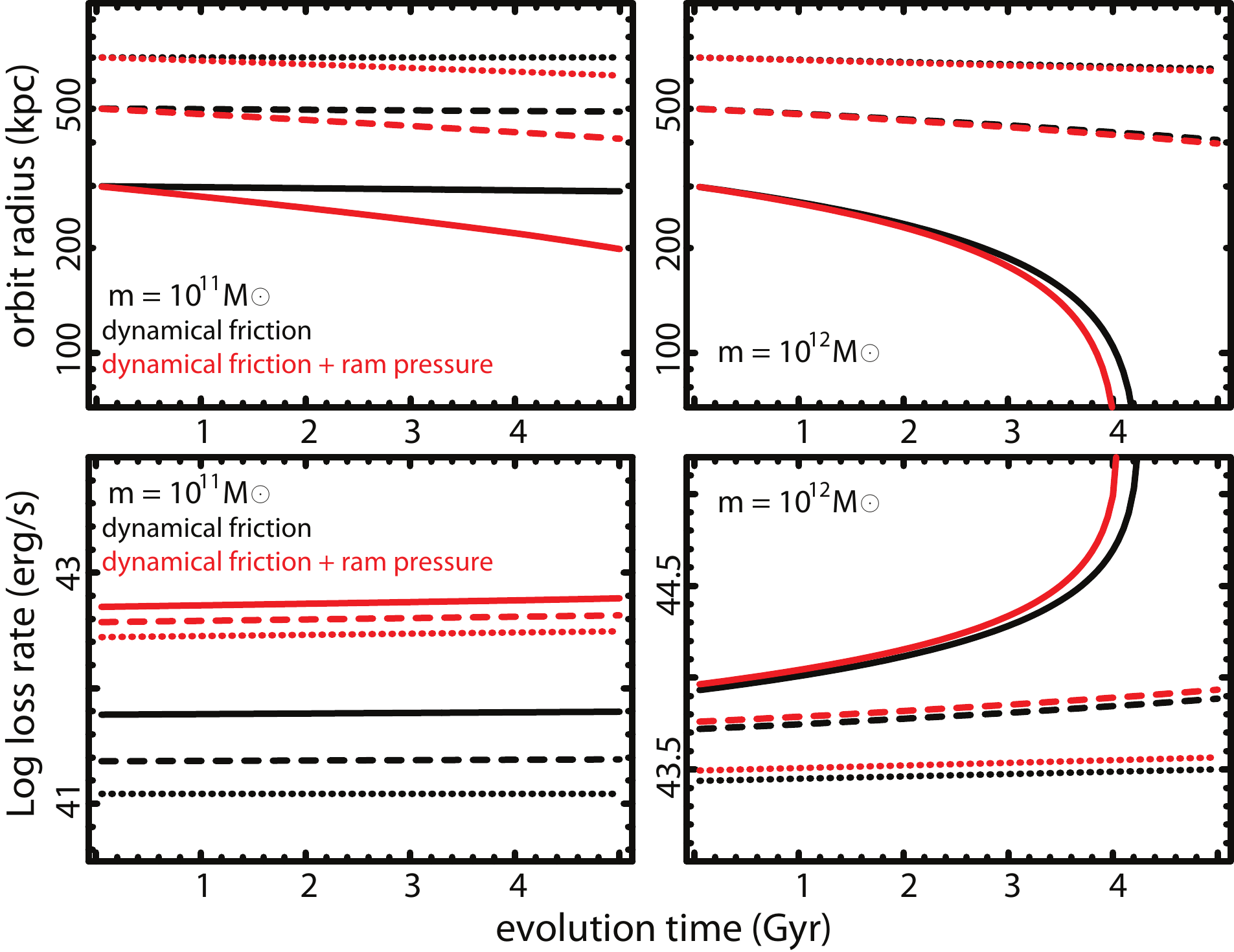}
\caption{({\it upper panels}) Predicted evolution of a circular orbit of a galaxy with a mass 
  of $m=1 \times 10^{11}$ $M_{\rm \odot}$ (left) and $m=1 \times 10^{12}$ $M_{\rm
    \odot}$ (right), starting from initial positions 
  of 300 kpc (solid line), 500 kpc (dashed line), and 700 kpc (dotted
  line). The black and red curves denote results when
  considering the dynamical friction alone and when also taking into
  account the ram pressure, respectively. ({\it lower panels})
Galaxy energy loss rate during the infall. The color and line
styles are the same as in the upper panel.}
\end{center}
\end{figure}

\clearpage

\begin{figure}
\begin{center}
\includegraphics[angle=-0,scale=.4]{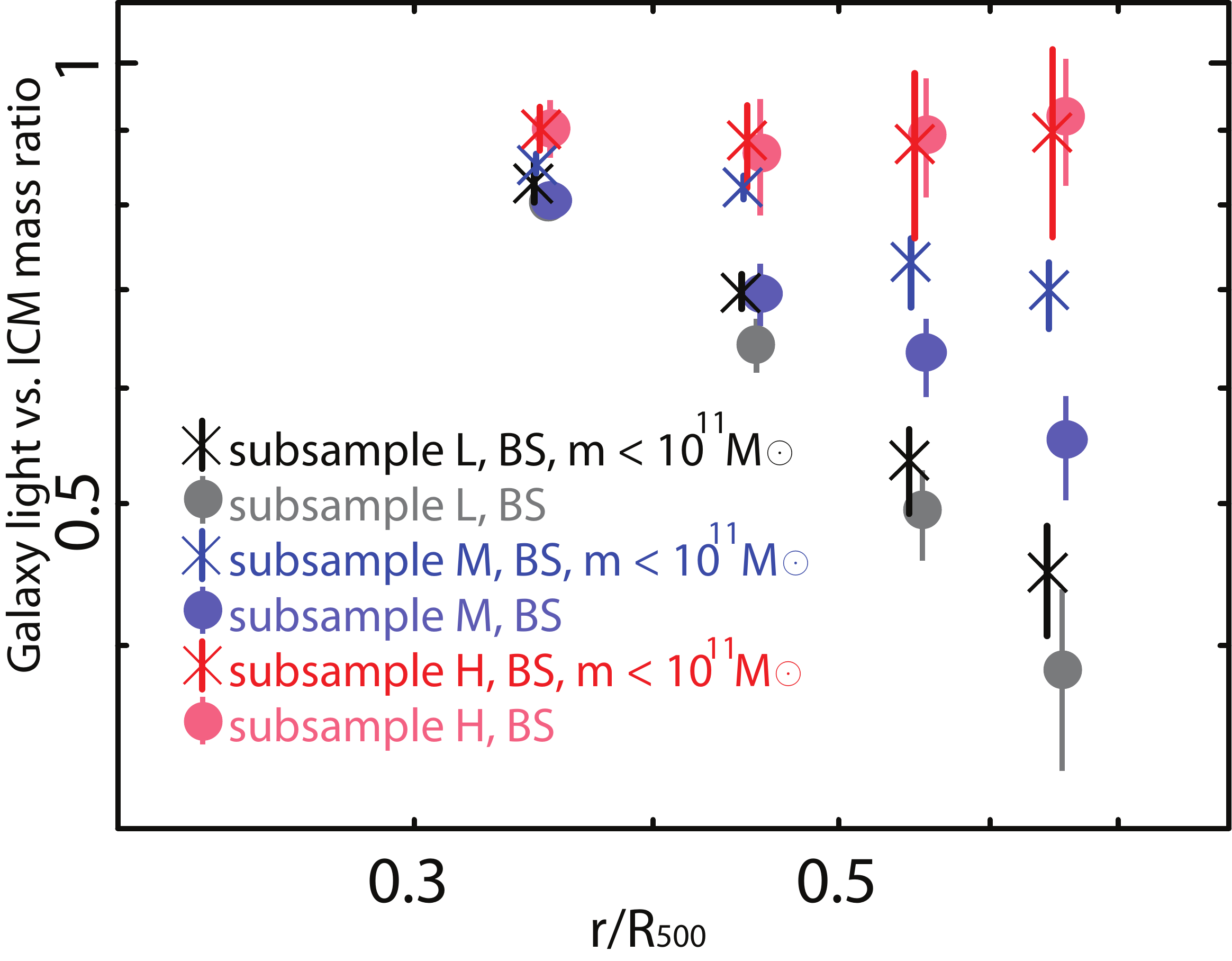}
\caption{The same as Figure 8, but calculated only using the low mass galaxies with
$m < 1 \times 10^{11}$ $M_{\rm \odot}$. The results before the mass discrimination are shown 
in light colors with filled circles.}
\end{center}
\end{figure}

\clearpage

\appendix
\section{Individual cluster profiles}
\setcounter{figure}{0} \renewcommand{\thefigure}{A.\arabic{figure}}
\begin{figure}
\begin{center}
\includegraphics[angle=-0,scale=.6]{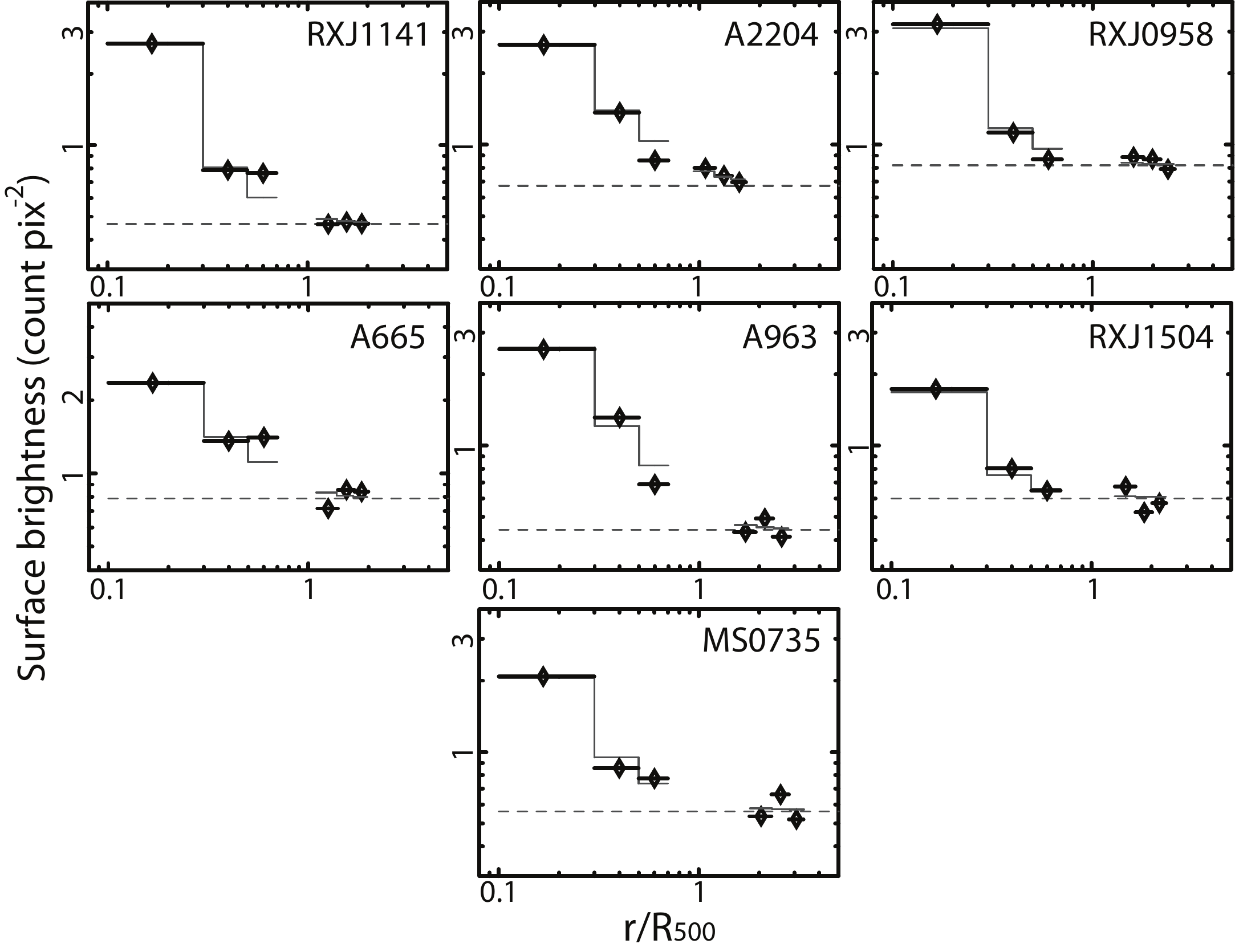}
\caption{Background-inclusive $I$- band surface brightness profiles of the L-subsample
  clusters, fitted with the king model plus constant background (solid
  line). The outermost three data points are from the offset pointings, while the rest are
 from the central pointing. The background values are shown in dashed line.  }
\end{center}
\end{figure}
\addtocounter{figure}{-1}

\begin{figure}
\begin{center}
\includegraphics[angle=-0,scale=.6]{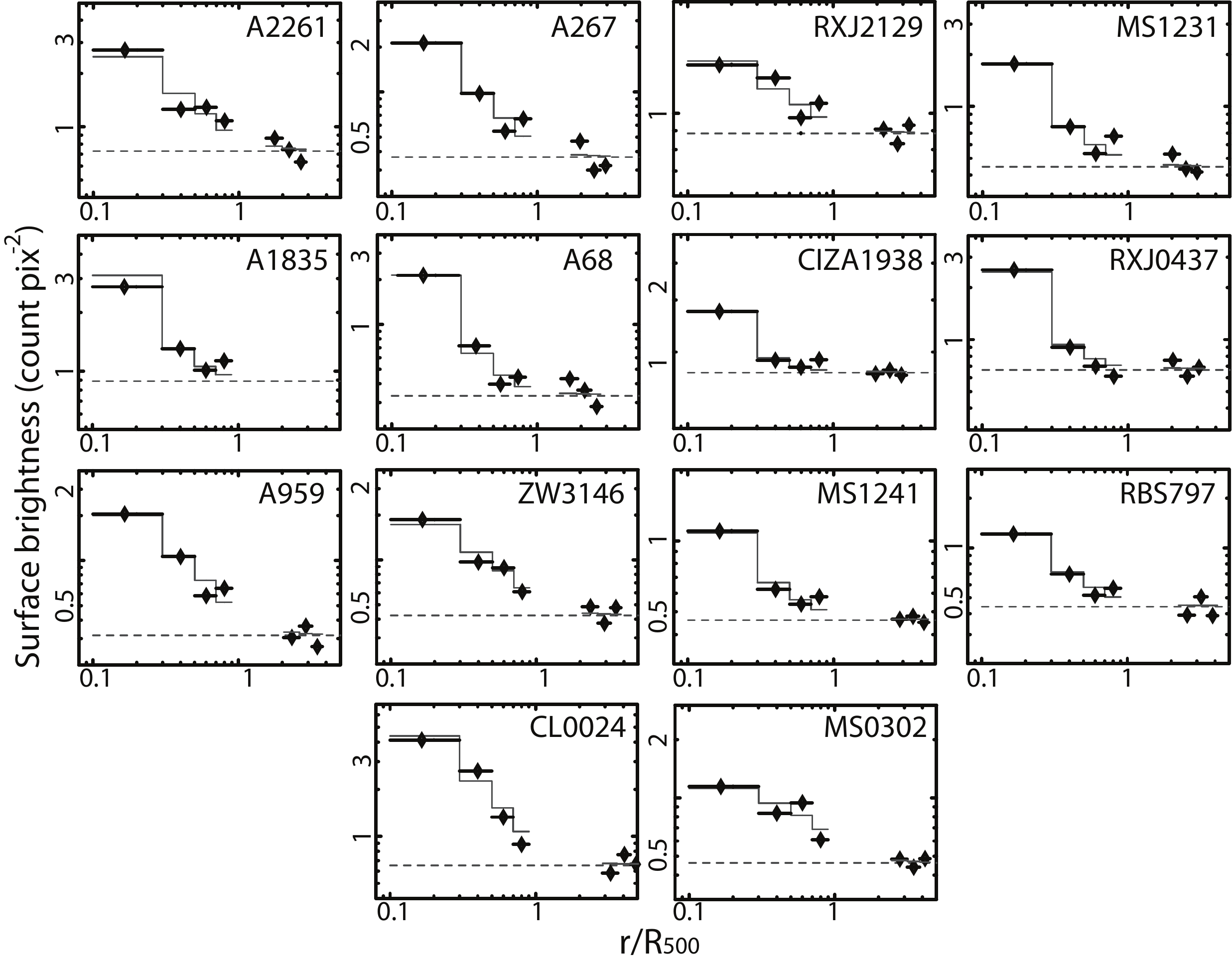}
\caption{({\it continue}) Background-inclusive $I$- band surface brightness profiles of the
  M-subsample
  clusters, fitted with the king model plus constant background (solid
  line). The outermost three data points are from the offset pointings, while the rest are
 from the central pointing. The background values are shown in dashed line.  }
\end{center}
\end{figure}
\addtocounter{figure}{-1}

\begin{figure}
\begin{center}
\includegraphics[angle=-0,scale=.6]{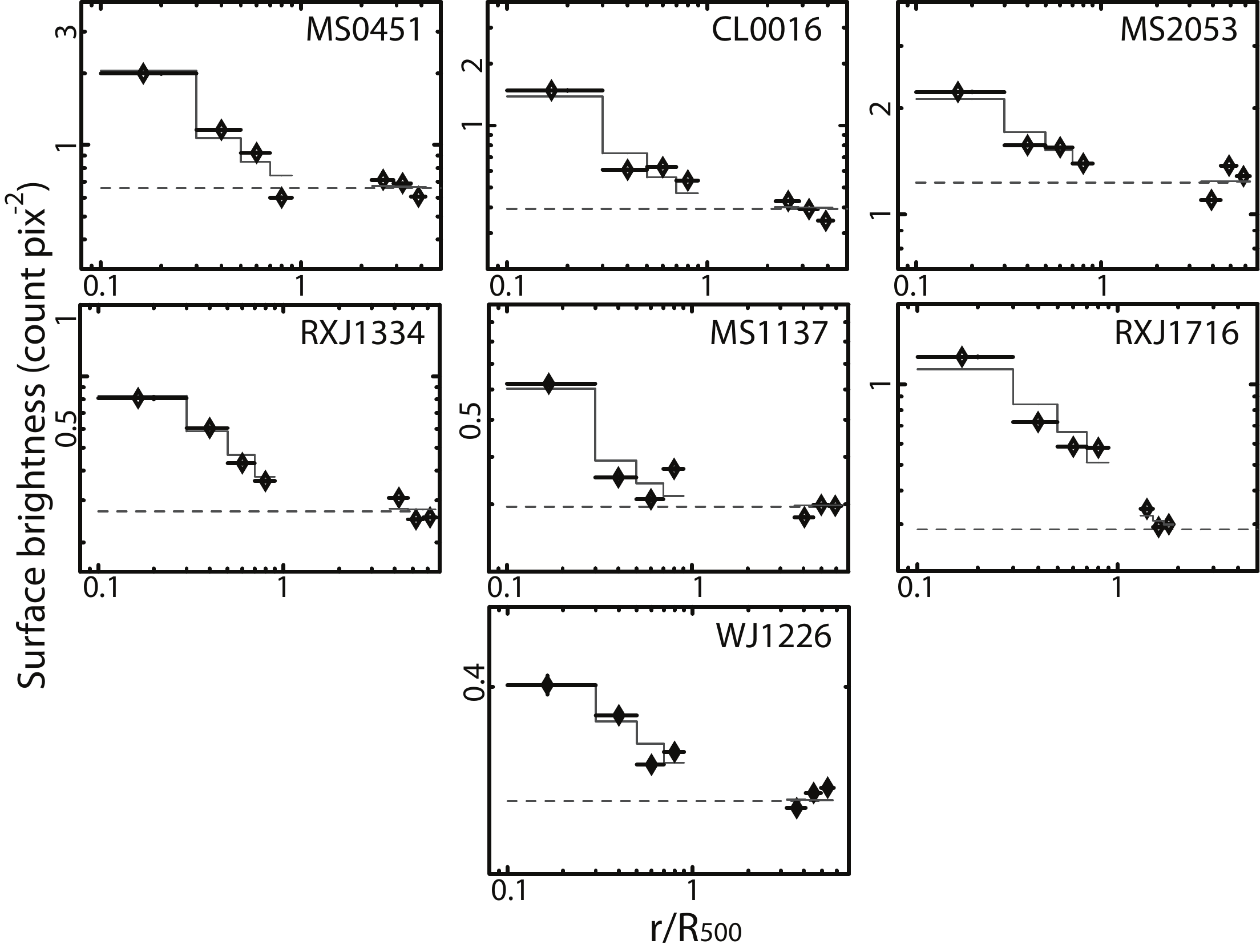}
\caption{({\it continue}) Background-inclusive $I$- band surface brightness profiles of the
  H-subsample
  clusters, fitted with the king model plus constant background (solid
  line). The outermost three data points are from the offset pointings, while the rest are
 from the central pointing. The background values are shown in dashed line.  }
\end{center}
\end{figure}

\begin{figure}
\begin{center}
\includegraphics[angle=-0,scale=.8]{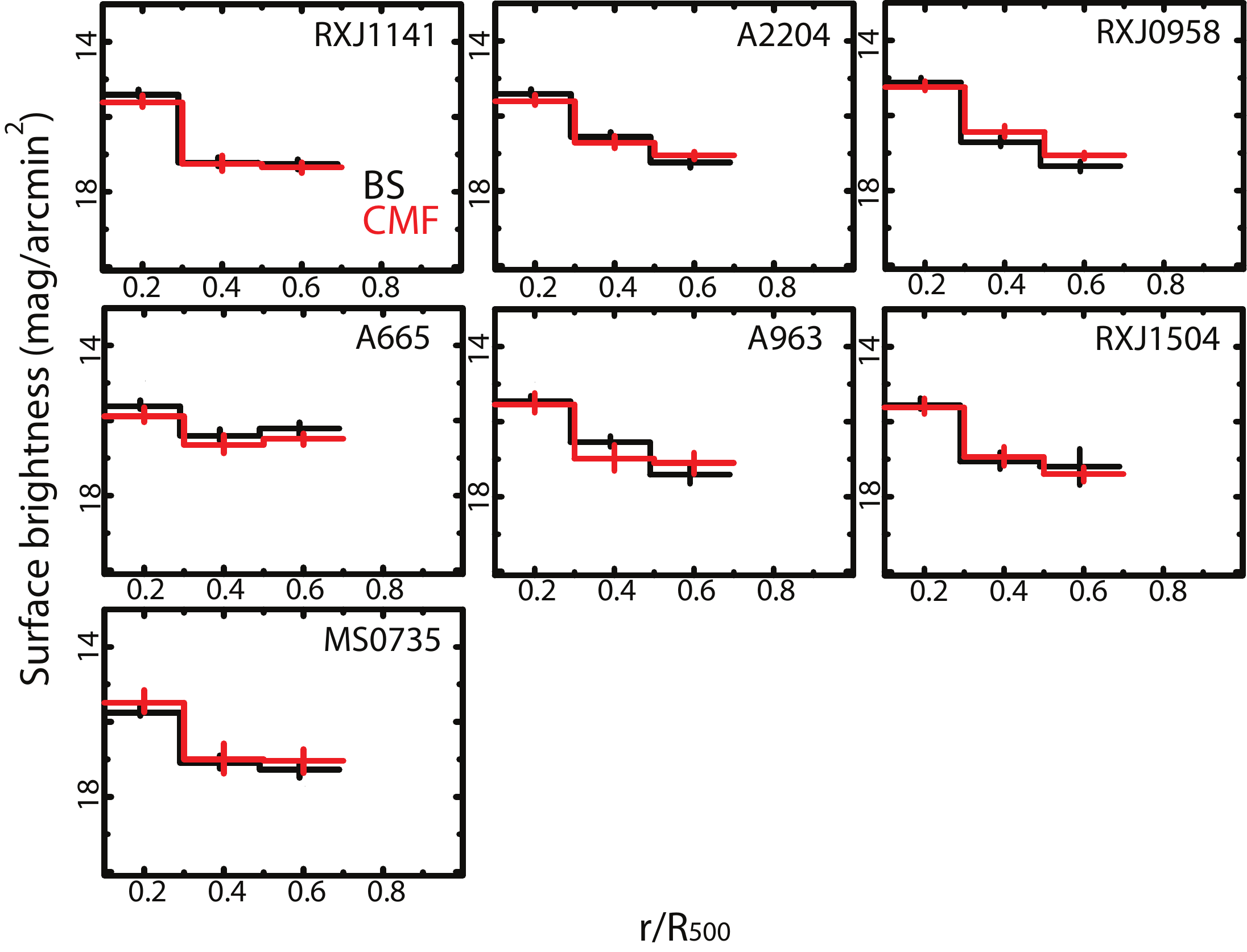}
\caption{$I$-band galaxy surface brightness profiles for the L-subsample clusters
  obtained with
  the BS (black) and CMF (red) methods. }
\end{center}
\end{figure}

\addtocounter{figure}{-1}
 \begin{figure}
\begin{center}
\includegraphics[angle=-0,scale=.6]{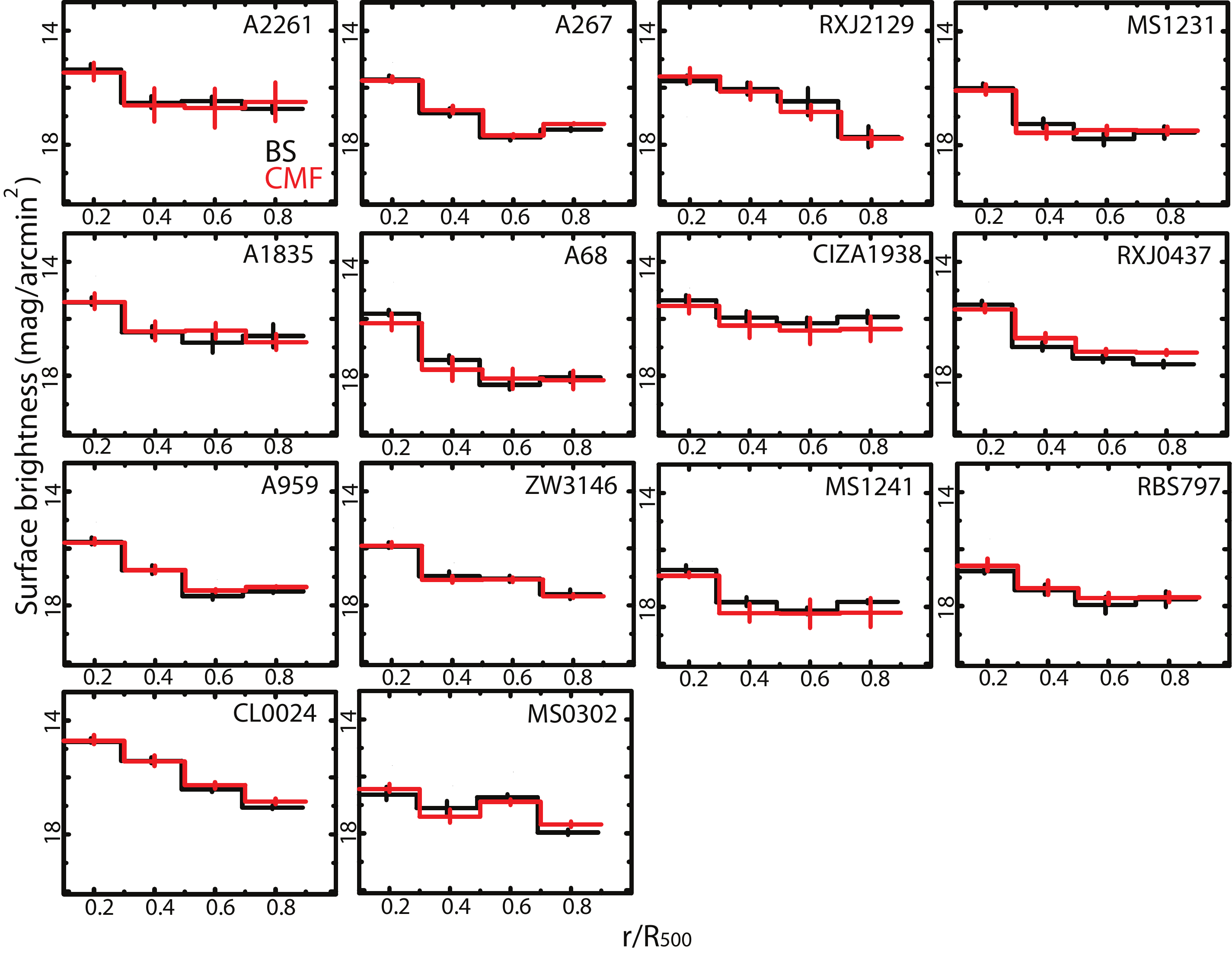}
\caption{({\it continue}) $I$-band galaxy surface brightness profiles
  for the M-subsample clusters
  obtained with
  the BS (black) and CMF (red) methods.}
\end{center}
\end{figure}

\addtocounter{figure}{-1}

\begin{figure}
\begin{center}
\includegraphics[angle=-0,scale=.8]{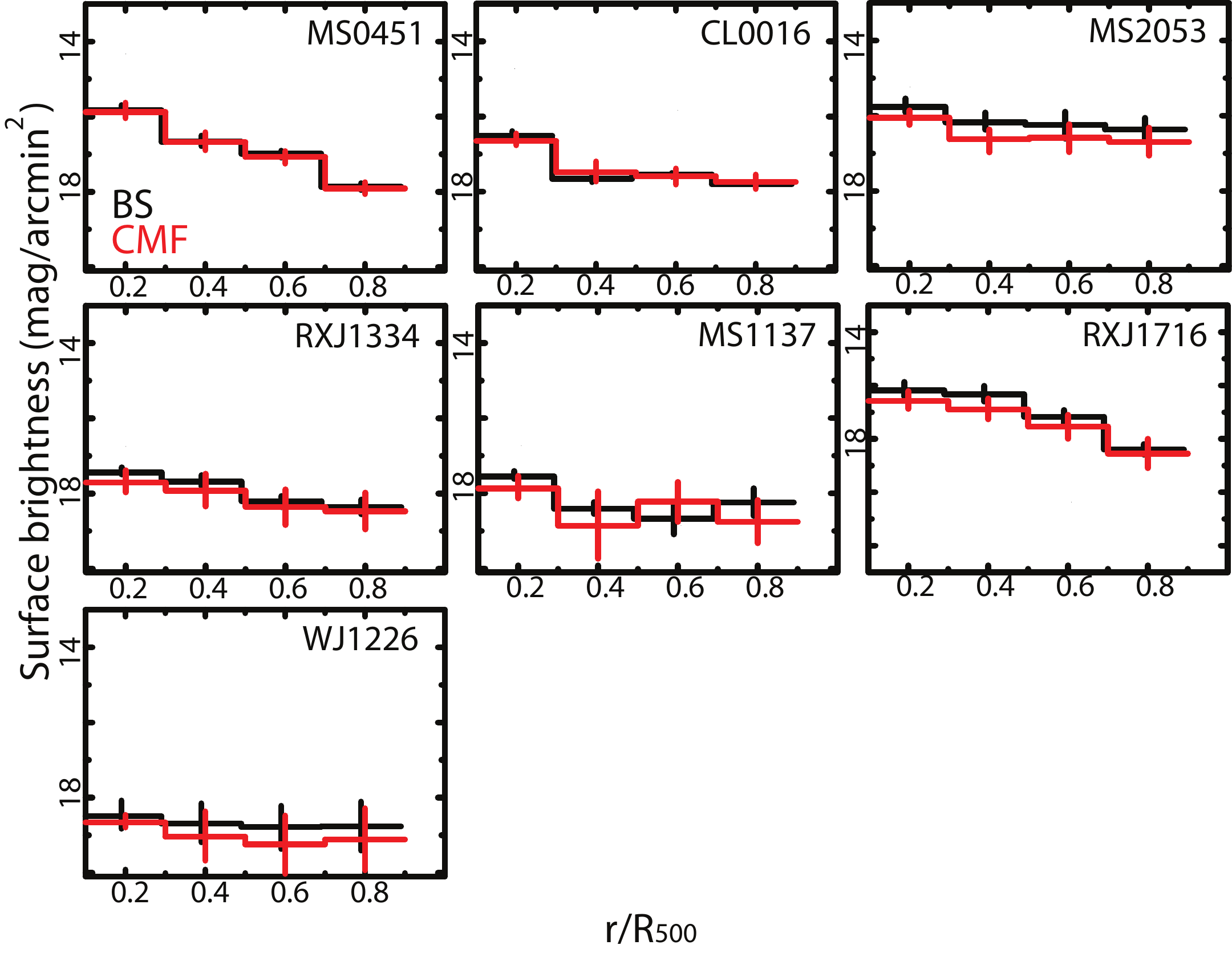}
\caption{({\it continue}) $I$-band galaxy surface brightness profiles
  for the H-subsample clusters
  obtained with
  the BS (black) and CMF (red) methods.}
\end{center}
\end{figure}

\begin{figure}
\begin{center}
\includegraphics[angle=-0,scale=.6]{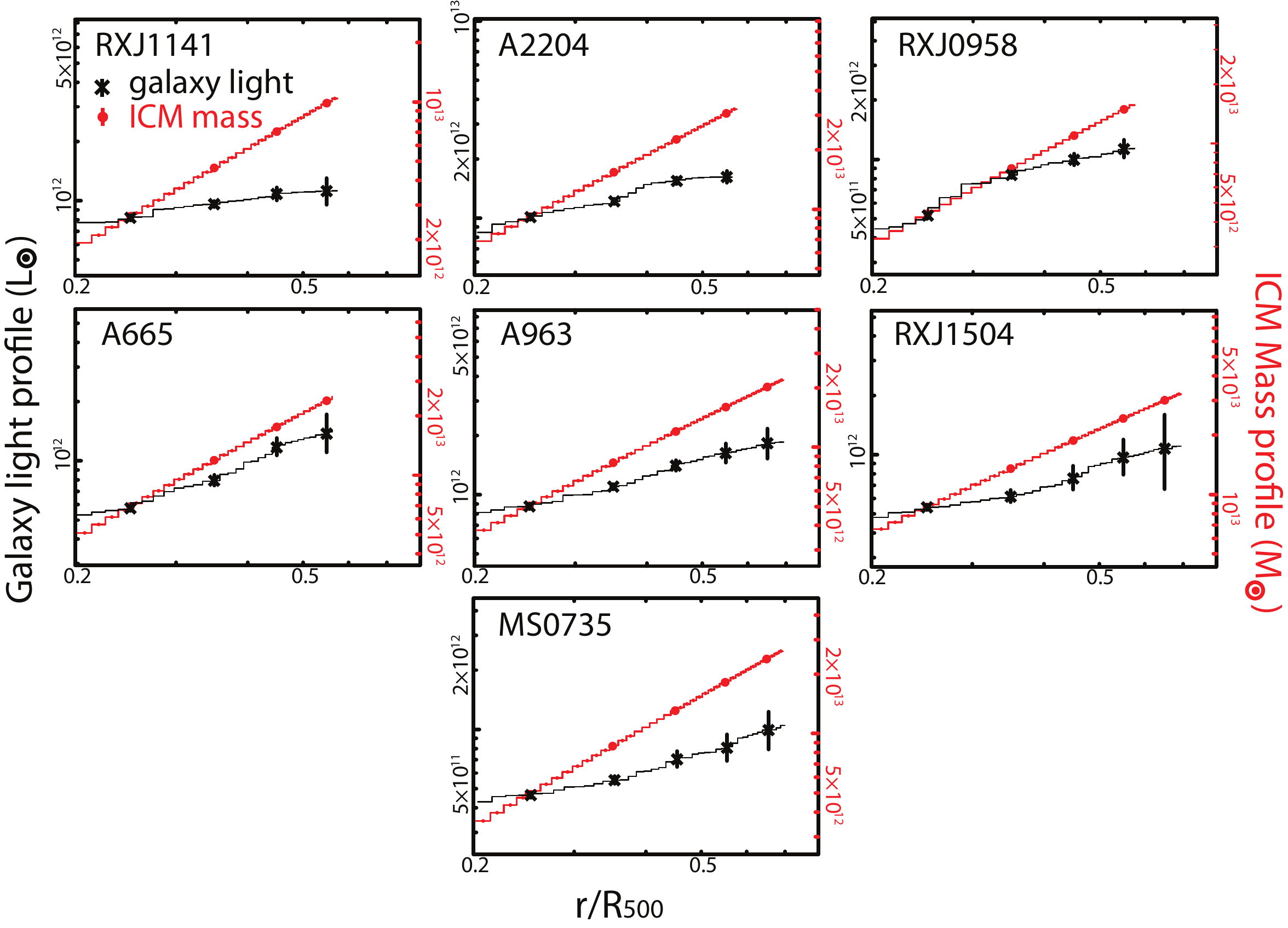}
\caption{2-D Optical light profiles obtained with the BS method (black; see text) of the L-subsample
clusters, compared with their 2-D ICM mass profiles (red). }
\end{center}
\end{figure}

\addtocounter{figure}{-1}

\begin{figure}
\begin{center}
\includegraphics[angle=-0,scale=.6]{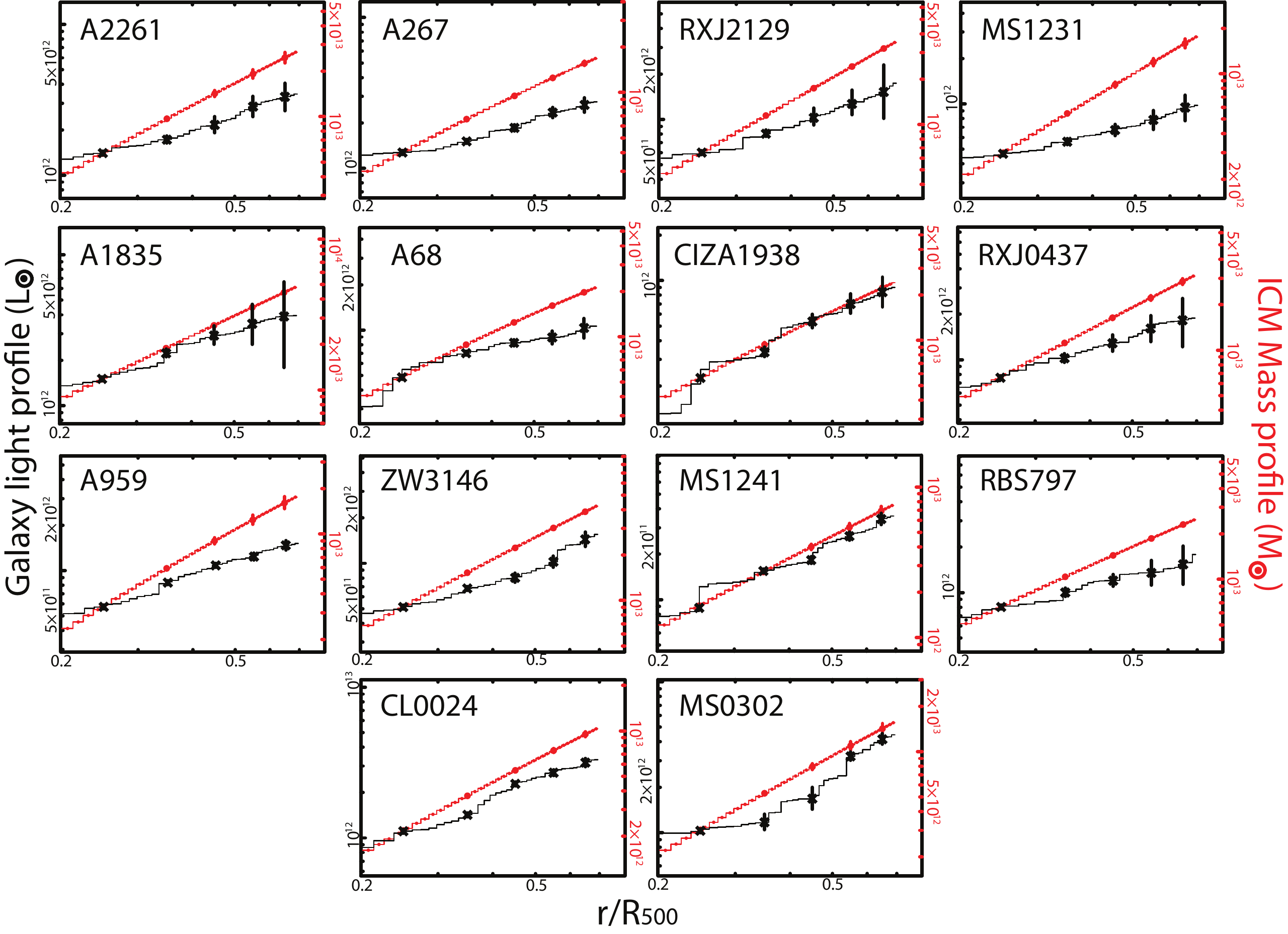}
\caption{({\it continue}) 2-D Optical light profiles obtained with the BS method (black; see text) of the M-subsample
clusters, compared with their 2-D ICM mass profiles (red).  }
\end{center}
\end{figure}

\addtocounter{figure}{-1}

\begin{figure}
\begin{center}
\includegraphics[angle=-0,scale=.6]{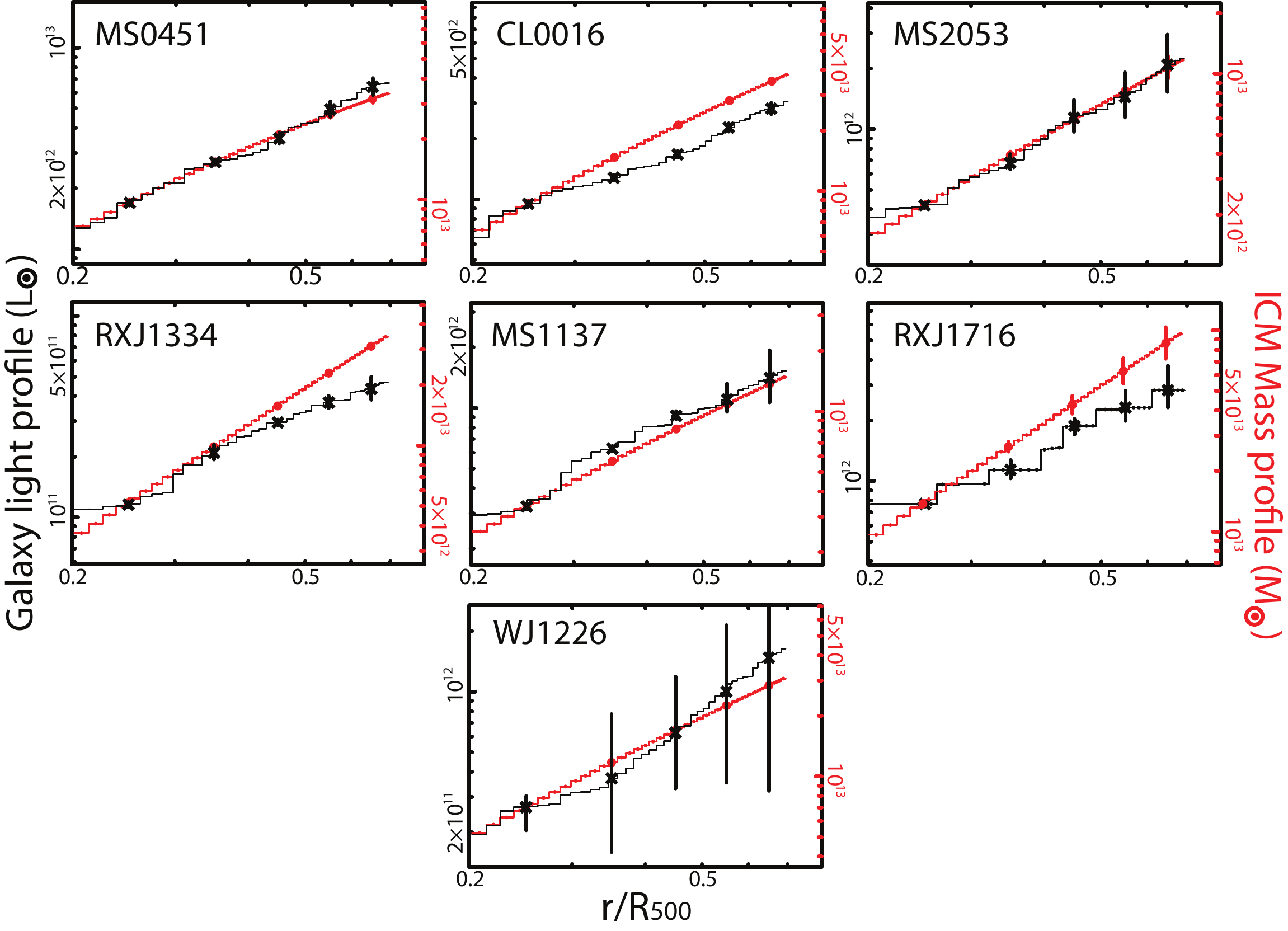}
\caption{({\it continue}) 2-D Optical light profiles obtained with the BS method (black; see text) of the H-subsample
clusters, compared with their 2-D ICM mass profiles (red).  }
\end{center}
\end{figure}

\begin{figure}
\begin{center}
\includegraphics[angle=-0,scale=.62]{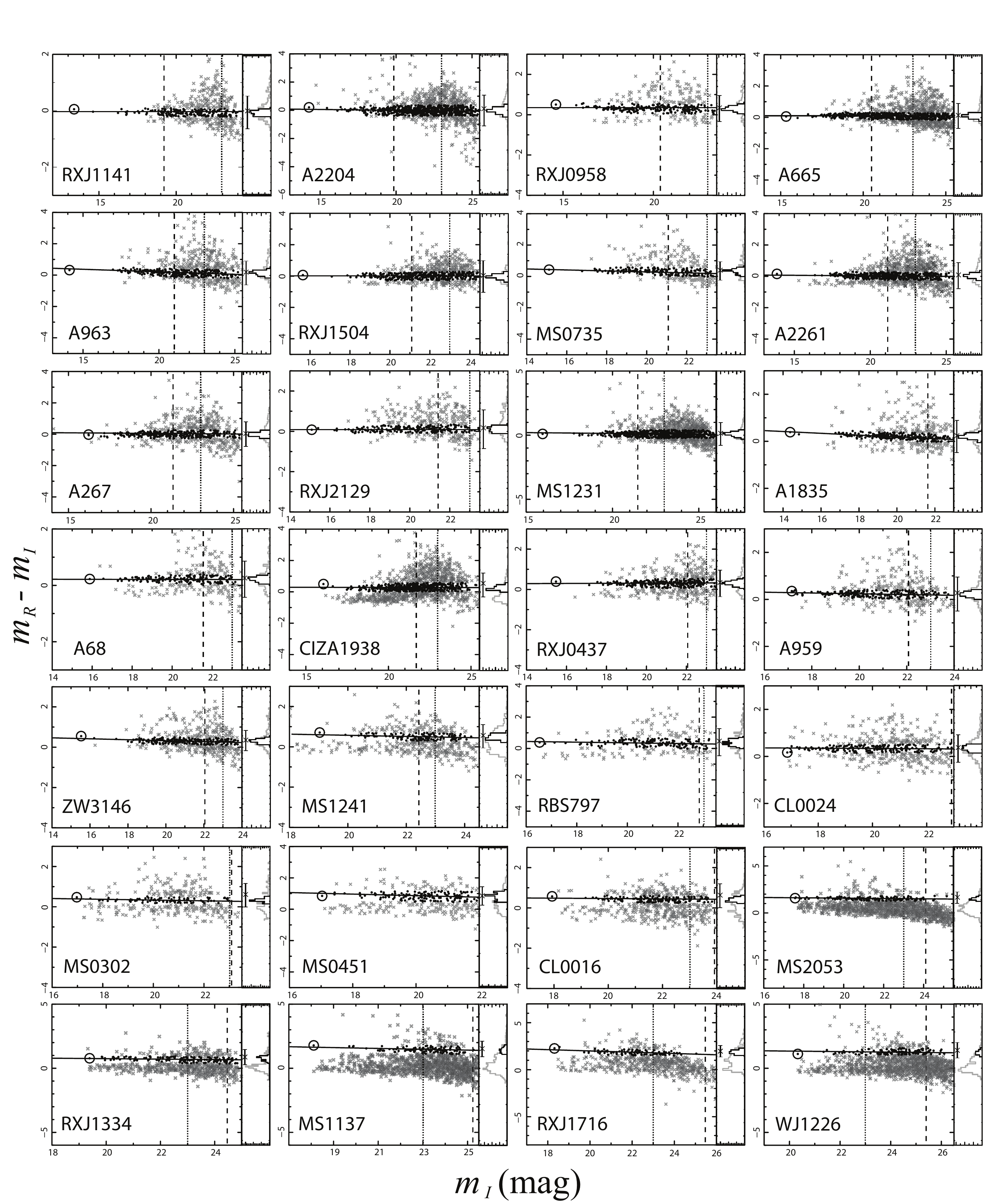}

\caption{Color-magnitude diagram of galaxies obtained within $\sim
  4^{\prime}$ region of each cluster. See the caption of Figure 5 for
details. }

\end{center}
\end{figure}

\begin{figure}
\begin{center}
\includegraphics[angle=-0,scale=.7]{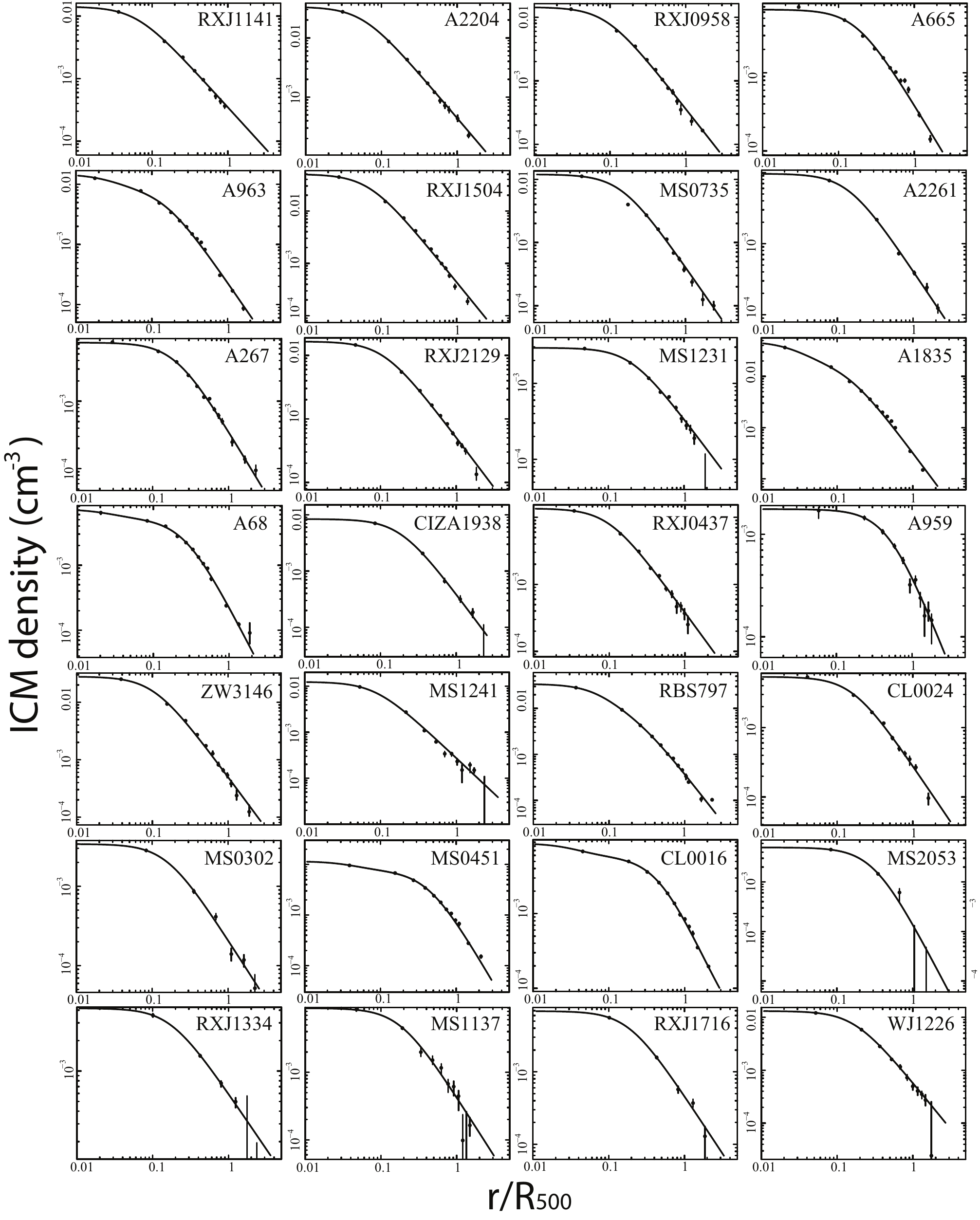}
\caption{ICM density profiles obtained with the {\it XMM-Newton} and {\it
    Chandra} data. See the caption of Figure 7 for details.}
\end{center}
\end{figure}

\end{document}